\documentstyle[epsfig,aps,axodraw_2,float,aps10]{revtex}

\begin{document}
\title{Quark Models of Baryon Masses and Decays}
\author{S. CAPSTICK}
\address{
Department of Physics, Florida State University, 
Tallahassee, FL,
32306-4350, USA
}
\author{W. ROBERTS}
\address{
National Science Foundation, 4201 Wilson Boulevard, Arlington, VA 22230\\
{\rm on leave from }\\
Department of Physics, Old Dominion University,
Norfolk, VA 23529, USA\\
{\rm and from}\\
Thomas Jefferson National Accelerator Facility, 12000
Jefferson Avenue, Newport News, VA 23606, USA}

\maketitle
\begin{abstract}
\begin{center}
{\bf \small Abstract}
\end{center}
The application of quark models to the spectra and strong and
electromagnetic couplings of baryons is reviewed. This review focuses
on calculations which attempt a global description of the masses and
decay properties of baryons, although recent developments in applying
large $N_c$ QCD and lattice QCD to the baryon spectrum are
described. After outlining the conventional one-gluon-exchange
picture, models which consider extensions to this approach are
contrasted with dynamical quark models based on Goldstone-boson
exchange and an algebraic collective-excitation approach. The spectra
and electromagnetic and strong couplings that result from these models
are compared with the quantities extracted from the data and each
other, and the impact of various model assumptions on these properties
is emphasized. Prospects for the resolution of the important issues
raised by these comparisons are discussed.
\end{abstract}

\def\tp0{^3P_0}
\def\beq{\begin{equation}}
\def\eeq{\end{equation}}
\def\beqa{\begin{eqnarray}}
\def\eeqa{\end{eqnarray}}
\def\nii{\noindent}
\def\veca#1{\mbox{\boldmath $#1$}}
\def\pbi{{\bf p}_i}
\def\pbj{{\bf p}_j}
\def\pbb{{\bf p}}
\def\Pbb{{\bf P}}
\def\qbb{{\bf q}}
\def\kbb{{\bf k}}
\def\Kbb{{\bf K}}
\def\yc{{\cal Y}}
\def\he{\hat \ell}
\def\hs{\hat S}
\def\hl{\hat L}
\def\hj{\hat J}
\def\rmb#1{{\bf #1}}
\def\lpmb#1{\mbox{\boldmath $#1$}}
\def\half{{\textstyle{1\over2}}}
\def\thalf{{\textstyle{3\over2}}}
\def\fhalf{{\textstyle{5\over2}}}
\def\shalf{{\textstyle{7\over2}}}
\def\nhalf{{\textstyle{9\over2}}}
\def\lhalf{{\textstyle{11\over2}}}
\def\tthalf{{\textstyle{13\over2}}}
\def\fthalf{{\textstyle{15\over2}}}
\def\ua{\uparrow}
\def\da{\downarrow}
\def\nn{\nonumber}
\def\>{\rangle}
\def\<{\langle}
\def\pr#1#2#3{ {Phys. Rev.\/} {\bf#1} (#2) #3}
\def\prl#1#2#3{ {Phys. Rev. Lett.\/} {\bf#1} (#2) #3}
\def\np#1#2#3{ {Nucl. Phys.\/} {\bf#1} (#2) #3}
\def\cmp#1#2#3{ {Comm. Math. Phys.\/} {\bf#1} (#2) #3}
\def\pl#1#2#3{ {Phys. Lett.\/} {\bf#1} (#2) #3}
\def\apj#1#2#3{ {Ap. J.\/} {\bf#1} (#2) #3}
\def\aop#1#2#3{ {Ann. Phy.\/} {\bf#1} (#2) #3}
\def\nc#1#2#3{ {Nuovo Cimento }{\bf#1} (#2) #3}
\def\cjp#1#2#3{ {Can. J. Phys. }{\bf#1} (#2) #3}
\def\zp#1#2#3{ {Z. Phys. }{\bf#1} (#2) #3}
\newcommand{\sfrac}[2]{\mbox{$\textstyle \frac{#1}{#2}$}}
\unitlength 0.6cm 
\thicklines

\tableofcontents
\parindent 0pt
\parskip 12pt
\section{Introduction}
\vskip -12pt
The quark model of hadron spectroscopy predated quantum chromodynamics
(QCD) by about twenty years and, in a very real sense, led to its
development. The puzzle surrounding the ground state decuplet led to a
number of postulates, including Greenberg's `parastatistics of order
three'~\cite{Greenberg:1964pe}, a precursor to the hypothesis of color
and, eventually, to the development of QCD. In addition, the early
quark model has led to what is sometimes regarded with some contempt,
the quark potential model. This model is an attempt to go beyond the
global symmetries of the original quark model, which was, in essence,
a symmetry-based classification scheme for the proliferating
hadrons. The potential model attempted to include dynamics, with the
dynamics arising from an inter-quark potential.

Despite the widespread acceptance of QCD as the theory of the strong
interaction, there is, as yet, no obviously successful way to go from
the QCD Lagrangian to a complete understanding of the multitude of
observed hadrons and their properties. Lattice QCD calculations offer
a bright promise, but calculations for anything but spectra and static
properties like magnetic moments are still a long way off, and even
for the spectra of excited states such calculations are just getting
under way. Other methods, such as large $N_c$ QCD, and effective field
theory approaches, also appear to be somewhat limited in their scope,
and would seem, by their very nature, to be unable to provide the
global yet detailed understanding that is necessary for strong
interaction physics. It is in this sense and in this context that the
quark potential model is currently an indispensable tool for guiding our
understanding. As an illustration, note that the very successful heavy
quark effective theory had its origins in a quark model calculation.

In its various forms, the quark potential model has had a large number
of successes. The spectra of mesons and baryons, as well as their
strong, weak and electromagnetic decays, have all been treated within the
framework of this model and the successes of this program can not be
dismissed as completely spurious. In addition, the predictions of the
model are always to be compared with expectations from QCD, in the
continuing effort to understand the intra-hadron dynamics of
non-perturbative QCD and confinement.

The study of baryon spectroscopy raises two very important global
questions that hint at the interplay between the quark model and
QCD. The first of these is the fact that, in many of its forms, the
quark model predicts a substantial number of `missing' light baryons
which have not so far been observed. There have been two possible
solutions postulated for this problem. One solution is that the dynamical
degrees of freedom used in the model, namely three valence quarks, are
not physically realized. Instead, a baryon consists of a quark and a
diquark, and the reduction of the number of internal degrees of
freedom leads to a more sparsely populated spectrum. Note, however,
that even this spectrum contains more states than observed. A second
possible solution is that the missing states couple weakly to the
formation channels used to investigate the states, and so give very
small contributions to the scattering cross sections. Investigation of
other formation channels should lead to the discovery of
some of these missing states in this scenario.

The second outstanding question is the fact that, in addition to the
usual three quark states, QCD predicts the existence of baryons with
`excited glue', the so-called hybrid baryons. Unlike mesons, all
half-integral spin and parity quantum numbers are allowed in the
baryon sector, so that experiments may not simply search for baryons
with exotic quantum numbers in order to identify such hybrid
states. Furthermore, no decay channels are {\it a priori}
forbidden. These two facts make identification of a baryonic hybrid
singularly difficult. If new baryon states are discovered at any of
the experimental facilities around the world, they can be interpreted
either as one of the missing baryons predicted by the quark model, or
as one of the undiscovered hybrid states predicted by QCD. Indeed,
some authors have suggested that a few of the known baryons could be
hybrid states.

These questions about baryon physics are fundamental. If no new
baryons are found, both QCD and the quark model will have made
incorrect predictions, and it would be necessary to correct the
misconceptions that led to these predictions. Current understanding of QCD
would have to be modified, and the dynamics within the quark model
would have to be changed. If a substantial number of new baryons are
found, it would be necessary to determine whether they were
three-quark states or hybrids or, as is likely, some admixture. It is
only by comparing the experimentally measured properties of these
states with model (or lattice QCD) predictions that this determination
can be made. As lattice calculations are only now beginning to address
the questions of light hadron spectroscopy in a serious fashion, the
quark model may be expected to play a vital role for many years to
come.

An excellent review article summarizing the situation for baryon
physics in the quark model in 1983 was written by Hey and
Kelly~\cite{Hey:1983aj}. This review will, therefore, concentrate on
the developments in the field which have taken place since then. It
will be necessary, however, to go into some detail about earlier
calculations in order to put modern work into perspective. For
example, the successes and failures of one-gluon-exchange based models
in describing baryon properties need to be understood before
examining alternative models. It is also not enough to describe one
aspect of baryon physics such as masses, or one type of baryon, since
there have existed for some time models which describe with reasonable
success the masses and strong and electromagnetic couplings of all
baryons. Furthermore, these models are equally useful in a description
of meson physics. Within the context of models, insight can only be gained
into nonperturbative QCD by comparing predictions to
experiment, in as many aspects as possible. Spectra pose only mild
tests of models; strong and electromagnetic decays and electromagnetic
form factors pose more stringent tests.

By far the most baryon states have been extracted from $\pi $N and
$\bar{K}N$ elastic and inelastic scattering data (24 and 22 well
established 3 and 4* states~\cite{Caso:1998tx} with spin assignments,
respectively), and so most of the available information on baryon
resonances is for nonstrange and strangeness $-1$ baryons. For this
reason this article will mainly concentrate on models which make
predictions for the static properties and the strong and
electromagnetic transitions of all of these states. Such models can
also be extended to multiply-strange baryons and heavy-quark baryons,
and the best of them are successful in predicting the masses and
magnetic moments of the ground states and the masses of the few
excited states of these baryons that have been seen. Models which
describe the spectrum in the absence of a strong decay model are of
limited usefulness. This is because the predicted spectrum must be
compared to the results of an experiment which examines excited
baryons using a specific formation and decay channel. If a model
predicts a baryon state which actually couples weakly to either or
both of these channels, this state cannot be identified with a
resonant state from analyses of that experiment.

Considerations of space do not allow a proper treatment here of the
electromagnetic transition form factors of excited nonstrange baryons,
or of the magnetic moments or other electromagnetic properties of the
nucleon and other ground states, even though intense theoretical and
experimental effort has been brought to bear on these issues. In
particular, it has been shown in a substantial literature that the
static properties of ground state baryons such as magnetic moments,
axial-vector couplings, and charge radii, are subject to large
corrections from relativistic effects and meson-loop couplings. This
is also true of the transition form factors. In order to avoid a
necessarily superficial treatment, the description of these quantities
in the various models of the spectrum dealt with here is not
discussed.
\vskip -12pt
\section{Ground state baryon masses}
\vskip -12pt
\subsection{mass formulae based on symmetry}
\vskip -12pt
The interactions of quarks in baryons depend weakly on the
masses of the quarks involved in the interaction, and depend very
weakly on their electric charges. The small size of the strange-light
quark mass difference compared to the confinement scale means that the
properties of baryon states which differ only by the interchange of
light and strange quarks are quite similar. The much smaller
difference between the masses of the up and down quarks means that the
properties of baryon states which differ only by the interchange of up
and down quarks are very similar. The further observation that the
properties of such states should be independent of an arbitrary
continuous rotation of each quark's flavor in the $(u,d,s)$ flavor
space led to the notion of the approximate SU(3)$_{\rm f}$ symmetry
and the much better isospin symmetry which is found in the spectrum of
the ground-state baryons. Although this continuous symmetry exists, of
course physical states are discrete with an integer number of quarks
of each flavor.

The most important explicit SU(3)$_{\rm f}$-symmetry breaking effect
is the strange-light quark mass difference; isospin-symmetry breaking
is small by comparison, and has roughly equally important
contributions from the up-down quark mass difference and from
electromagnetic interactions between the quarks. The charge averaged
masses of the octet of $J^P=\half^+$ ground state baryon masses
should, therefore, be related by the addition of $m_s-m_{u,d}$ for
each light quark replaced by a strange quark. This assumes that each
state has the same kinetic and potential energy, which is only
approximately true due to the dependence of their expectation values
on the quark masses. This observation leads to mass relations between
baryons which differ only by the addition of a strange quark, for
example the Gell-Mann--Okubo mass formula, which eliminates the
parameters $M_0$, $m_{u,d}$ and $m_s$ from the assumed masses
\beqa 
M_N&=&M_0+3m_{u,d}\nonumber \\
M_\Lambda=M_\Sigma&=&M_0+2m_{u,d}+m_s\nonumber \\
M_\Xi&=&M_0+m_{u,d}+2m_s \eeqa of the $J^P=\half^+$ baryons to find
\beq {M_\Sigma+3M_\Lambda\over 2}=M_N+M_\Xi.
\label{GMOk}
\eeq
With this simple argument any linear combination of $M_\Sigma$ and
$M_\Lambda$ should work; however, the particular linear combination
chosen here makes this relation good to about 1\%. This is based on
the assumption of a baryon mass term in an effective Lagrangian of the
form $\bar{\psi}(a+b\lambda_8)\psi$, where $\psi$ is a baryon field
which represents SU(3)$_{\rm f}$ and $\lambda_8$ is a Gell-Mann
matrix. This form contains only the SU(3)$_{\rm f}$ singlet and a
symmetry-breaking octet term. Using this assumption, called the octet
dominance rule, and the calculation~\cite{Itzykson:1980} of some
SU(3)$_{\rm f}$ recoupling constants for $8\otimes 8=1\oplus
8^\prime\oplus 8^{\prime\prime}$, one obtains
\beqa
M_N&=&M_1-2M_{8^\prime}+M_{8^{\prime\prime}}\nonumber \\
M_\Lambda&=&M_1-M_{8^\prime}-M_{8^{\prime\prime}}\nonumber \\
M_\Sigma&=&M_1+M_{8^\prime}+M_{8^{\prime\prime}}\nonumber \\
M_\Xi&=&M_1+M_{8^\prime}-2M_{8^{\prime\prime}}
\eeqa
which yields the desired relation.

An analysis of $J^P=\thalf^+$ ground-state decuplet baryon masses
taking into account only the quark mass differences yields Gell-Mann's
equal spacing rule
\beq
M_{\Sigma^*}-M_\Delta=M_{\Xi^*}-M_{\Sigma^*}=M_\Omega-M_{\Xi^*},
\label{decuplet}
\eeq
which compare well to the charge-averaged experimental mass splittings
of 153, 149 and 139 MeV respectively. The systematic
deviations can be attributed to a quark-mass-dependent hyperfine
interaction which splits these states from their $J^P=\half^+$
partners. This topic will be revisited below.

Okubo~\cite{Okubo:1962jc} and G\"ursey and
Radicati~\cite{Gursey:1964dc} derive a mass formula for the
ground-state baryons using matrix elements of a mass operator with
$J^P=0^+$, isospin, hypercharge ($Y=2[Q-I_3]$), and strangeness zero,
and with the assumptions that it contains only one and two-body
operators and octet dominance they find
\beq
M=a+bY+c\left[I(I+1)-Y^2/4\right]+dJ(J+1),
\label{GR}
\eeq
which contains the above relations for the $J^P=\half^+$ octet and
$\thalf^+$ decuplet. Equation~(\ref{GR}) also contains the SU(6)
relation~\cite{Beg:1964nm}
\beq
M_{\Sigma^*}-M_\Sigma=M_{\Xi^*}-M_\Xi
\label{BB*split}
\eeq
which takes into account energy differences due to the spin of the
states and their flavor structure. This compares well to the
experimental differences of 192 and 215 MeV, respectively. The
deviation from this rule can be explained by a quark-mass dependent
hyperfine interaction and wave function size.

An example of a mass relation which takes into account both the
approximate SU(3)$_{\rm f}$ and isospin symmetries of the strong
interaction is the Coleman-Glashow relation, which can be expressed as
a linear combination of $I=1$ mass splittings which should be
zero~\cite{Jenkins:1995td},
\beq
N_1+\Xi_1-\Sigma_1=0,
\eeq
where
\beqa
N_1&=&M_p-M_n\nonumber \\
\Sigma_1&=&M_{\Sigma^+}-M_{\Sigma^-}\nonumber \\
\Xi_1&=&M_{\Xi^0}-M_{\Xi^-}.
\eeqa
This relation holds to within experimental errors on the determination
of the masses. Jenkins and Lebed~\cite{Jenkins:1995td} analyze this
and many other ground-state baryon octet and decuplet isospin
splittings in a $1/N_c$ expansion which is combined with perturbative
flavor breaking. They find that this relation should receive
corrections at the level of $\epsilon \epsilon^\prime/N_c^2$, where
$\epsilon$ is a small SU(3)$_{\rm f}$ symmetry breaking, and where
$\epsilon^\prime$ is a much smaller isospin-symmetry breaking
parameter. This pattern of symmetry breaking is what is observed
experimentally. This relation is found to be preserved by the
explicitly isospin and flavor-symmetry breaking interactions in a
dynamical potential model to be described
later~\cite{Capstick:1999ab}. Jenkins and Lebed's work establishes
clear evidence for a $1/N_c$ hierarchy of the observed mass
splittings, which differs from that based on SU(6) symmetry.
\vskip -12pt
\subsection{QCD-based dynamical models}
\vskip -12pt
Mass-formula approaches based solely on symmetry are not able to
explain the sign of the $\Delta-N$ and $\Sigma-\Lambda$ splittings;
this requires a dynamical model. Development of QCD led to models of
these splittings based on one-gluon exchange, such as the
nonrelativistic potential model of De Rujula, Georgi, and Glashow
(DGG)~\cite{DeRujula:1975ge} and the MIT bag model of
Refs.~\cite{Chodos:1974je,Bardeen:1975wr}. The DGG model uses
unperturbed eigenstates which are SU(6) multiplets with unknown
masses, and then mass formulae are derived using the Fermi-Breit
interaction between colored quarks from one-gluon exchange. For ground
states the important interaction is the Fermi contact term, which
leads to ground-state baryon masses
\beq M=\sum_{i=1}^3 m_i + {2\alpha_s\over 3}{8\pi\over 3}\langle
\delta^3({\bf r})\rangle \sum_{i<j=1}^3{{\bf S}_i\cdot {\bf S}_j\over
m_im_j},
\label{OGEcontact} 
\eeq
where the $m_i$ are the ``constituent'' quark effective masses and
where, by exchange symmetry of the ground state spatial wave
functions, $\langle \delta^3({\bf r}_{ij})\rangle$ is the same for any
one of the relative coordinates ${\bf r}_{ij}={\bf r}_i-{\bf r}_j$. In
the limit of SU(3)$_f$ symmetry this expectation value will also be
the same for all ground-state baryons and so the coefficient of the
spin sum will be a constant. Taking this coefficient and the
constituent quark masses $m_u=m_d\not= m_s$ as three parameters, a
three parameter formula for the ground state masses is
obtained. Equations~(\ref{GMOk}) and~(\ref{BB*split}) follow to first
order in $(m_s-m_{u,d})/m_{u,d}$, and the interaction naturally
explains the sign of the decuplet-octet mass difference, since the
decuplet states have all three quark spins aligned and so have higher
energy with this interaction. The size of the splitting is related to
the QCD coupling constant $\alpha_s$, albeit at low values of $Q^2$
where it need not be small.

The mass dependence of the spin-spin interaction above leads to the
relation
\beq
{M_\Sigma-M_\Lambda\over M_\Delta-M_N}={2\over
3}\left(1-{m_{u,d}\over m_s}\right)
\label{SigLam}
\eeq
between the $\Sigma-\Lambda$ and $\Delta-N$ mass splittings. With the
rough values of the effective quark masses $m_s\simeq 550$ MeV and
$m_{u,d}\simeq$ 330 MeV implied by a naive additive nonrelativistic
quark model fit to the size of the ground-state octet magnetic
moments, an apparently very good explanation of the relative size of
these two splittings is obtained. This is, however, subject to
substantial corrections. Distortions of the wave functions due to the
heavier strange quark mass increase the size of the contact
interactions between the strange quark and the light quarks, which
tends to compensate for the lowered size of the interaction due to the
mass factor in Eq.~(\ref{OGEcontact}), lowering the $\Sigma-\Lambda$
splitting~\cite{LeYaouanc:1978bf,Isgur:1979be} by about 30 MeV.
Equation~(\ref{SigLam}) also holds only in first-order (wave function)
perturbation theory, and in the calculation of
Ref.~\cite{Isgur:1979be} second-order effects due to a stronger mixing
with N=2 levels in the $\Sigma$ than the $\Lambda$ open up the
splitting again by about the same amount, so that this relation
happens to survive these corrections.

The bag model calculation of DeGrand, Jaffe, Johnson, and
Kiskis~\cite{DeGrand:1975cf,DeGrand:1976kx} is relativistic, and
confines the quarks to the interior of hadrons by a bag pressure term
with a parameter $B$ used to set the scale of the baryon masses. The
consequence of this confinement is that the equations of motion inside
the bag are supplemented by a homogeneous boundary condition which
eliminates quark currents across the bag surface, and a quadratic
boundary condition that balances the pressure from the quarks and
gluons with that of the bag locally on the surface. The quarks
interact among themselves relatively weakly by gluon exchange which is
treated in lowest order, and which gives a short-distance interaction
between the quarks of the same kind as that of DGG. The strange quark
mass is larger than the light-quark mass, to which the spectrum is
rather insensitive and which can be taken to be zero. As a
consequence, there is flavor dependence to the short-distance
interaction which explains the sign of the $\Sigma-\Lambda$ splitting,
although the calculated size is about half of the observed
value. Assuming that the bag radius parameter $R$ is the same for each
state also leads to the Gell-Mann-Okubo and equal spacing relations in
Eq.~(\ref{GMOk}) and Eq.~(\ref{BB*split}). The model has four
parameters, which are $B$, the quark-gluon coupling constant
$\alpha_c$, the strange quark mass, and a parameter $Z_0$ associated
with the zero-point energy for quantum modes in the bag. The cavity
radius $R$ is fixed by the quadratic boundary condition for each
hadron considered. A reasonable fit to ground state masses and static
properties is attained.

The authors of Refs.~\cite{DeGrand:1975cf,DeGrand:1976kx} point out
that in their model the momentum of a 300 MeV constituent quark is
large, of the order of 500 MeV, and conclude that the nonrelativistic
models which do not properly take into account the quark kinetic energy are
inconsistent. They also point out that the quarks are confined by a
dynamical mechanism which must carry energy (in their case it is the
bag energy), and that a prescription for reconciling the confining
potential locally with relativity is necessary. On the other hand, it
is difficult to ensure that the center of mass of a hadron state is
not moving when the dynamics of each of the constituents is the
solution of a one-body Dirac equation. Such spurious states are
therefore mixed in with the states of interest and will affect their
masses and properties. Nonrelativistic models have no such
difficulty. Some of these points will be revisited below.

Other models of ground state baryons include the cloudy bag model of
Th\'eberge, Thomas and Miller~\cite{Theberge:1980ye}, which
incorporates chiral invariance by allowing a cloud of pion fields to
couple to the confined quarks only at the surface of the MIT bag, and
is used to describe the pion-nucleon scattering cross section in the
$P_{33}$ partial wave in the region of the $\Delta$ resonance. The
result is that both pion-nucleon scattering (Chew-Low) diagrams and an
elementary $\Delta$ diagram contribute, with about 20\% and 80\%
contributions to the $\pi N$ resonance strength, respectively.

Ground-state baryon masses have also been examined in the chiral
perturbation theory~\cite{Jenkins:1992ts} and in large $N_c$
QCD~\cite{Jenkins:1993zu} by Jenkins. The former work shows that the
Gell-Mann-Okubo relation Eq.~(\ref{GMOk}) and Gell-Mann's equal
spacing rule Eq.~(\ref{decuplet}) are consistent with chiral
perturbation theory up to incalculable corrections of order the
strange quark current mass squared, or of order 20 MeV. Both
relations hold to better than this accuracy experimentally. The latter
work shows that baryon hyperfine mass splittings, which were known from
previous work to first appear at order $1/N_c$, are identical to those
produced by the operator ${\bf J}^2$, which is consistent with the
Gursey-Radicati formula Eq.~(\ref{GR}) and with the Skyrme
model\cite{Adkins:1983ya}. The masses of the nucleon, $\Delta$, and
the lowest-lying $D_{13}$ $(J^P=3/2^-)$ $N^*$ resonance are also calculated
with QCD sum rules by Belyaev and Ioffe~\cite{Belyaev:1982sa}, which
leads to a 10-15\% overestimate with the parameters used there.

Recent progress in lattice calculations with improved quark and gluon
actions and inevitable improvements in computer speed have led to
results for $N$, $\Delta$, and vector meson masses for full
(unquenched) QCD with two dynamical quark
flavors~\cite{Allton:1999gi,Aoki:2000yr}. The results are mainly in a
dynamical regime where the pion mass is above 500 MeV, with one recent
calculation on a larger lattice at a lower (300-400 MeV) pion
mass. These results therefore require extrapolation to the physical
pion mass, which Leinweber, Thomas, Tsushima, and
Wright~\cite{Leinweber:2000ig} show is complicated by non-linearity in
the pion mass squared due to the $N\pi$ threshold, and by non-analytic
behavior associated with dynamical chiral symmetry breaking.  They
examine the corrections to the extrapolated light baryon masses from
the pion-induced self-energies implied by chiral perturbation
theory. They defer making conclusions about the agreement between
extrapolated lattice results and experiment until systematic errors in
the full QCD calculations are reduced from the current 10\% level and
lattice measurements are made at lower pion masses.

\begin{figure}[t]
\vskip 0cm
\vbox{
\hskip 1.0cm
\epsfig{file=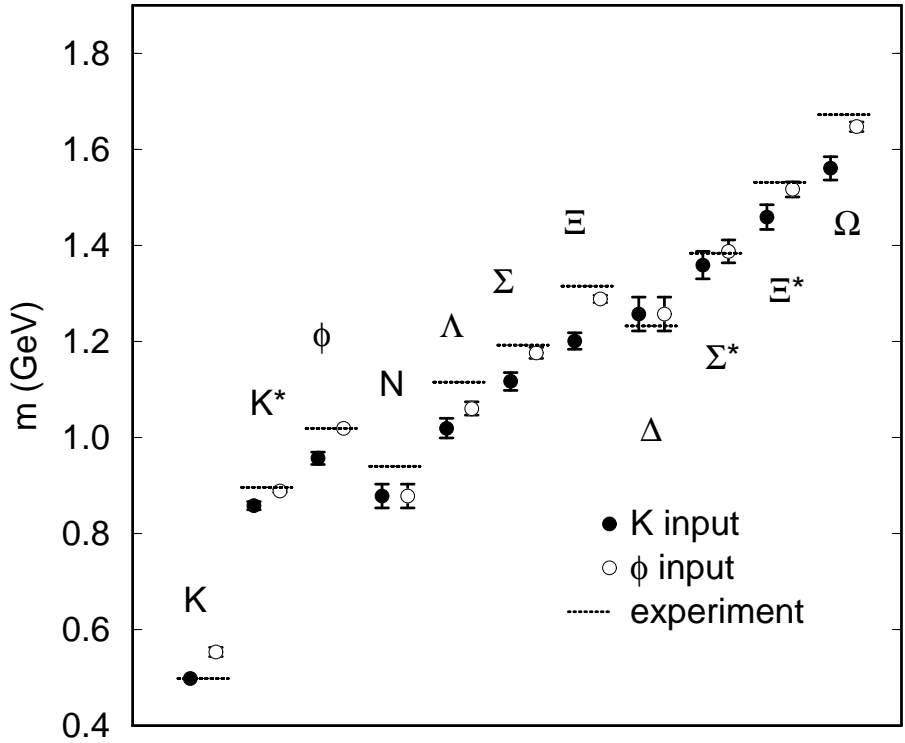,width=12cm,angle=0}}
\vskip 1.0 cm
\caption{Results of CP-PACS simulation of the flavor 
non-singlet light hadron spectrum in quenched lattice QCD with Wilson 
quark action.}
\label{lattice}
\end{figure}

As an example of quenched lattice results, where the reaction
of dynamical sea quarks is turned off, the CP-PACS
collaboration~\cite{Aoki:2000yr} spectrum is shown in
Figure~\ref{lattice}. 
Although the quenched results show systematic deviations
from the experimental masses, the spectrum shows a good pattern of
SU(3)$_f$ breaking mass splittings, and if the strange-quark mass is
fixed by fitting to the $\phi$ rather than the kaon mass, baryon
masses are fit to an error of at most -6.6\% (the nucleon mass).
Lattice QCD calculations should eventually be able to describe the
properties of the lowest-lying baryon states of any flavor of baryon
with small angular momentum, such as the P-wave baryons. However, for
technical reasons it may remain difficult to extract signals for the
masses of excited states beyond the second recurrence of a particular
set of flavor, spin, and parity quantum numbers, and for high angular
momentum states. It is, therefore, likely that models of the kind
described below will remain useful for the description of excited
baryons.
\vskip -12pt
\section{Excited baryon masses in the nonrelativistic model}
\vskip -12pt
\subsection{negative-parity excited baryons in the Isgur-Karl model}
\vskip -12pt
Many studies of excited baryon states were made based on the
observations of De Rujula, Georgi, and Glashow~\cite{DeRujula:1975ge};
the most detailed and phenomenologically successful of these is the
model of Isgur and Karl and their
collaborators~\cite{Isgur:1977ef,Isgur:1978xb,Isgur:1978xi,Isgur:1978xj,Isgur:1979wd,Copley:1979wj,Isgur:1979be,Isgur:1980ee,Chao:1981em,Isgur:1982yz}. 
This and other potential models have implicit assumptions which should
be made clear here. If one probes the proton with an electron which
transfers an amount of energy and momentum modest compared to the
confinement scale, then it may be useful to describe the proton in
this `soft' region as being made up of three constituent quarks. In
such models the light ($u$ and $d$) constituent (effective) quarks
have masses of roughly 200 to 350 MeV, and can be thought of as
extended objects, {\it i.e.} `dressed' valence quarks. Strange quarks
are about 150-200 MeV heavier. These are not the partons required to
describe deep inelastic scattering from the proton; their interactions
are described in an effective model which is not QCD but which is
motivated by it.

It is also assumed that the gluon fields affect the quark dynamics by
providing a confining potential in which the quarks move, which is
effectively pair-wise linear at large separation of the quarks, and at
short distance one-gluon exchange provides a Coulomb potential and the
important spin-dependent potential. Otherwise the effects of the gluon
dynamics on the quark motion are neglected. This model will obviously
have a limited applicability: to `soft' (low-$Q^2$ or coarse-grid)
aspects of hadron structure, and to low-mass hadrons where gluonic
excitation is unlikely. Similarly, since only three quarks components
have been allowed in a baryon [other Fock-space components like
$qqq(q\bar{q})$ have been neglected], it will strictly only be
applicable to hadrons where large mass shifts from couplings to decay
channels are not expected. Recent progress
in understanding these mass shifts, and the effects of including the
dynamics of the glue, will be described below.

Isgur and Karl solve the Schr\"odinger equation $H\Psi=E\Psi$ for the
three valence-quark system baryon energies and wave functions, with a
Hamiltonian
\begin{equation}
H=\sum_i \left( m_i+{\rmb{p}_i^2 \over 2m_i} \right) 
+ \sum_{i<j} \left( V^{ij}+H^{ij}_{\rm hyp} \right),
\label{IKHam}
\end{equation}
where the spin-independent potential $V^{ij}$ has the form
$V^{ij}=C_{qqq}+br_{ij}-2\alpha_s/3r_{ij}$, with $r_{ij}=\vert
\rmb{r}_i-\rmb{r}_j \vert$. In practice, $V^{ij}$ is written as a
harmonic-oscillator potential $Kr_{ij}^2/2$ plus an
unspecified anharmonicity $U_{ij}$, which is treated as a perturbation.
The hyperfine interaction $H^{ij}_{\rm hyp}$ is the sum
\begin{equation}
H^{ij}_{\rm hyp}={2\alpha_s\over 3 m_i m_j} \left\{
{8\pi\over 3} \rmb{S}_i\cdot\rmb{S}_j\delta^3(\rmb{r}_{ij})
+{1\over r_{ij}^3}
\left[ {3(\rmb{S}_i\cdot\rmb{r}_{ij})(\rmb{S}_j\cdot\rmb{r}_{ij})
         \over r_{ij}^2} - \rmb{S}_i\cdot\rmb{S}_j \right]
\right\}
\label{Hhyp}
\end{equation}
of contact and tensor terms arising from the color magnetic
dipole-magnetic dipole interaction. In this model the spin-orbit
forces which arise from one-gluon exchange and from Thomas precession
of the quark spins in the confining potential are deliberately
neglected; their inclusion spoils the agreement with the spectrum,
since the resulting splittings tend to be too large. The relative
strengths of the contact and tensor terms are as determined from the
Breit-Fermi limit [the expansion to O$(p^2/m^2)$] of the one-gluon
exchange potential. Note that there are also spin-independent,
momentum-dependent terms in the expansion to this order, such as
Darwin and orbit-orbit interaction terms, which are neglected by
Isgur and Karl.

Baryon states are written as the product of a color wave function $C_A$
which is totally antisymmetric under the exchange group $S_3$, and a
sum $\sum \psi \chi \phi$, where $\psi$, $\chi$, and $\phi$ are the
spatial, spin, and flavor wave functions of the quarks
respectively. The spatial wave functions are chosen to represent $S_3$
in the case of baryons with equal-mass quarks, and the spin
wave functions automatically do so.  The sum is constructed so that it
is symmetric under exchange of equal mass quarks, and also implicitly
includes Clebsch-Gordan coefficients for coupling the quark orbital
angular momentum $\rmb L$ with the total quark spin $\rmb S$.

$S_3$ has three irreducible representations. The two one-dimensional
representations are the antisymmetric (denoted $A$), and symmetric
($S$) representations. The two-dimensional mixed-symmetry ($M$)
representation has states $M^\rho$ and $M^\lambda$ which transform
among themselves under exchange transformations. For example,
$(12)M^\rho=-M^\rho$, $(12)M^\lambda=M^\lambda$, and
$(13)M^\rho=M^\rho/2 - \sqrt{3} M^\lambda/2$, $(13)M^\lambda=-\sqrt{3}
M^\rho/2 - M^\lambda/2$. The spin wave functions $\chi$ found from
coupling three spins-$\half$ are an example of such a representation,
\begin{equation}
\half\otimes\half\otimes\half=(1\oplus 0)\otimes \half=(\thalf
\oplus[\half]_\rho)\oplus[\half]_\lambda.
\end{equation}
The spin-$\thalf$ wave function $\chi^S$ is totally symmetric, and the
two spin-$\half$ wave functions $\chi^{M_\lambda}$ (from $1\otimes
\half$) and $\chi^{M_\rho}$ (from $0\otimes \half$) form a
mixed-symmetry pair,
\begin{eqnarray}
\chi^S_{\thalf,\thalf}&=&|\ua\ua\ua\,>,\ \
\chi^{\rho}_{\half,\half}
={1\over \sqrt{2}}
\left\{|\ua\da\ua\,>-|\da\ua\ua\,> \right\}, \nonumber\\
\chi^{\lambda}_{\half,\half}&=&-{1\over \sqrt{6}}
\left\{|\da\ua\ua\,>+|\ua\da\ua\,>-2|\ua\ua\da\,> \right\},
\label{spin}
\end{eqnarray}
where the subscripts are the total spin and its projection, and other
projections can be found by applying a lowering operator.

The construction of the flavor wave functions $\phi$ for nonstrange
states proceeds exactly analogously to that of the spin wave functions,
using isospin. They either have mixed symmetry ($N, I=\half$) or are
totally symmetric ($\Delta, I=\thalf$),
\begin{eqnarray}
\phi^S_{\Delta^{++}}&=&uuu,\ \ 
\phi^S_{\Delta^{+}}={1\over \sqrt{3}}\left\{uud+udu+
duu \right\},\ {\rm etc.}, \nonumber\\
\phi^{M_\rho}_{p}&=&{1\over \sqrt{2}}
\left\{udu-duu \right\},\ \ \phi^{M_\rho}_{n}={1\over \sqrt{2}}
\left\{udd-dud \right\}, \nonumber\\
\phi^{M_\lambda}_{p}&=&-{1\over \sqrt{6}}
\left\{duu+udu-2\,uud \right\},\ \ \phi^{M_\lambda}_{n}={1\over \sqrt{6}}
\left\{udd+dud-2\,ddu \right\}.
\label{NDphi}
\end{eqnarray}
For strange (and heavy-quark) baryons it is advantageous to use a
basis which makes explicit SU(3)$_f$ breaking and so does not
antisymmetrize under exchange of the unequal mass quarks. Since the
mass differences $m_s-(m_u+m_d)/2$ (and $m_c-[m_u+m_d]/2$, etc.) are
substantial on the scale of the average quark momentum, there are
substantial SU(3)$_f$ breaking differences between, for example, the
spatial wave functions of the ground state of the nucleon and that of
the ground state $\Lambda$. For such states the `$uds$ basis' is used,
which imposes symmetry only under exchange of equal mass quarks, and
adopt the flavor wave functions
\begin{eqnarray}
\phi_{\Lambda}&=&{1\over\sqrt{2}}(ud-du)s, \nonumber\\
\phi_{\Sigma^{+,0,-}}&=&uus,{1\over\sqrt{2}}(ud+du)s,dds, \nonumber\\
\phi_{\Xi^{0,-}}&=&ssu,ssd, \nonumber\\
\phi_{\Omega^-}&=&sss.
\end{eqnarray}
Note that these flavor wave functions are all either even or odd under
(12) exchange, as are the spin wave functions in Eq.~(\ref{spin}). This
allows sums $\sum \psi \chi \phi$ to be easily built for these states
which are symmetric under exchange of the equal mass quarks, taken as
quarks 1 and 2.

In zeroth order in the anharmonic perturbation $U$ and hyperfine
perturbation $H_{\rm hyp}$, the spatial wave functions $\psi$ are the
harmonic-oscillator eigenfunctions
$\psi_{NLM}(\lpmb{\rho},\lpmb{\lambda})$, where
\beq
\lpmb{\rho}=(\rmb{r}_1-\rmb{r}_2)/\sqrt{2},\ \ 
\lpmb{\lambda}={1\over \sqrt{6}}(\rmb{r}_1+\rmb{r}_2-2\rmb{r}_3)
\label{rholam}
\eeq
are the Jacobi coordinates which separate the Hamiltonian in
Eq.~(\ref{IKHam}) into two independent three-dimensional oscillators
when $U=H_{\rm hyp}=0$. The $\psi_{NLM}(\lpmb{\rho},\lpmb{\lambda})$
can then be conveniently written as sums of products of
three-dimensional harmonic oscillator eigenstates with quantum numbers
$(n,l,m)$, where $n$ is the number of radial nodes and $|l,m\rangle$
is the orbital angular momentum, and where the zeroth order energies are
$E=(N+\thalf)\omega=(2n+l+\thalf)\omega$, with $\omega^2=3K/m$.

For nonstrange states these sums are arranged so that the
resulting $\psi_{NLM}$ have their orbital angular momenta coupled to
$\rmb {L}=\lpmb {l_{\rho}} + \lpmb {l_{\lambda}}$, and so that the
result represents the permutation group $S_3$. The resulting
combined six-dimensional oscillator state has energy $E=(N+3)\omega$,
where $N=2(n_\rho+n_\lambda)+l_\rho+l_\lambda$, and parity
$P=(-1)^{l_\rho+l_\lambda}$.

Ground states are described, in zeroth order in the perturbations, by
basis states with $N=0$. For example the nucleon and $\Delta$ ground
states have, in zeroth order, the common spatial wave function
\beq
\psi^S_{00}={\alpha^3\over \pi^\thalf}
e^{-\alpha^2(\rho^2+\lambda^2)/2},
\eeq
where the notation is $\psi^\pi_{LM}$ with $\pi$ labeling the exchange
symmetry, and the harmonic oscillator scale constant is
$\alpha=(3Km_u)^{1\over 4}$. For $\Lambda$ and $\Sigma$ states the
generalization is to allow the wave function to have an asymmetry
between the $\lpmb{\rho}$ and $\lpmb{\lambda}$ oscillators
\beq
\psi_{00}={\alpha_\rho^\thalf\alpha_\lambda^\thalf\over \pi^\thalf}
e^{-({\alpha_\rho^2}\rho^2+{\alpha_\lambda^2}\lambda^2)/2},
\eeq
where $\alpha_\rho=(3Km)^{1\over 4}$, with $m=m_1=m_2=(m_u+m_d)/2$,
and $\alpha_\lambda=(3Km_\lambda)^{1\over 4}$, with
$m_\lambda=3mm_3/(2m+m_3) > m$. Note that the orbital degeneracy is
now broken, since $\omega_\rho^2=3K/m >
\omega_\lambda^2=3K/m_\lambda$.

In zeroth order, the low-lying negative-parity excited $P$-wave
resonances have $N=1$ spatial wave functions with either $l_{\rho}=1$
or $l_{\lambda}=1$. The wave functions used for strange baryons are
\beqa
\psi^\rho_{1\pm 1}&=&\mp 
{\alpha_\rho^\fhalf\alpha_\lambda^\thalf\over \pi^\thalf}
\rho_{\pm}
e^{-({\alpha_\rho^2}\rho^2+{\alpha_\lambda^2}\lambda^2)/2}, 
\nonumber \\
\psi^\lambda_{1\pm 1}&=&\mp 
{\alpha_\rho^\thalf\alpha_\lambda^\fhalf\over \pi^\thalf}
\lambda_{\pm}
e^{-({\alpha_\rho^2}\rho^2+{\alpha_\lambda^2}\lambda^2)/2}, 
\nonumber \\
\psi^\rho_{10}&=&
{\alpha_\rho^\fhalf\alpha_\lambda^\thalf\over \pi^\thalf}
\sqrt{2}\rho_0
e^{-({\alpha_\rho^2}\rho^2+{\alpha_\lambda^2}\lambda^2)/2},
\label{psi-P}
\eeqa
and similarly for $\psi^\lambda_{10}$, where $\rho_{\pm}\equiv\rho_x
\pm i\rho_y$, $\rho_0\equiv \rho_z$, etc. For the equal mass case the
mixed symmetry states $\psi^{M_\rho}$ and $\psi^{M_\lambda}$ found
from Eq.~(\ref{psi-P}) by equating $\alpha_\rho$ and $\alpha_\lambda$
are used.

Using the following rules for combining a pair of mixed-symmetry
representations,
\beqa
{1\over \sqrt{2}}(M^\rho M^\rho + M^\lambda M^\lambda)&=&S, \nonumber\\
{1\over \sqrt{2}}(M^\rho M^\lambda - M^\lambda M^\rho)&=&A, \nonumber\\
{1\over \sqrt{2}}(M^\rho M^\lambda + M^\lambda M^\rho)&=&M^\rho, \nonumber\\
{1\over \sqrt{2}}(M^\rho M^\rho - M^\lambda M^\lambda)&=&M^\lambda,
\eeqa
the two nonstrange ground states are, in zeroth order in the
perturbations, represented by
\beqa
\vert N^2S_S \half^+\rangle&=&C_A \psi^S_{00}
{1\over \sqrt{2}}
(\phi^\rho_N \chi^\rho_\half+\phi^\lambda_N \chi^\lambda_\half),\nonumber\\
\vert \Delta^4S_S \thalf^+\rangle&=&C_A\phi^S_\Delta\psi^S_{00}
\chi^S_\thalf.
\eeqa
The states are labeled by $\vert X^{2S+1}L_\pi J^P\rangle$, where
$X=N$ or $\Delta$, $S$ is the total quark spin, $L=S,P,D...$ is the
total orbital angular momentum, $\pi=S,M$ or $A$ is the permutational
symmetry (symmetric, mixed symmetry, or antisymmetric respectively) of
the spatial wave function, and $J^P$ is the state's total angular
momentum and parity. Here and in what follows the spin projections
$M_S$ and Clebsch-Gordan coefficients for the ${\bf J}={\bf L}+{\bf S}$
coupling are suppressed. Another popular notation for classification
of baryon states is that of SU(6), which would be an exact symmetry of
the interquark Hamiltonian if it was invariant under SU(3)$_f$, {\it
and} independent of the spin of the quarks (so that, for example, the
ground state octet and decuplet baryons would be degenerate). In this
notation there is an SU(3)$_f$ octet of ground state baryons with
$(J=)S=\half$, giving $(2S+1)\cdot 8=16$ states, plus an SU(3)$_f$
decuplet of ground state baryons with $(J=)S=\thalf$, giving
$(2S+1)\cdot 10=40$ states, for 56 states in total. These ground
states all have $L^P=0^+$, so the SU(6) multiplet to which they belong
is labelled $[56,0^+]$.

The negative-parity $P$-wave excited states occur at $N=1$ in the
harmonic oscillator, and have the compositions
\beqa
&\vert &N^4P_M (\half^-,\thalf^-,\fhalf^-)\rangle=C_A \chi^S_\thalf
{1\over \sqrt{2}}
(\phi^\rho_N \psi^{M_\rho}_{1M}
+\phi^\lambda_N \psi^{M_\lambda}_{1M}),\nonumber\\
&\vert &N^2P_M (\half^-,\thalf^-)\rangle=C_A {1\over 2}
\left\{\phi^\rho_N[\psi^{M_\rho}_{1M} \chi^\lambda_\half 
+ \psi^{M_\lambda}_{1M} \chi^\rho_\half] 
+\phi^\lambda_N[\psi^{M_\rho}_{1M} \chi^\rho_\half 
- \psi^{M_\lambda}_{1M} \chi^\lambda_\half]\right\}\nonumber,\\
&\vert &\Delta^2P_M (\half^-,\thalf^-)\rangle=C_A \phi^S_\Delta
{1\over \sqrt{2}}
(\psi^{M_\rho}_{1M} \chi^\rho_\half+\psi^{M_\lambda}_{1M} \chi^\lambda_\thalf).
\label{NDnegwvfns}
\eeqa
where the notation $(\half^-,\thalf^-,\fhalf^-)$ lists all of the
possible $J^P$ values from the ${\bf L}+{\bf S}$ coupling. These are
members of the $[70,1^-]$ SU(6) multiplet, where the 70 is made up of
two octets of $S=\thalf$ and $S=\half$ states and a decuplet of
$S=\half$ states as above, plus a singlet $\Lambda$
state\cite{PLambda} with $S=\half$.

Isgur and Karl use five parameters to describe the nonstrange and
strangeness $S=-1$ P-wave baryons: the unperturbed level of
the nonstrange states, about 1610 MeV; the quark mass difference
$\Delta m=m_s-m_u=280$ MeV; the ratio $x=m_u/m_s=0.6$; the nonstrange
harmonic oscillator level spacing $\omega\simeq 520$ MeV; and the
strength of the hyperfine interaction, determined by fitting to the
$\Delta-N$ splitting which is
\beq
\delta={4 \alpha_s \alpha^3\over 3\sqrt{2\pi}m_u^2}\simeq 300\ {\rm
MeV},
\eeq
where $\alpha$, given by $\alpha^2=m_u\omega=\sqrt{3Km_u}$, is the
harmonic-oscillator constant used in the nonstrange wave functions. In
the strange states two constants $\alpha_\rho=\alpha$ and
$\alpha_\lambda$ are used; their ratio is determined by the quark
masses and by separation of the center of mass motion in the harmonic
oscillator problem. Note that the effective value of $\alpha_s$ implied by
these equations is $\alpha_s\simeq 0.95$. The value of $\Delta m$ used
here is larger than that implied by a simple SU(3)$_f$ analysis of the
ground state baryon masses where distortions to the wave functions due
to the heavier strange quark are ignored. Taking these effects into
account increases the size of $\Delta m$ required to fit the ground
states to 220 MeV~\cite{Isgur:1979be}.

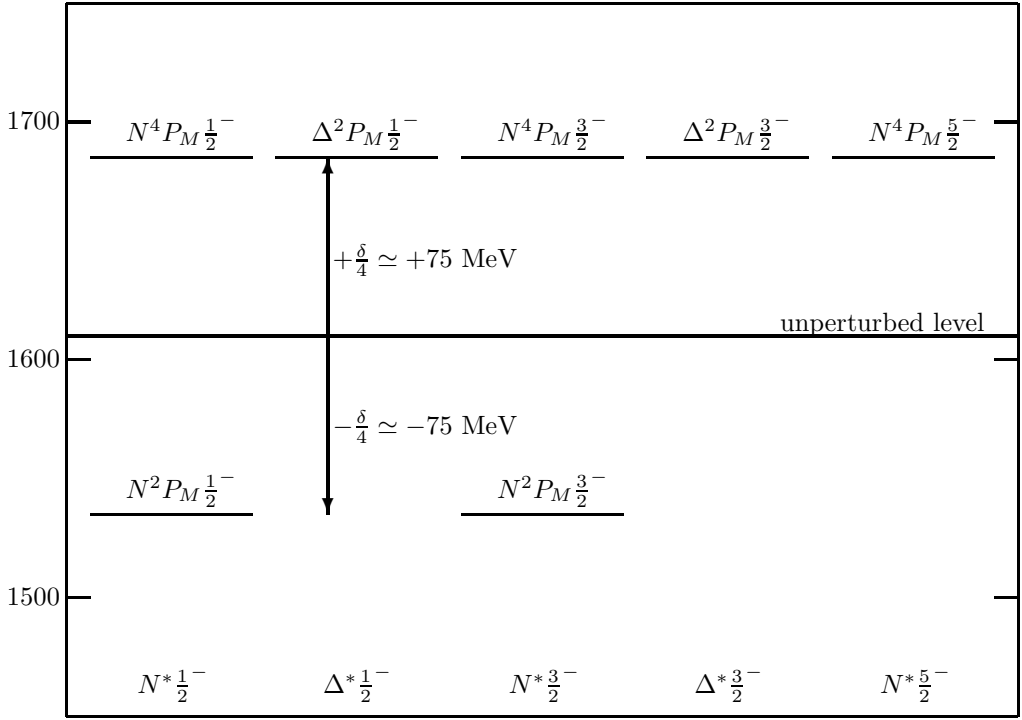
\begin{figure}[t]
\unitlength 0.9pt
\def\half{{\textstyle{1\over2}}}
\def\thalf{{\textstyle{3\over2}}}
\def\fhalf{{\textstyle{5\over2}}}
\def\shalf{{\textstyle{7\over2}}}
\def\nhalf{{\textstyle{9\over2}}}
\def\lhalf{{\textstyle{11\over2}}}
\def\tthalf{{\textstyle{13\over2}}}
\def\fthalf{{\textstyle{15\over2}}}

\begin{center}
\begin{picture}(400,300)
\put(0,0){\line(1,0){400}}
\put(0,0){\line(0,1){300}}
\put(400,300){\line(-1,0){400}}
\put(400,300){\line(0,-1){300}}
\put(0,50){\line(1,0){10}}
\put(390,50){\line(1,0){10}}
\put(-25,47){1500}
\put(0,150){\line(1,0){10}}
\put(390,150){\line(1,0){10}}
\put(-25,147){1600}
\put(0,250){\line(1,0){10}}
\put(390,250){\line(1,0){10}}
\put(-25,247){1700}
\put(0,160){\line(1,0){400}}
\put(300,162){unperturbed level}
\put(110,160){\vector(0,-1){75}}
\put(112,119){$-{\delta\over 4}\simeq -75$ MeV}
\put(110,160){\vector(0,1){75}}
\put(112,189){$+{\delta\over 4}\simeq +75$ MeV}
\thicklines	
\put(30,10){$N^*\half^-$}
\put(10,85){\line(1,0){68}}
\put(25,92){$N^2P_M\half^-$}
\put(10,235){\line(1,0){68}}
\put(25,242){$N^4P_M\half^-$}
\put(108,10){$\Delta^*\half^-$}
\put(88,235){\line(1,0){68}}
\put(103,242){$\Delta^2P_M\half^-$}
\put(186,10){$N^*\thalf^-$}
\put(166,85){\line(1,0){68}}
\put(181,92){$N^2P_M\thalf^-$}
\put(166,235){\line(1,0){68}}
\put(181,242){$N^4P_M\thalf^-$}
\put(264,10){$\Delta^*\thalf^-$}
\put(244,235){\line(1,0){68}}
\put(259,242){$\Delta^2P_M\thalf^-$}
\put(342,10){$N^*\fhalf^-$}
\put(322,235){\line(1,0){68}}
\put(337,242){$N^4P_M\fhalf^-$}
\end{picture}
\end{center}
\caption{The hyperfine contact perturbation applied to
the P-wave nonstrange baryons.}
\label{contact}
\end{figure}
The result of evaluating the contact perturbation in the nonstrange
baryons is the pattern of splittings shown in
Figure~\ref{contact}. The tensor part of the hyperfine interaction has
generally smaller expectation values, and so is not as important to
the spectroscopy as the contact interaction. Isgur and Karl argue,
however, that it does cause significant mixing in some states which
are otherwise unmixed; this has important consequences for the strong
decays of these states.  Note that the tensor interaction, like the
contact interaction, is a total angular momentum $J$ and isospin
scalar, so it can only mix states with the same flavor and $J$ but
different total quark spin $S$. For the set of states considered in
Figure~\ref{contact} this means mixing between the two states of
$J^P=\half^-$, and also between the two states with $J^P=\thalf^-$.

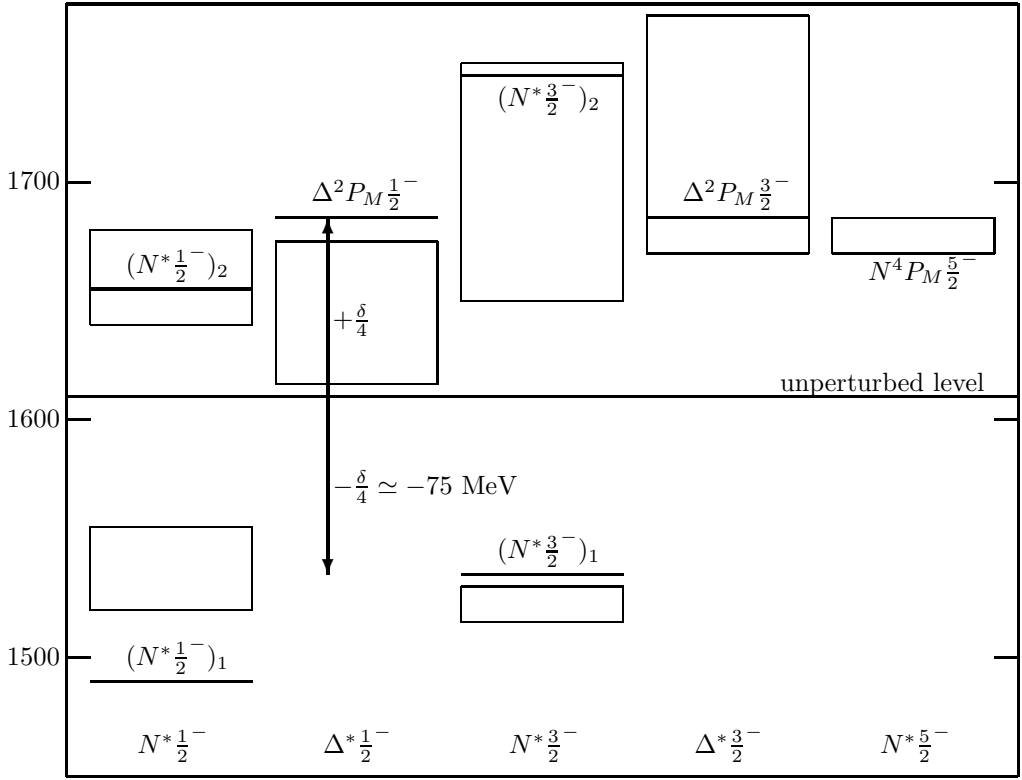
\begin{figure}[t]
\unitlength 0.9pt
\begin{center}
\begin{picture}(400,325)
\put(0,0){\line(1,0){400}}
\put(0,0){\line(0,1){325}}
\put(400,325){\line(-1,0){400}}
\put(400,325){\line(0,-1){325}}
\put(0,50){\line(1,0){10}}
\put(390,50){\line(1,0){10}}
\put(-25,47){1500}
\put(0,150){\line(1,0){10}}
\put(390,150){\line(1,0){10}}
\put(-25,147){1600}
\put(0,250){\line(1,0){10}}
\put(390,250){\line(1,0){10}}
\put(-25,247){1700}
\put(0,160){\line(1,0){400}}
\put(300,162){unperturbed level}
\put(110,160){\vector(0,-1){75}}
\put(112,119){$-{\delta\over 4}\simeq -75$ MeV}
\put(110,160){\vector(0,1){75}}
\put(112,189){$+{\delta\over 4}$}
\thicklines	
\put(30,10){$N^*\half^-$}
\put(10,40){\line(1,0){68}}
\put(25,47){$(N^*\half^-)_1$}
\put(10,205){\line(1,0){68}}
\put(25,212){$(N^*\half^-)_2$}
\put(108,10){$\Delta^*\half^-$}
\put(88,235){\line(1,0){68}}
\put(103,242){$\Delta^2P_M\half^-$}
\put(186,10){$N^*\thalf^-$}
\put(166,85){\line(1,0){68}}
\put(181,92){$(N^*\thalf^-)_1$}
\put(166,295){\line(1,0){68}}
\put(181,282){$(N^*\thalf^-)_2$}
\put(264,10){$\Delta^*\thalf^-$}
\put(244,235){\line(1,0){68}}
\put(259,242){$\Delta^2P_M\thalf^-$}
\put(342,10){$N^*\fhalf^-$}
\put(322,220){\line(1,0){68}}
\put(337,209){$N^4P_M\fhalf^-$}
\thinlines	
%
%
\put(10,70){\line(1,0){68}}
\put(10,70){\line(0,1){35}}
\put(78,70){\line(0,1){35}}
\put(10,105){\line(1,0){68}}
%
%
\put(10,190){\line(1,0){68}}
\put(10,190){\line(0,1){40}}
\put(78,190){\line(0,1){40}}
\put(10,230){\line(1,0){68}}
%
%
\put(166,65){\line(1,0){68}}
\put(166,65){\line(0,1){15}}
\put(234,65){\line(0,1){15}}
\put(166,80){\line(1,0){68}}
%
%
\put(166,200){\line(1,0){68}}
\put(166,200){\line(0,1){100}}
\put(234,200){\line(0,1){100}}
\put(166,300){\line(1,0){68}}
%
%
\put(322,220){\line(1,0){68}}
\put(322,220){\line(0,1){15}}
\put(390,220){\line(0,1){15}}
\put(322,235){\line(1,0){68}}
%
%
\put(88,165){\line(1,0){68}}
\put(88,165){\line(0,1){60}}
\put(156,165){\line(0,1){60}}
\put(88,225){\line(1,0){68}}
%
%
\put(244,220){\line(1,0){68}}
\put(244,220){\line(0,1){100}}
\put(312,220){\line(0,1){100}}
\put(244,320){\line(1,0){68}}

\end{picture}
\end{center}
\caption{The hyperfine contact and tensor perturbations applied to the
P-wave nonstrange baryons. Isgur-Karl model predictions are shown as
bold lines, the range of central values of the masses quoted by the
PDG are shown as boxes.}
\label{hyperf}
\end{figure}

Figure~\ref{hyperf} shows the results of Isgur and Karl's calculation
of the hyperfine contact and tensor interactions for these states. For
the $N^*\half^-$ and $N^*\thalf^-$ states this involves diagonalizing
a 2x2 matrix, and the result is that the eigenstates of the Hamiltonian
are admixtures of the (quark)-spin-$\half$ and -$\thalf$ basis
states. Note also that the quark-spin part of the tensor interaction
is a second-rank tensor, which has a zero expectation value in the
purely quark-spin-$\half$ states $\Delta^2P_M\half^-$ and
$\Delta^2P_M\thalf^-$. The boxes show the approximate range of central
values of the resonance mass extracted from various fits to partial
wave analyses and quoted by the Particle Data
Group~\cite{Caso:1998tx}.
It is arguable whether the tensor part of the hyperfine interaction
has improved the agreement between the model predictions for the
masses of these states and those extracted from the partial wave
analyses. Isgur and Karl~\cite{Isgur:1977ef,Isgur:1978xj} argue,
however, that there is evidence from analysis of strong decays of
these states for the mixing in the wave functions caused by the tensor
interaction. If the tensor mixing between the two states $\vert
N^2P_MJ^-\rangle$ and $\vert N^4P_MJ^-\rangle$ (for $J=\half$ and
$\thalf$) are written in terms of an angle, the result is strong
mixing in the $N\half^-$ sector
\beq
\vert (N^*\half^-)_1 \rangle=\cos\theta_S\vert N^2P_M\half^-\rangle 
- \sin\theta_S\vert N^4P_M\half^-\rangle,
\eeq
with $\theta_S \simeq -32^{\rm o}$, and very little mixing in the
$N\thalf^-$ sector
\beq
\vert (N^*\thalf^-)_1 \rangle=\cos\theta_D\vert N^2P_M\thalf^-\rangle 
- \sin\theta_D\vert N^4P_M\thalf^-\rangle,
\eeq
with $\theta_D \simeq +6^{\rm o}$. Empirical mixing angles were
determined from an SU(6)$_W$ analysis of decay data~\cite{Hey:1975nc} to be
$\theta_S\simeq -32^{\rm o}$ and $\theta_D\simeq +10^{\rm o}$. It is
important to note that these mixing angles are independent of the
size of the strength $2\alpha_s/m_u^2$ of the hyperfine interaction in
Eq.~(\ref{Hhyp}) and the harmonic oscillator parameter $\alpha$. They
depend only on the presence of the tensor term and its size relative
to the contact term, here taken to be prescribed by the assumption
that the quarks interact at short distances by one-gluon exchange.

The $uds$-basis states for the $S=-1$ are simpler than those of
Eq.~(\ref{NDnegwvfns}), since antisymmetry needs only to be imposed
between the two light quarks (1 and 2). The seven $J^P=\half^-$,
$\thalf^-$ and $\fhalf^-$ $\Lambda$ baryons have zeroth order
wave functions with their space-spin wave function product odd under
(12) exchange,
\beqa
\vert \Lambda^4P_\rho (\half^-,\thalf^-,\fhalf^-)\rangle&=&C_A
\psi^\rho_{1M}\chi^S_\thalf \phi_{\Lambda},\nonumber \\
\ \vert \Lambda^2P_\rho (\half^-,\thalf^-)\rangle&=&C_A
\psi^\rho_{1M}\chi^\lambda_\half \phi_{\Lambda},\nonumber \\
\ \vert \Lambda^2P_\lambda (\half^-,\thalf^-)\rangle&=&C_A
\psi^\lambda_{1M}\chi^\rho_\half \phi_{\Lambda},
\label{Lnegwvfns}
\eeqa 
where once again the spin projections $M_S$ and Clebsch-Gordan
coefficients for the ${\bf J}={\bf L}+{\bf S}$ coupling have been
suppressed. The corresponding $\Sigma$ baryons have their space-spin
wave function product even under (12) exchange and involve the (12)
symmetric flavor wave functions $\phi_\Sigma$. 

\begin{figure}[t]
\vskip 0cm
\vbox{
\hskip 0.0cm
\epsfig{file=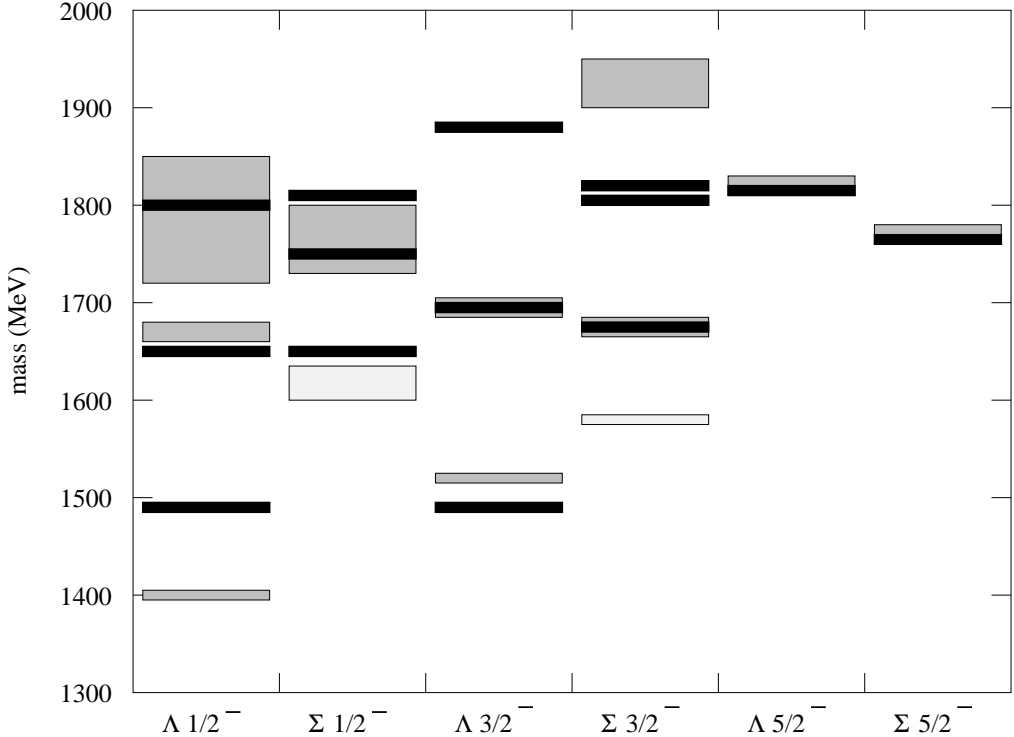,width=13.5cm,angle=0}}
\vskip 0.5 cm
\caption{The hyperfine perturbation applied to the P-wave 
strangeness $-1$ baryons. Isgur-Karl model
predictions are shown as bars, the range of central values of the 
masses quoted by the PDG are shown as shaded boxes with 3 and 4*
states shaded darker than 2* states.}
\label{LSneg}
\end{figure}

The results of Isgur and Karl's calculation of the hyperfine contact
and tensor interactions for these states are shown as solid bars in
Figure~\ref{LSneg}, along with the PDG quoted range in central mass
values, shown as shaded boxes. For the $J^P=\half^-$ and $\thalf^-$
states this involves diagonalizing a 3x3 matrix, so that in addition
to the energies this model also predicts the admixtures in the
eigenstates of the basis states in Eq.~(\ref{Lnegwvfns}) and their
equivalents for the $\Sigma$ states, which will affect their strong
and electromagnetic decays. Note that although deviations of the
spin-independent part of the potential from the harmonic potential
will affect the spectrum of strange baryon states, the anharmonic
perturbation is not applied here.

The accurate prediction of the mass difference between the two
$J^P=5/2^-$ states shows that the quark mass difference is correctly
affecting the oscillator energies, since $\Lambda\fhalf^-$ and
$\Sigma\fhalf^-$ differ only by having either the $\lpmb{\rho}$ or
$\lpmb{\lambda}$ oscillators orbitally excited, respectively. Because
of the heavier strange quark involved in the $\lpmb{\lambda}$
oscillator, the frequency $\omega_\lambda$ is smaller and so the
$\Sigma\fhalf^-$ is lighter. This splitting is reduced by about 20 MeV
by hyperfine interactions. Isgur and Karl point out that the $\Lambda
\thalf^-$ sector is well reproduced, with the third state at around
1880 MeV being mainly $\Lambda^4P_\rho \thalf^-$. This state is
expected to decouple from the $\bar{K}N$ formation
channel~\cite{Faiman:1972np}, which may explain why it has not been
observed. This is explained by a simple selection
rule~\cite{Isgur:1979wd} for excited hyperon strong decays based on
the assumption of a single-quark transition operator and approximately
harmonic-oscillator wave functions; those states which have the
$\lpmb{\rho}$-oscillator between the two nonstrange quarks excited
cannot decay into $\bar{K}N, \bar{K}\Delta, \bar{K}^*N$, or
$\bar{K}^*\Delta$. This is because such an operator cannot
simultaneously de-excite the nonstrange quark pair and emit the
strange quark into the $\bar{K}$ or $\bar{K}^*$.
The $\Sigma\thalf^-$ sector is problematic; Isgur and Karl find a
dominantly (quark) spin-$\half$ state coinciding with a well
established $D_{13}$ state at 1670 MeV which is found to also be
dominantly spin-$\half$ in the analyses of Faiman and
Plane~\cite{Faiman:1972np} and Hey, Litchfield and
Cashmore~\cite{Hey:1975nc}. The model also predicts a pair of
degenerate states at around 1810 MeV. In addition to the $\Sigma
(1670)$, there are two $D_{13}$ states extracted from the
analyses~\cite{Caso:1998tx}, one poorly established state with two
stars at 1580 MeV, and another with three stars at 1940 MeV. The
situation appears better for the $\Sigma\half^-$ sector, although only one
state is resolved at higher mass mass where the model predicts two,
and there is no information from the analyses of
Refs.~\cite{Hey:1975nc,Faiman:1972np} about the composition of the lighter
state.

The largest problem with the spectrum in Figure~\ref{LSneg} is the
prediction for the mass of the well-established $\Lambda\half^-(1405)$
state, {\it i.e.} essentially degenerate with that of the lightest
$\Lambda\thalf^-$ state which corresponds to the
$\Lambda\thalf^-(1520)$.  Isgur and Karl note that the composition
essentially agrees with the analyses of
Refs.~\cite{Hey:1975nc,Faiman:1972np}. It is possible that the mass of
the lightest $\Lambda\half^-$ bound state should be shifted downwards
by its proximity to the $\bar{K}N$ threshold, essentially by mixing
with this virtual decay channel to which it is predicted to strongly
couple. One interaction which can split the lightest
$\Lambda\half^-$ and $\Lambda\thalf^-$ states is the spin-orbit
interaction. Isgur and Karl deliberately neglect these interactions,
as their inclusion spoils the agreement with the spectrum in other
sectors, for example the masses of the light P-wave nonstrange
baryons.

Dalitz\cite{Caso:1998tx} reviews the details of the analyses leading
to the properties of the $\Lambda\half^-(1405)$ resonance, leading to
the conclusion that the data can only be fit with an $S$-wave pole in
the reaction amplitudes (30 MeV) below $\bar{K}N$ threshold.  He
discusses two possibilities for the physical origin of this pole: a
three quark bound state similar to the one found above, coupled with
the S-wave meson-baryon systems; or an unstable $\bar{K}N$ bound
state. If the former the problem of the origin of the splitting
between this state and the $\Lambda(1520)$ is
unresolved. Flavor-dependent interactions between the quarks such as
those which result from the exchange of an octet of pseudoscalar
mesons between the quarks~\cite{Glozman:1996xy} do not explain the
$\Lambda(1520)-\Lambda(1405)$ splitting. If the latter, another state
at around 1520 MeV is required and this region has been explored
thoroughly in $\bar{K}N$ scattering experiments with no sign of such a
resonance. 

An interesting possibility is that the reality is somewhere between
these two extremes. The cloudy bag model calculations of Veit,
Jennings, Thomas and Barrett\cite{Veit:1984an,Veit:1985jr} and
Jennings~\cite{Jennings:1986yg} allow these two types of configuration
to mix and find an intensity of only 14\% for the quark model bound
state in the $\Lambda(1405)$, and predict another $\Lambda\half^-$
state close to the $\Lambda(1520)$. However,
Leinweber~\cite{Leinweber:1990hh} obtains a good fit to this splitting
using QCD sum rules. Kaiser, Siegel and Weise~\cite{Kaiser:1995eg}
examine the meson-baryon interaction in the $K^-p$, $\Sigma\pi$, and
$\Lambda\pi$ channels using an effective chiral Lagrangian. They
derive potentials which are used in a coupled-channels calculation of
low-energy observables, and find good fits to the low-energy physics
of these channels, including the mass of the $\Lambda(1405)$.  This
important issue will be revisited later when discussing spin-orbit
interactions.
\vskip -12pt
\subsection{positive-parity excited baryons in the Isgur-Karl model}
\vskip -12pt
In order to calculate the spectrum of excited positive-parity
nonstrange baryon states, Isgur and Karl~\cite{Isgur:1979wd} construct
zeroth-order harmonic-oscillator basis states from spatial states of
definite exchange symmetry at the N=2 oscillator level. The details of
this construction are given in Appendix~A. In this sector both the
anharmonic perturbation and the hyperfine interaction cause splittings
between the states.

The anharmonic perturbation is assumed to be a sum of two-body forces
$U=\sum_{i<j}U_{ij}$, and since it is flavor independent and a
spin-scalar, it is SU(6) symmetric. This means that it causes no
splittings within a given SU(6) multiplet, which is one of the reasons
for classifying the states using SU(6). It does, however, break the
initial degeneracy within the N=2 harmonic oscillator band. The
anharmonicity is treated as a diagonal perturbation on the energies of
the states, and so is not allowed to cause mixing between the $N=0$
and $N=2$ band states. It causes splittings between the $N=1$ band
states only when the quark masses are unequal; the diagonal
expectations of $U$ in the $N=0$ band (and $N=1$ band for the $N$ and
$\Delta$ states) are lumped into the band energies.

\begin{figure}[t]
\begin{center}
\unitlength 1.0pt
\begin{picture}(400,220)
\put(10,10){\vector(0,1){89}}
\put(0,50){$\Omega$}
\put(10,101){\vector(0,1){89}}
\put(0,140){$\Omega$}
\put(14,10){\line(1,0){100}}
\put(116,8){$[56,0^+],\ E_0$}
\put(14,100){\line(1,0){100}}
\put(116,100){$[70,1^-]\ E_0+\Omega$}
\put(14,90){\line(1,0){100}}
\put(116,86){$[56^\prime,0^+]\ E_0+2\Omega-\Delta$}
\put(64,91){\vector(0,1){48}}
\put(66,111){$\Delta/2$}
\put(14,140){\line(1,0){100}}
\put(116,136){$[70,0^+]\ E_0+2\Omega-\Delta/2$}
\put(64,141){\vector(0,1){8}}
\put(66,141){$\Delta/10$}
\put(14,150){\line(1,0){100}}
\put(116,150){$[56,2^+]\ E_0+2\Omega-2\Delta/5$}
\put(64,151){\vector(0,1){18}}
\put(66,156){$\Delta/5$}
\put(14,170){\line(1,0){100}}
\put(116,168){$[70,2^+]\ E_0+2\Omega-\Delta/5$}
\put(64,171){\vector(0,1){18}}
\put(66,176){$\Delta/5$}
\put(14,190){\line(1,0){100}}
\put(116,188){$[20,1^+]\ E_0+2\Omega$}
\end{picture}
\end{center}
\caption{The first-order anharmonic perturbation $U$ applied to
the nonstrange oscillator basis. The parameters $E_0$, $\Omega$, and
$\Delta$ are defined in the text.}
\label{anharm}
\end{figure}
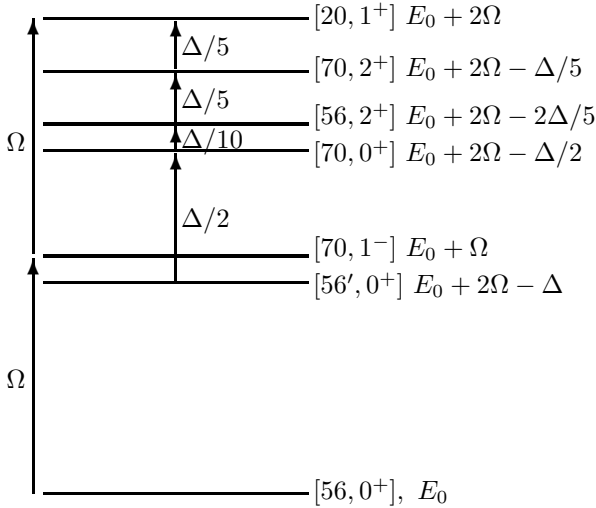

\begin{figure}[t]
\vskip 0cm
\vbox{
\hskip 0.0cm
\epsfig{file=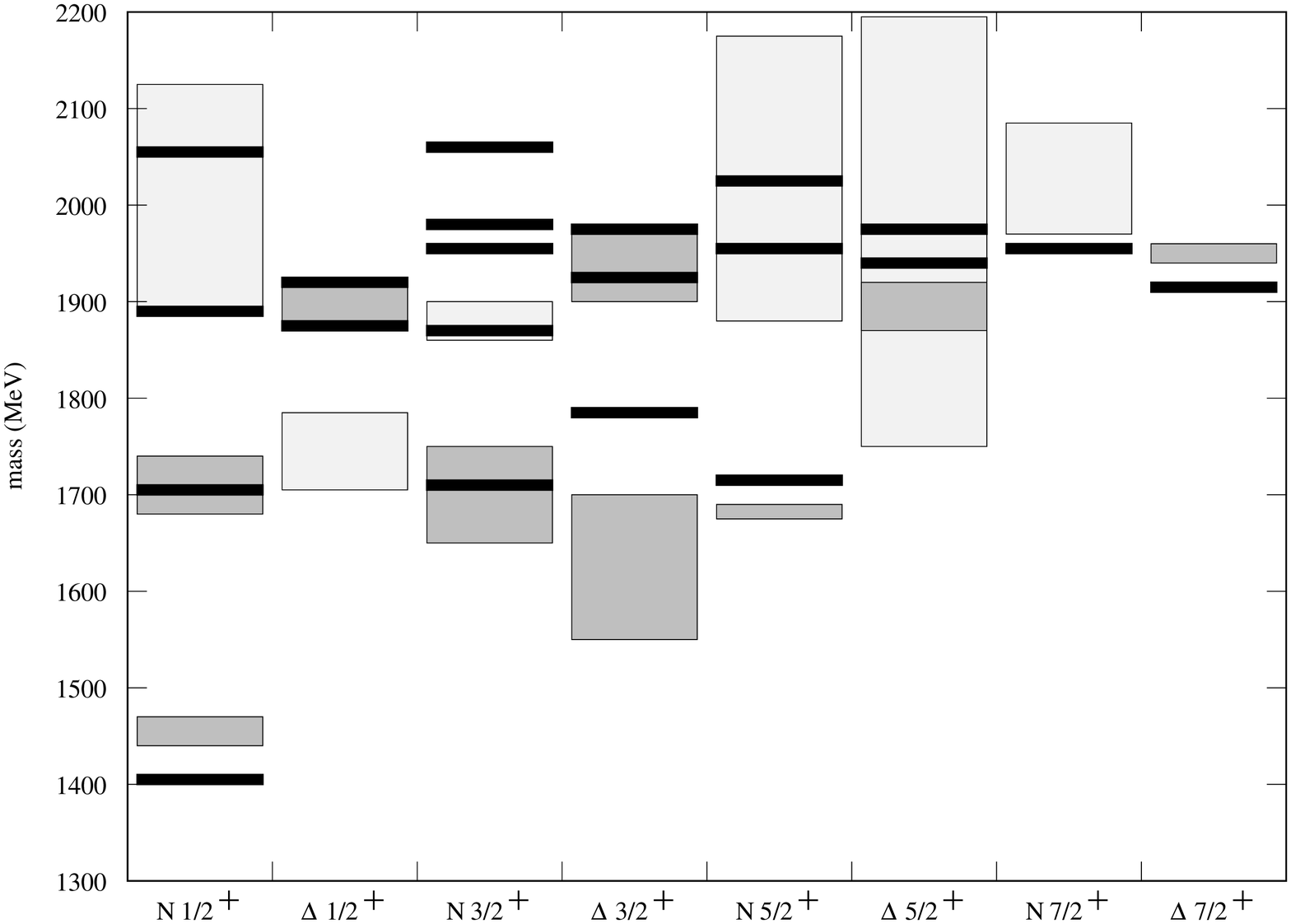,width=13.5cm,angle=0}}
\vskip 0.5 cm
\caption{The anharmonic and hyperfine perturbations applied to the
positive-parity excited nonstrange baryons. Isgur-Karl model
predictions are shown as bars, the range of central values of the 
masses quoted by the PDG are shown as shaded boxes with 3 and 4*
states shaded darker than 1 and 2* states.}
\label{NDpos}
\end{figure}

For $N$ and $\Delta$ states the symmetry of the states can be used to
replace $U=\sum_{i<j}U_{ij}$ by $3U(r_{12})$, where ${\bf r}_{12}={\bf
r}_1-{\bf r_2}=\sqrt{2}\lpmb{\rho}$, and the spin-independent
potential is assumed to be a sum of two-body terms. Defining the
moments $a$, $b$, and $c$ of the anharmonic perturbation by
\beq
\left\{a,b,c\right\}:=3{\alpha^3\over \pi^\thalf} 
\int d^3\lpmb{\rho}\left\{1,\alpha^2\rho^2,\alpha^4\rho^4\right\}
U(\sqrt{2}\rho)e^{-\alpha^2\rho^2},
\eeq
it is straightforward to show that, the $[56,0^+]$ ground states are
shifted in mass by $a$, and the $[70,1^-]$ states are shifted by
$a/2+b/3$ in first order in the perturbation $U$. The states in the
$N=2$ band are known~\cite{Horgan:1973ww,Gromes:1976cr} to be split
into a pattern which is independent of the form of the anharmonicity
$U$. This pattern is illustrated in Figure~\ref{anharm}, with the
definitions $E_0=3m+3\omega+a$, $\Omega:=\omega-a/2+b/3$ and
$\Delta:=-5a/4+5b/3-c/3$, where $m$ the nonstrange quark mass and
$\omega$ the oscillator energy. Isgur and Karl~\cite{Isgur:1979wd} fit
$\Omega$ and $\Delta$ to the spectrum of the positive-parity excited
states, using $E_0=1150$ MeV, $\Omega\simeq 440$ MeV, $\Delta\simeq
440$ MeV, which moves the first radial excitation in the
$[56^\prime,0^+]$, assigned to the Roper resonance $N(1440)$, {\it
below} the (unperturbed by $H^{\rm hyp}$) level of the P-wave
excitations in the $[70,1^-]$. Note that $\omega$ is 250 MeV, which
makes this first order perturbative treatment questionable (since
$\Delta\simeq\Omega>\omega$). Also, the off-diagonal matrix elements
between the $N=0$ and $N=2$ band states which are present in second
order tend to increase the splitting between these two bands. Richard
and his collaborators~\cite{Richard:1992uk} have shown that, within a
certain class of models, it is impossible to have this state lighter
than the $P$-wave states of Fig.~\ref{hyperf} in a calculation which
goes beyond first order wave function perturbation theory.
\vskip -36pt
Diagonalization of the hyperfine interaction of Eq.~(\ref{Hhyp})
within the harmonic oscillator basis described in Appendix~A results
in the energies for nonstrange baryons states shown in
Figure~\ref{NDpos}.  The agreement with the masses of the states
extracted from data analyses is rather good, with some
exceptions. Obviously the model predicts too many states compared with
the analyses--several predicted states are `missing'. This issue will
be discussed in some detail in the section on decays that follows. It
will be shown there that Koniuk and Isgur's model~\cite{KI} of strong
decays, which uses the wave functions resulting from the
diagonalization procedure above, establishes that the states seen in
the analyses are those which couple strongly to the $\pi N$ (or
$\bar{K}N$) formation channel, an idea examined in detail
earlier by Faiman and Hendry~\cite{Faiman:1968js}. It will be shown
that this is also verified in some other models which are able to
simultaneously describe the spectrum and strong decays.
Figure~\ref{NDpos} shows that the predicted mass for the lightest
excited $\Delta\thalf^+$ state is higher than that of the lightest
$P_{33}$ excited state seen in the analyses, although two recent
multi-channel analyses~\cite{MANSA,Vrana:2000nt} have found central
mass values between about 1700 and 1730 MeV, at the high end of the
range shown in Fig.~\ref{NDpos}. The weakly established (1*)
$\Delta\half^+$ state $\Delta(1750)$ also does not fit well with the
predicted model states. Elsewhere the agreement is good.

\begin{figure}[t]
\vskip 0cm
\vbox{
\hskip 0.0cm
\epsfig{file=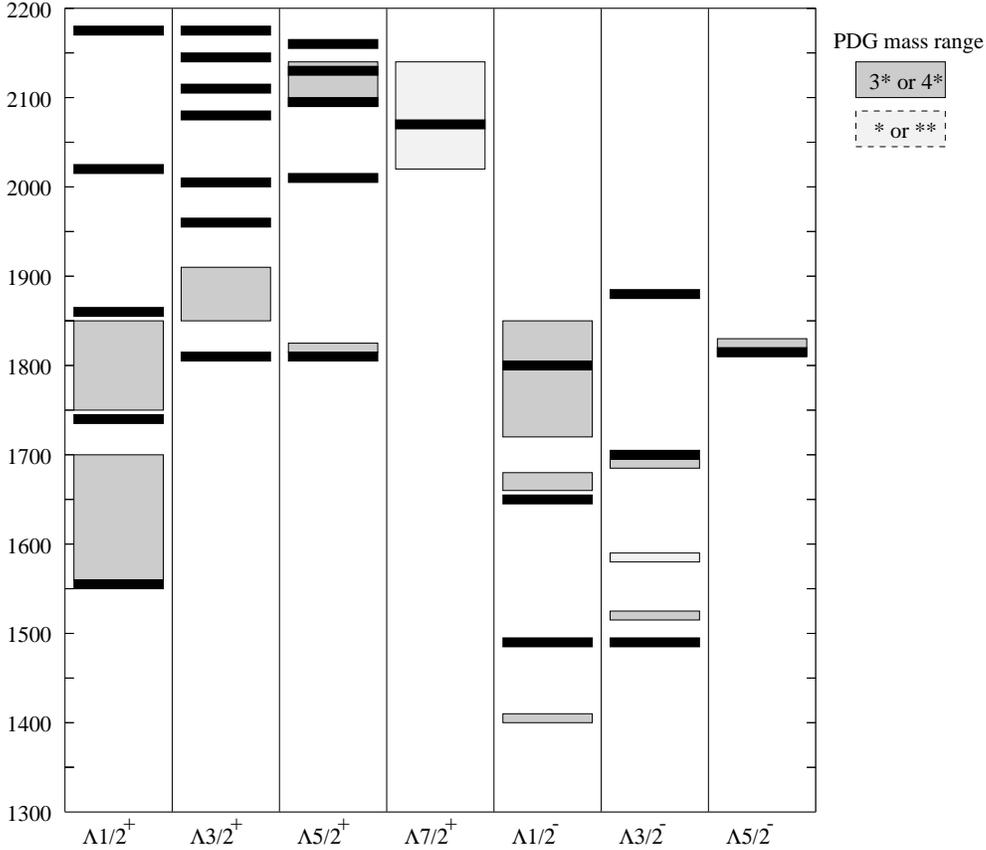,width=13.5cm,angle=0}}
\vskip 0.5 cm
\caption{The anharmonic and hyperfine perturbations applied to the
negative and positive-parity excited $\Lambda$ baryons. Isgur-Karl model
predictions are shown as bars, the range of central values of the 
masses quoted by the PDG are shown as shaded boxes with 3 and 4*
states shaded darker than 1 and 2* states. Note that the
negative-parity and positive-parity states were fit independently.}
\label{LIK}
\end{figure}

\begin{figure}[t]
\vskip 0cm
\vbox{
\hskip 0.0cm
\epsfig{file=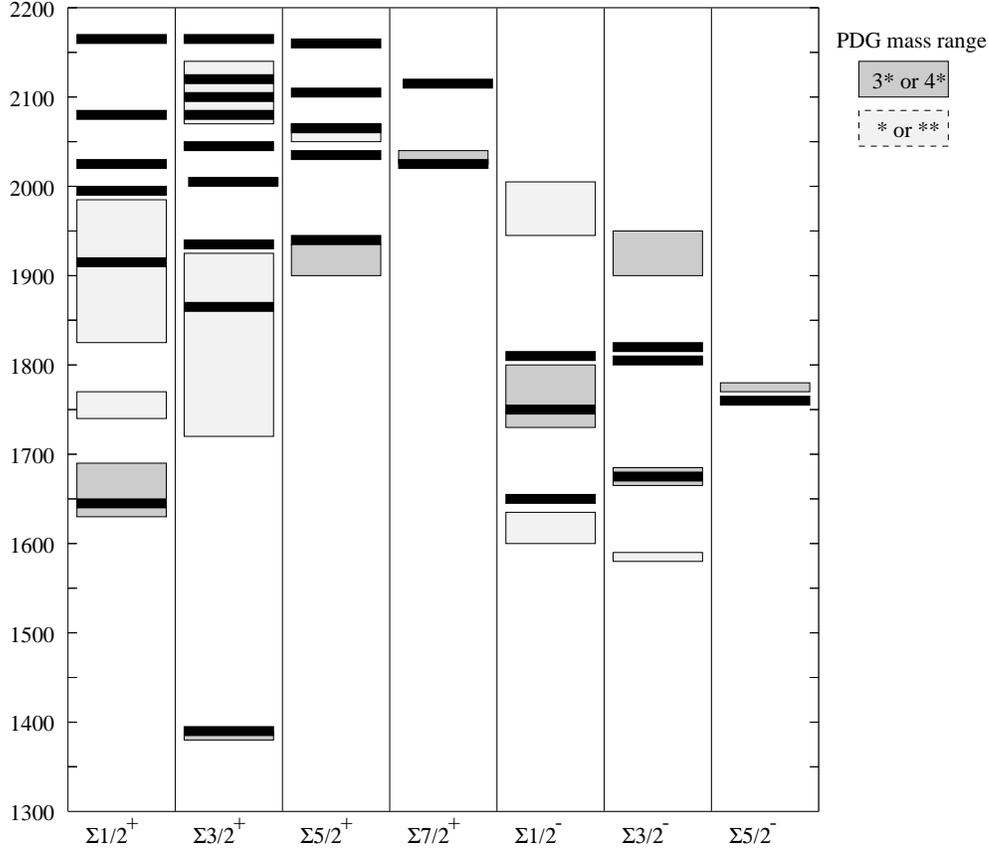,width=13.5cm,angle=0}}
\vskip 0.5 cm
\caption{The anharmonic and hyperfine perturbations applied to the
negative and positive-parity excited $\Sigma$ baryons. Caption as in Fig.~\protect\ref{LIK}.}
\label{SIK}
\end{figure}

Masses for positive-parity excited $\Lambda$ and $\Sigma$ states are
also calculated by Isgur and Karl in Ref.~\cite{Isgur:1979wd}. The
construction of the wave functions for these states is simpler than
that of the symmetrized states detailed in Appendix A, as antisymmetry
is only required under exchange of the two light quarks.
Figures~\ref{LIK} and~\ref{SIK}, which show the Isgur-Karl model fit
to the the masses of all $\Lambda$ and $\Sigma$ resonances extracted
from the data below 2200 MeV, show that the fit for strange
positive-parity states is of similar quality to that of the nonstrange
positive-parity states, although there are considerably fewer well
established experimental states. Once again many more states are
predicted than are seen in the analyses. Using the simple selection
rule for excited hyperon strong decays described above as a rough
guide, Isgur and Karl show~\cite{Isgur:1979wd} that there is a good
correspondence in mass between states found in analyses of $\bar{K}N$
scattering data and those predicted to couple to this channel. This
strong decay selection rule is modified by configuration mixing in the
initial and final states, and so these decays have been examined using
subsequent strong decay models which take into account this
mixing~\cite{KI}.
\vskip -12pt
\subsection{criticisms of the nonrelativistic quark model}
\vskip -12pt
The previous sections show that the main features of the spectrum of
the low-lying baryon resonances are quite well described by the
nonrelativistic model. Just as importantly, the mixing of the states
caused by the hyperfine tensor interaction is crucial to explaining
their observed strong decays, {\it e.g.}  the $N\eta$ decays of the
$N^*\half^-$ states. However, there are several criticisms which can
be made of the model. In strongly bound systems of light quarks such
as the baryons considered above, where $p/m\simeq 1$, the
approximation of nonrelativistic kinematics and dynamics is not
justified. For example, if one forms the one-gluon exchange T-matrix
element {\it without} performing a nonrelativistic reduction, factors
of $m_i$ in Eq.~(\ref{Hhyp}) are replaced, roughly, with factors of
$E_i=\sqrt{\rmb {p}_i^2+m_i^2}$. In a potential model picture there
should also be `kinematic' smearing of the interquark coordinate $\rmb
{r}_{ij}$ due to relativistic effects, with a characteristic size
given by the Compton wavelength of the quark $1/m_q$. A partially
relativistic treatment like that of the MIT bag model would at first
seem preferable, but problems imposed by restriction to motion within
a spherical cavity and in dealing with center of mass motion in this
model have not allowed progress to be made in understanding details of
the physics of the majority of excited baryon states.

Neglecting the scale dependence of a cut-off field theory (QCD) has
resulted in non-fundamental values of parameters like the quark mass,
the string tension (implicit in the size of the anharmonic
perturbations) and the strong coupling $\alpha_s\simeq 1$. A
consistent theory with constituent quarks should give those quarks a
commensurate size, which will smear out the interactions between the
quarks. The string tension should be consistent with meson
spectroscopy, and the relation between the anharmonicity and the meson
string tension should be explored. If there are genuine three-body
forces in baryons, they have been neglected. The neglect of spin-orbit
interactions in the Hamiltonian is also inconsistent, independent of
the choice of {\it ansatz} for the short-distance and confining
physics. There is some evidence in the {\it observed} spectrum for
spin-orbit splittings, {\it e.g.} that between the states
$\Delta\half^-(1620)$ and $\Delta\thalf^-(1700)$, and between
$\Lambda\thalf^-(1520)-\Lambda\half^-(1405)$, although the latter is
likely complicated by decay channel couplings. It is also inconsistent
to neglect the various spin-independent but momentum-dependent terms
in the Breit-Fermi reduction of the one-gluon-exchange potential.

The model also uses a first-order perturbative evaluation of large
perturbations. The contact term is, unless the above smearing is
implemented, formally infinite. As shown above, the size of the
first-order anharmonic splitting of the $N=2$ band must be larger than
the 0-th order harmonic splitting, to get the lightest $N=2$ band
nucleon [identified with the Roper resonance $N(1440)$] below the
$P$-wave non-strange states. This calls into question the usefulness
of first order perturbation theory. It also means that the
wave functions of states like the Roper resonance should have a large
anharmonic {\it mixing} with the ground states. There are also some
inconsistencies between the parameters used in describing the
negative and positive-parity excited states, which presumably can be
traced back to this source.
\vskip -12pt
\section{Recent models of excited baryon spectra}
\vskip -12pt
\subsection{relativized quark model}
\vskip -12pt
Several potential model
calculations~\cite{Stanley:1980fe,Forsyth:1983dq,Forsyth:1981qy,Carlson:1983xi,Sartor:1985ss}
which retain the one-gluon exchange picture of the quark-quark
interactions have gone beyond Isgur and Karl's model for the spectrum
and wave functions of baryons in an attempt to correct the flaws in
the nonrelativistic model described above. A representative example
is the relativized quark model, first applied to meson spectroscopy by
Godfrey and Isgur~\cite{Godfrey:1985xj}, and later adapted to
baryons~\cite{Capstick:1986bm}. In this model the Schr\"odinger
equation is solved in a Hilbert space made up of dressed valence quarks
with finite spatial extent and masses of 220 MeV for the light quarks,
and 420 MeV for the strange quark. The Hamiltonian is now given by
\beq
H=\sum_i \sqrt{\rmb {p}_i^2+m_i^2} + V,
\label{CIHam}
\eeq
where $V$ is a relative-position and -momentum dependent potential which
tends, in the nonrelativistic limit (which is {\it not} taken here) to
the sum 
\beq
\lim_{p_i/m_i\to 0}V=V_{\rm string}+V_{\rm Coul}+V_{\rm hyp}
+V_{\rm so(cm)}+V_{\rm so(Tp)}.
\eeq
The terms are a confining string potential, a pairwise color-Coulomb
potential, a hyperfine potential, and spin-orbit potentials associated
with one-gluon exchange and Thomas precession in the confining
potential. The momenta $p_i$ are written in terms of the momenta ${\bf
p}_\rho$, ${\bf p}_\lambda$ conjugate to the Jacobi coordinates of
Eq.~(\ref{rholam}) and the total momentum ${\bf P}$, in order to
separate out the center of mass momentum. Note that spin-independent
but momentum-dependent terms present in O$(p^2/m^2)$ reduction of the
one-gluon-exchange potential are omitted here.

The gluon fields are taken to be in their adiabatic ground state, and
generate a confining potential $V_{\rm string}$ in which the quarks
move. This is effectively linear at large separation and can be
written as the sum $V_{\rm string}=\sum_i b\,l_i +C$ of the energies
of strings of length $l_i$ connecting quark $i$ to a string junction
point, where $b$ is the meson string tension. The string is assumed to
adjust infinitely quickly to the motion of the quarks so that it is
always in its minimum length configuration; this generates a
three-body adiabatic potential for the
quarks~\cite{Dosch:1976gf,Cutkosky:1977if,Stanley:1980fe,Carlson:1983rw,Sartor:1985ss}
which includes genuine three-body forces.

The Coulomb, hyperfine, color-magnetic and Thomas-precession
spin-orbit potentials are as they were in the nonrelativistic
model~\cite{Isgur:1977ef,Isgur:1978xj,Isgur:1979wd} except that: (1)
the inter-quark coordinate $\rmb {r}_{ij}$ is smeared out over
mass-dependent distances, as suggested by relativistic kinematics and
QCD; (2) the momentum dependence away from the $p/m\to 0$ limit is
parametrized, as suggested by relativistic dynamics.  In practice, (1)
is brought about by convoluting the potentials with a smearing
function $\rho_{ij}(\rmb {r}_{ij})=\sigma_{ij}^3 e^{-\sigma_{ij}^2
r_{ij}^2}/\pi^{3\over 2}$, where the $\sigma_{ij}$ are chosen to smear
the inter-quark coordinate over distances of O(1/$m_Q$) for Q heavy,
and approximately 0.1 fm for light quarks; (2) is brought about by
introducing factors which replace the quark masses $m_i$ in the
nonrelativistic model by, roughly, $E_i$. For example the contact part
of $H^{ij}_{\rm hyp}$ becomes
\begin{equation}
V^{ij}_{\rm cont}=
\left({m_im_j\over E_iE_j}\right)^{{1\over 2}+\epsilon_{\rm cont}} 
{8\pi\over 3}\alpha_s(r_{ij}){2\over 3}{\rmb {S}_i\cdot\rmb {S}_j\over m_im_j}
\left[{\sigma_{ij}^3\over \pi^{3\over 2}}e^{-\sigma_{ij}^2 r_{ij}^2}\right]
\left({m_im_j\over E_iE_j}\right)^{{1\over 2}+\epsilon_{\rm cont}},
\end{equation}
where $\epsilon_{\rm cont}$ is a constant parameter, and
$\alpha_s(r_{ij})$ is a running-coupling constant which runs according
to the lowest-order QCD formula, saturating to 0.6 at $Q^2=0$.

The energies and wave functions of all the baryons are then solved for
by expanding the states in a large harmonic oscillator basis. Wave
functions are expanded to $N\le 7$ ($N\le 8$ for $J^P=\half^+$). The
Hamiltonian matrix is diagonalised in this basis, and energy
eigenvalues are then crudely minimized in $\alpha$, the
harmonic-oscillator size parameter. To avoid construction of
symmetrized wave functions in this large basis, the non-strange wave
functions are not explicitly antisymmetrized in $u$ and $d$
quarks. Instead, the strong Hamiltonian (which cannot distinguish $u$
and $d$ quarks) is allowed to sort the basis into $N$ and $\Delta$
states.
\begin{figure}[t]
\vskip 0cm
\vbox{
\hskip 0.2cm
\epsfig{file=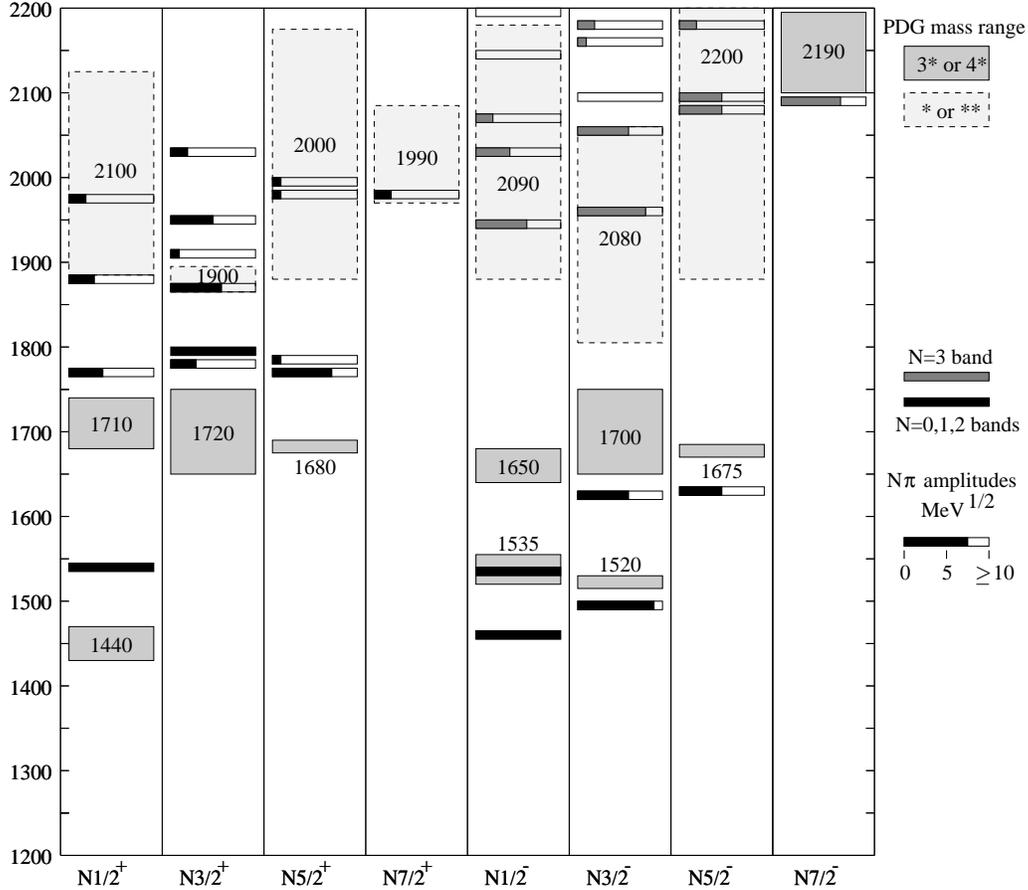,width=14cm,angle=0}}
\vskip 0.5 cm
\caption{Mass predictions and $N\pi$ decay amplitudes (whose square is
the $N\pi$ decay width) for nucleon resonances below 2200 MeV from
Refs.~\protect\cite{Capstick:1986bm,Capstick:1993th}, compared to the
range of central values for resonances masses from the
PDG~\protect\cite{Caso:1998tx}, which are shown as boxes. The boxes
are lightly shaded for one and two star states and heavily shaded for 3
and four star states. Predicted masses are shown as a thin bar, with
the length of the black shaded region indicating the size of the
$N\pi$ amplitude. The ground state nucleon mass from this model is 960
MeV.}
\label{Nexpthr}
\end{figure}
The resulting fit to the masses of excited nucleon states below 2200
MeV is compared to the range of central values for resonances masses
in that mass range from the PDG~\cite{Caso:1998tx} in
Figure~\ref{Nexpthr}. In addition, Fig.~\ref{Nexpthr} illustrates the
results for the $N\pi$ decay amplitudes of these states of a strong
decay calculation based on the creation of a pair of quarks with
$^3P_0$ quantum numbers~\cite{Capstick:1993th}. The length of the
shaded region in the bar representing the model state's mass is
proportional to the size of the $N\pi$ decay amplitudes, which when
squared gives the partial width to decay to $N\pi$. This is so that
states which are likely to have been seen in analysis of elastic and
inelastic $\pi N$ scattering can be identified among the states
predicted by the model. Figure~\ref{Dexpthr} illustrates the same
quantities for excited $\Delta$ states. Figures~\ref{LCI}
and~\ref{SCI} compare the fit to the excited $\Lambda$ and $\Sigma$
states below 2200 MeV to the range of central values for resonance
masses in that mass range, which are extracted from analyses of
$\bar{K}N$ and other scattering experiments listed in the
PDG~\cite{Caso:1998tx}.
\begin{figure}[t]
\vskip 0cm
\vbox{
\hskip 0.2cm
\epsfig{file=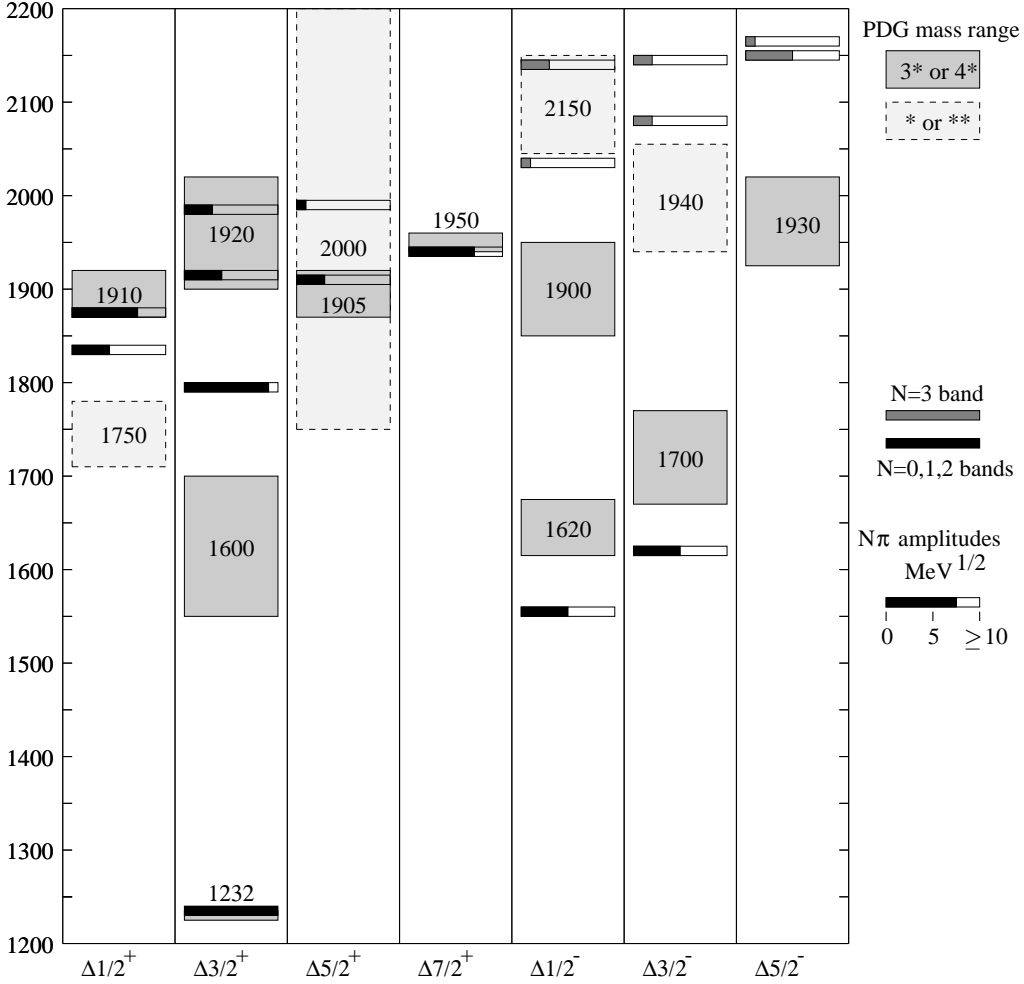,width=14cm,angle=0}}
\vskip 0.5 cm
\caption{Model masses and $N\pi$ decay amplitudes for $\Delta$
resonances below 2200 MeV from
Refs.~\protect\cite{Capstick:1986bm,Capstick:1993th}, compared to
the range of central values for resonances masses from the
PDG~\protect\cite{Caso:1998tx}. Caption as in
Fig.~\protect\ref{Nexpthr}.}
\label{Dexpthr}
\end{figure}
The pattern of splitting in the negative and positive-parity bands of
excited nonstrange states is reproduced quite well, although the
centers of the bands are missed by about $-50$ MeV and $+50$ MeV,
respectively. The Roper resonance mass is about 100 MeV too high
compared to the nucleon ground state, but fits well into the pattern
of splitting of the positive-parity band. This problem is slightly
worse in the case of the state $\Delta\thalf^+(1600)$. The
negative-parity $\Delta$ states in the $N=3$ band are too high when
compared to the masses of a well-established pair of resonant states
$\Delta\half^-(1900)$ and $\Delta\thalf^-(1930)$. It is clear from
Figs.~\ref{LCI} and~\ref{SCI} that the relativized model does not have
the freedom that the nonrelativistic model has to fit the mass
splittings caused by the strange-light quark mass difference in the
negative-parity strangeness $-1$ baryons. The pattern of splitting in
the positive-parity strangeness $-1$ baryons is reproduced well, with
again the band being predicted too heavy by about 50 MeV.

It has been shown by Sharma, Blask, Metsch and
Huber~\cite{Sharma:1989na} that the inclusion of the spin-independent
but momentum-dependent terms present in an O($p^2/m^2$) reduction of
the one-gluon exchange potential reduces the energy of certain
positive-parity excited states, and raises that of the P-wave excited
states. It is possible that the inclusion of such terms in the
Hamiltonian of Eq.~(\ref{CIHam}) could explain the roughly $\pm 50$
MeV discrepancy between the relativized model masses of these bands of
states and those extracted from the analyses.
\begin{figure}[t]
\vskip 0cm
\vbox{
\hskip 0.2cm
\epsfig{file=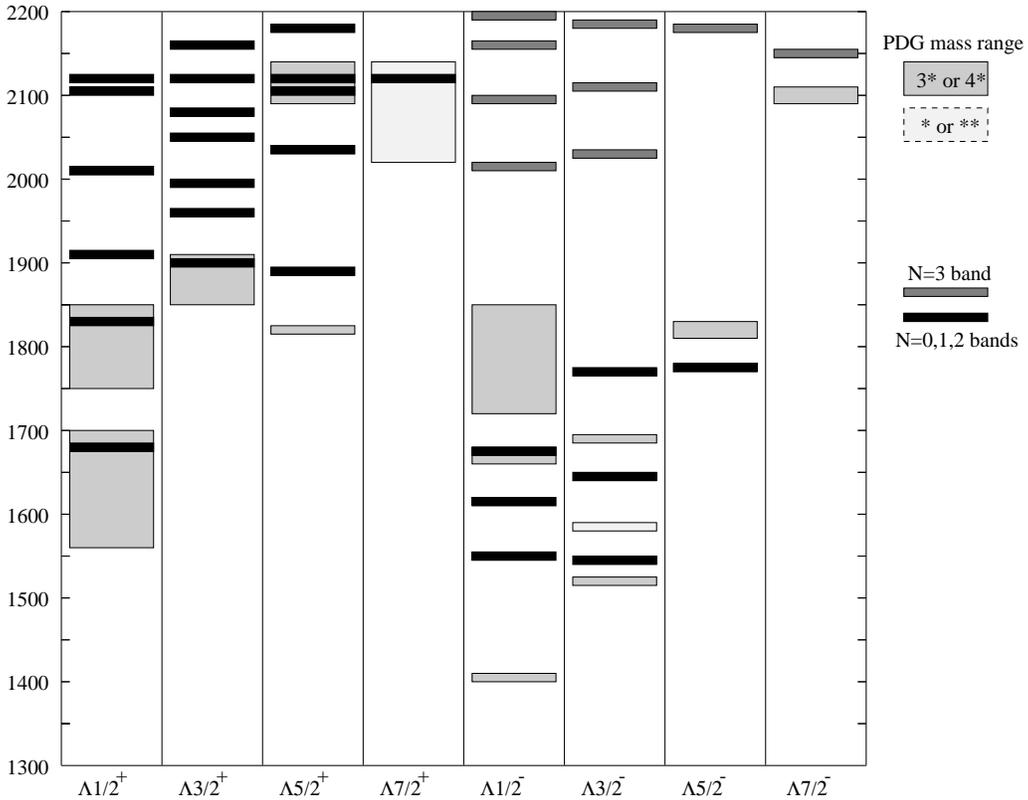,width=14cm,angle=0}}
\vskip 0.5 cm
\caption{Model masses for $\Lambda$ resonances below 2200 MeV from
Ref.~\protect\cite{Capstick:1986bm}, compared to the range of central
values for resonances masses from the PDG~\protect\cite{Caso:1998tx},
which are shown as boxes. The boxes are lightly shaded for one and two
star states and heavily shaded for 3 and four star states. Model
masses are shown as a thin bar. The ground state $\Lambda$ mass in
this model is 1115 MeV.}
\label{LCI}
\end{figure}
The model of Ref.~\cite{Capstick:1986bm} shows some improvements, and
some deterioration relative to the nonrelativistic model, largely
because it does not contain the freedom to separately fit the
negative-parity and positive-parity states present in the
nonrelativistic model.  The same set of parameters is used to fit all
mesons~\cite{Godfrey:1985xj} and baryons, except the string tension
$b$ is reduced about 15\% from the meson value in the best fit of
Ref.~\cite{Capstick:1986bm}. Spin-orbit interactions in this model are
small, for several reasons; a smaller $\alpha_s$ is used, while
retaining the same contact interaction splittings, by virtue of the
non-perturbative evaluation of the expectation value of the smeared
contact interaction. Perturbative evaluation of a $\delta$-function
contact term underestimates the size of the contact splittings, and so
requires a larger value of $\alpha_s$ to compensate, which increases
the size of the OGE spin-orbit interactions.  There is also, as
expected~\cite{Isgur:1978xj}, a partial cancellation of the
color-magnetic and Thomas-precession spin-orbit terms, and freedom
present in the model to choose $\epsilon_{\rm cont}<\epsilon_{\rm so}$
to suppress the spin-orbit interactions relative to the hyperfine
terms has been exploited.
\begin{figure}[t]
\vskip 0cm
\vbox{
\hskip 0.2cm
\epsfig{file=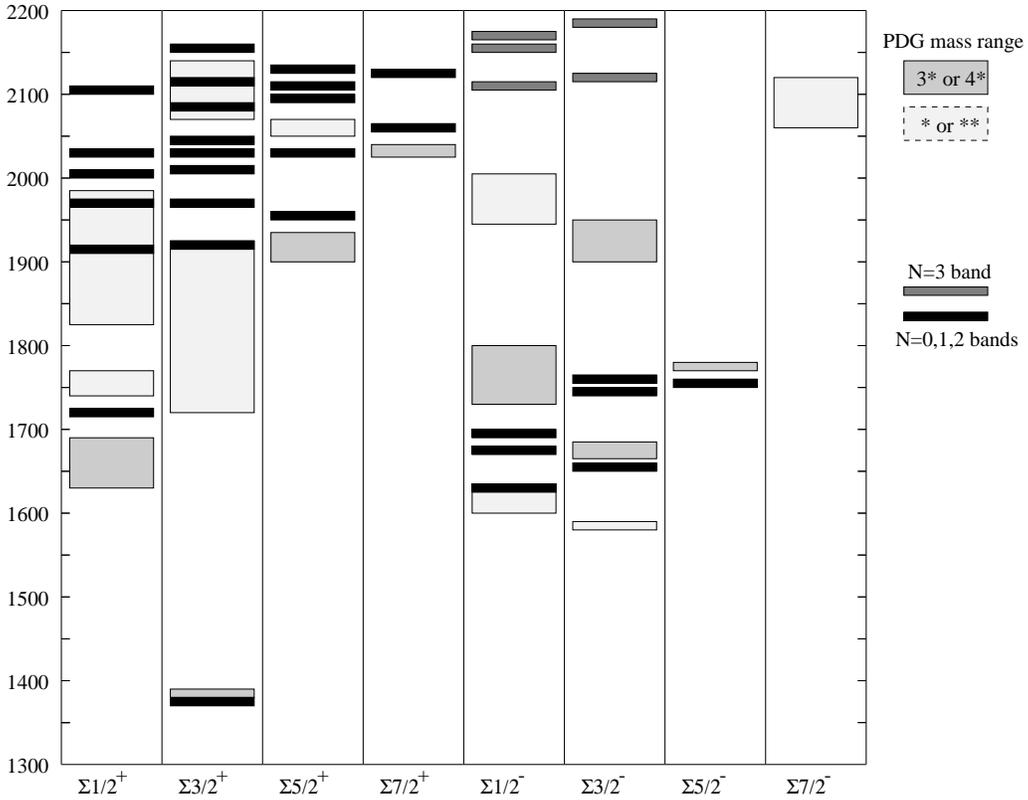,width=14cm,angle=0}}
\vskip 0.5 cm
\caption{Model masses for $\Sigma$
resonances below 2200 MeV from
Ref.~\protect\cite{Capstick:1986bm}, compared to
the range of central values for resonances masses from the
PDG~\protect\cite{Caso:1998tx}. Caption as in
Fig.~\protect\ref{LCI}. The ground state $\Sigma$ mass in
this model is 1190 MeV.}
\label{SCI}
\end{figure}
\vskip -12pt
\subsection{semirelativistic flux-tube model}
\vskip -12pt
A parallel extension of one-gluon-exchange (OGE) based models was
carried out by Sartor and Stancu~\cite{Sartor:1985ss}, and by Stancu
and Stassart~\cite{Stancu:1990ht,Stancu:1991cz}. The spin-independent
part of the interquark Hamiltonian contains a relativistic kinetic
energy term and the string confining potential, as well as a pairwise
color-Coulomb interaction between the quarks. The hyperfine
interaction has a spin-spin contact term and a tensor interaction
which are properly smeared out over the finite size of the constituent
quark. In contrast to the use of a harmonic-oscillator basis in the
relativized model, this work uses a variational wave function basis
first employed by Carlson, Kogut and Pandharipande
(CKP)~\cite{Carlson:1983rw}. This basis essentially interpolates
between Coulomb and linear potential solutions, and contains a factor
which decreases as the length of the Y-shaped string connecting the
quarks increases. This allows the use of a significantly smaller set
of basis states than the oscillator basis used by other authors,
because the large distance behavior of the wave functions is closer to
that of the true eigenstates. On the other hand, it is somewhat more
complicated to work with, especially in momentum space. Sartor and
Stancu~\cite{Sartor:1985ss} extend the calculation of CKP by
calculation of the hyperfine interaction between the quarks. This
model is used by these authors to carry out an extensive survey of
baryon strong decay couplings, which is described in detail below.
\vskip -12pt
\subsection{models based on instanton-induced interactions}
\vskip -12pt
An alternate QCD-based candidate for the short-range interactions
between quarks is that calculated by
't~Hooft~\cite{'tHooft:1976fv,Shifman:1980uw,Petry:1985mn} from instanton
effects. This interaction is flavor-dependent, and was originally
designed to solve the $\pi$-$\eta$-$\eta^\prime$ puzzle which exists
in quark models of mesons based on one-gluon exchange. An expansion of
the Euclidean action around single-instanton solutions of the gauge
fields assuming zero-mode dominance in the fermion sector leads to an
effective contact interaction between quarks, which acts only if the
quarks are in a flavor anti-symmetric state. The color structure
requires that the quarks be in an anti-triplet of color, which is
always true of a pair of quarks in the ground state antisymmetric 
color configuration of a baryon. The strength of the
interaction is proportional to a divergent interaction which must be
regularized, and so is usually taken to be an adjustable constant. In
the nonrelativistic approximation this leads to an interaction between
quarks in a baryon which has nonzero matrix elements~\cite{Blask:1990ez}
\beq
\<q^2; S, L , T\vert W \vert q^2; S,L,T\>=
-4g\,\delta_{S,0}\,\delta_{L,0}\,\delta_{T,0}\cal{W}
\eeq
of the interaction $W$, where ${\cal W}$ is the radial matrix element of
the contact interaction. Note that the interaction acts only on pairs
in a spin singlet, S-wave, isospin-singlet state. Although not present
in the one-loop calculation of 't~Hooft, higher-order calculations are
expected to regularize the $\delta$-function contact interaction to
yield
\beq
\delta^3({\bf r})\to {1\over \Lambda^3\pi^{3/2}}e^{-r^2/\Lambda^2},
\eeq
so that the radial matrix elements ${\cal W}$ are finite. 

This interaction can be extended to encompass the interactions between
three flavors of quark, with the result that they are completely
antisymmetric in flavor space. An additional parameter $g^\prime$ is
required for the strength of the effective interaction between
nonstrange and strange quarks. The result is an interaction between
quarks with three parameters $g$, $g^\prime$ and $\Lambda$ that causes
no shifts of the masses of decuplet baryon states, but shifts octet
flavor states downward if they contain spin-singlet, S-wave,
flavor-antisymmetric pairs. 

Ground-state baryon mass splittings with instanton-induced
interactions are explored in a simple model by Shuryak and
Rosner~\cite{Shuryak:1989bf}, and in the MIT bag model by Dorokhov and
Kochelev~\cite{Dorokhov:1990mv} and Klabu{\v
c}ar~\cite{Klabucar:1994qf}. Dorokhov and Kochelev's model also
included OGE-based interactions. An extensive study of the meson and
baryon spectra using a string-based confining interaction and
instanton-induced interactions is made by Blask, Bohn, Huber, Metsch,
and Petry (BBHMP)~\cite{Blask:1990ez}. Evidence for the role of
instantons in determining light-hadron structure and quark propagation
in the QCD vacuum is given by Chu, Grandy, Huang, and
Negele~\cite{Chu:1994vi} in their study based on lattice QCD.

The nonrelativistic model of BBHMP~\cite{Blask:1990ez} confines the
quarks in mesons and baryons using a string potential of the kind
proposed by Carlson, Kogut and Pandharipande~\cite{Carlson:1983xi}. In
addition, it has only instanton-induced interactions between the
quarks, and is applied to baryons and mesons of all flavors made up of
$u$, $d$, and $s$ quarks. The model has only eight parameters which
are the three quark masses, the string tension, two energy offset
parameters for mesons and baryons, and the three parameters $g$,
$g^\prime$ and $\Lambda$ of the instanton-induced interactions. It is
able to explain the sign and rough size of the splittings in the
ground state baryons, and also the size of the splittings in certain
P-wave baryons, such as the $N$ and $\Lambda$ states are described
well. The splittings in the $P$-wave $\Sigma$ states are smaller than
the (less certain) splittings extracted from
experiment. Positive-parity excited states tend to be too massive by
about 200-250 MeV. Nevertheless, given the simplicity of the model,
the authors have demonstrated the possibility that
instanton-induced interactions may play an important role in the
determination of the spectrum of light hadrons.
\vskip -12pt
\subsection{Goldstone-boson exchange models}
\vskip -12pt
Many authors have proposed that because of the special nature of the
pion, and to a lesser extent the other members of the octet of
pseudoscalar mesons, one should consider the exchange of pions between
light quarks in nucleon and $\Delta$ baryons as a source of hyperfine
interactions. The best developed early exploration of the consequences
of this for the baryon spectrum is by Robson~\cite{DR}.  Glozman and
Riska have popularized this idea by making an extensive analysis of
the baryon spectrum using a model based on a hyperfine interaction
arising {\it solely} from the exchange of a pseudoscalar octet. They
argue that, contrary to the substantial evidence presented in the
prior literature, there is no evidence for a one-gluon-exchange
hyperfine interaction. For a comprehensive review of this work, see
Ref.~\cite{Glozman:1996fu}. More recently Glozman, Plessas, Varga and
Wagenbrunn~\cite{Glozman:1998ag,Glozman:1998xs,Wagenbrunn:2000sg} (for
the most recent version of this model see Ref.~\cite{Glozman:2000vd})
have extended this model to include the exchange of a nonet of vector
mesons and a scalar meson, and the calculation of radial matrix elements
of the exchange potentials.

\begin{figure}[t]
\vskip 0cm
\vbox{
\hskip 0.2cm
\epsfig{file=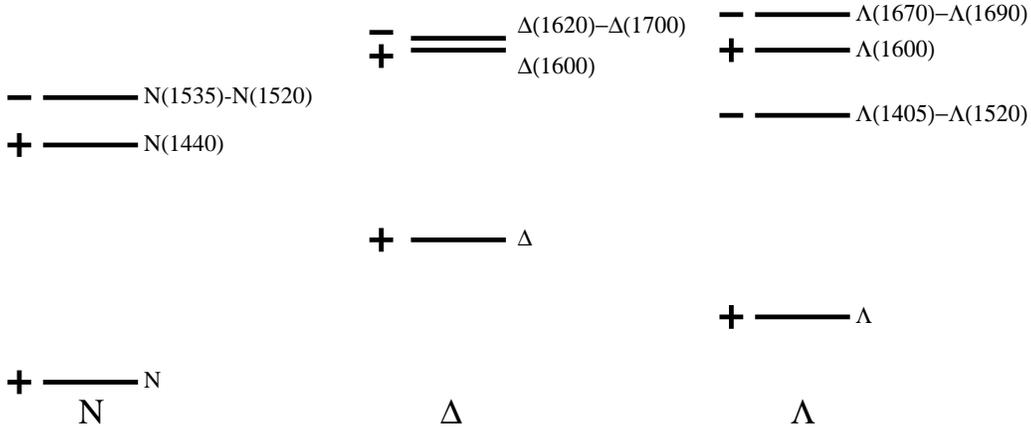,width=13.5cm,angle=0}}
\vskip 0.5 cm
\caption{Low-lying spectra of nucleon, $\Delta$ and $\Lambda$ states
from Ref.~\protect\cite{Glozman:2000vd}.}
\label{GBEorder}
\end{figure}

To illustrate the argument used by these authors, consider
Fig.~\ref{GBEorder} from Ref.~\cite{Glozman:2000vd}. 
The argument is that the order of the states extracted from the
analyses is inverted in the $N$, $\Delta$, and $\Lambda$ spectra
compared to the ($+,-,+$) ordering of the levels using either harmonic
or linear confinement. Of course this ordering can be perturbed by the
one-gluon exchange interaction, and in the case of linear confinement
the splitting between the negative-parity and positive-parity excited
states becomes smaller, but it is still not possible to lower with the
hyperfine contact interaction the mass of the lightest positive-parity
$N\half^+$ excited state below that of the lightest negative-parity
$N\half^-$ state (see, for example, Fig.~\ref{Nexpthr}). For the
$\Delta$ states the one-gluon exchange OGE interaction shifts the
radially excited $\Delta\thalf^+$ state up relative to the negative
parity excited states $\Delta\half^-$ and $\Delta\thalf^-$.  It is
argued that this points conclusively to a flavor-dependence of the
hyperfine interaction, and that the best candidate for such an
interaction is Goldstone-boson exchange (GBE). Note that the situation
becomes less clear if one admits that the masses of these resonances,
which are extracted from difficult analyses which do not always agree,
have a range of possible values. For example, the range in masses
quoted by the PDG~\cite{Caso:1998tx} for the state
`$\Delta\thalf^+(1600)$' is 1550 to 1700 MeV, with more recent
analyses~\cite{MANSA,Vrana:2000nt} at the upper end of this
range. Similarly the state $\Lambda\half^+(1600)$ has a mass range of
1560-1700 MeV. In both cases the greater uncertainty on the mass of
the positive-parity state means that it {\it could} have a mass
roughly the same as, or higher than, the negative parity states. As we
have seen above, there
is also another QCD-based candidate for a flavor-dependent force
between quarks, which is that arising from instanton
effects.

In Ref.~\cite{Glozman:1996fu} the flavor-dependent spin-spin force
has the form 
\beq
H_\chi\sim -\sum_{i<j}
\frac{V({\bf r}_{ij})}{m_i m_j}
\lpmb{\lambda}^{\rm F}_i \cdot \lpmb{\lambda}^{\rm F}_j\,
\lpmb{\sigma}_i \cdot \lpmb{\sigma}_j,
\label{GBE}
\eeq where $\lpmb{\lambda}^{\rm F}_i$ is a Gell-Mann matrix in the
flavor space and the radial dependence of the function $V({\bf
r}_{ij})$ is assumed to be unknown. This interaction can be made to
roughly fit the pattern of mass splittings of Figure~\ref{GBEorder} in
a model with harmonic confinement by choosing 5 parameters, which are
the harmonic oscillator excitation energy $\omega$, and four radial
matrix elements of the pion exchange radial function $V^\pi({\bf
r}_{ij})$. The fitted mass differences are those between the nucleon
and $\Delta\thalf^+(1232)$, $N\half^+(1440)$, and
$\Delta\thalf^+(1600)$, as well as the average masses of the two pairs
of states $N\half^-(1535)$--$N\thalf^-(1520)$, and
$N\thalf^+(1720)$--$N\fhalf^+(1680)$. Four further parameters are used
in the fit to other nonstrange baryon states (up to N=2 in the
harmonic oscillator spectrum), which are the radial matrix
elements of the $\eta$-exchange radial potential between light quarks
$V^{uu}({\bf r}_{ij})$, for a total of nine parameters.

The size of the resulting harmonic oscillator energy $\omega\simeq
160$ MeV is considerably smaller than that required in the
Isgur-Karl model. This acts to lower the splittings between the
harmonic-oscillator levels prior to application of the spin-spin
force. The difference between the oscillator frequencies in the
$\lpmb{\rho}$ and $\lpmb{\lambda}$ systems is neglected. 

In fitting to the strange baryon masses twelve new parameters are
required, which are the four radial matrix elements of the
kaon-exchange potential $V^{K}({\bf r}_{ij})$, and the
light-strange and strange-strange $\eta$-exchange radial potentials
$V^{us}({\bf r}_{ij})$ and $V^{ss}({\bf r}_{ij})$. The difference
between the oscillator frequencies in the $\lpmb{\rho}$ and
$\lpmb{\lambda}$ systems is neglected, which amounts to the adoption
of a flavor-dependent confining force. This gives a total of 23
parameters, including the values of the quark masses $m_u=340$ MeV
and $m_s=461$ MeV, used to fit the spectrum of $N$, $\Delta$,
$\Lambda$, $\Sigma$, $\Xi$, and $\Omega$ baryons up to the $N=2$
band. Given the large number of parameters the work reviewed in
Ref.~\cite{Glozman:1996fu} can be considered a demonstration that a
flavor-dependent contact interaction in Eq.~(\ref{GBE}) can be used to
fit the spectrum. 

Calculations now exist which go beyond a parametrization of the
spectrum in terms of GBE, and have included vector and scalar meson
exchanges~\cite{Glozman:1998xs,Glozman:1998ag,Wagenbrunn:2000sg}.  In
this work the relativistic quark kinetic energy $\sum_i\sqrt{{\bf
p}_i^2+m_i^2}$ is used, as well as a pairwise linear confining
interaction with a strength $C=0.46$ GeV/fm. It can be shown that the
sum of interquark separations $\sum_{i<j=1}^3r_{ij}$ and the sum of
the string lengths in a Y-shaped string connecting the quarks differ
by about a factor of 0.55 in $S$-states~\cite{Capstick:1986bm}. This
means that this linear potential strength is reasonable compared to the
string tension of 1 GeV/fm found in lattice calculations. Note,
however, that a string model of the confining potential contains
genuine three-body forces which are not in a pairwise linear
potential. The Schr\"odinger equation is solved using a variational
calculation in a large harmonic oscillator basis as in
Ref.~\cite{Capstick:1986bm}.  Tensor interactions which are associated
with the GBE contact interactions as well as scalar and vector meson
exchanges are now included. The primary motivation for this appears to
be that certain tensor mixings change sign when going from vector
exchange (like OGE) to GBE, and this makes problematic the strong
decays~\cite{Glozman:1998xs} and nucleon form
factors~\cite{Boffi:1999gt} in a GBE model based only on pseudoscalar
exchange. There are now separate exchange potentials for a
pseudoscalar nonet ($\pi$, $K$, $\eta$ and $\eta^\prime$), a vector
meson nonet ($\rho$, $K^*$, $\omega_8$ and $\omega_0$) and a singlet
scalar meson ($\sigma$).

The result is a complicated model with on the surface a large amount
of freedom to fit the spectrum, since associated with each of the
seventeen exchanged particles is a coupling constant $g_\gamma$, and a
(monopole) meson-quark form factor parameter $\Lambda_\gamma$. In
addition each of the vector mesons couples with a second coupling
constant, {\it i.e.} there are two constants $g^V_\gamma$ and
$g^T_\gamma$ for each vector meson. A range parameter $1/\mu_\gamma$
is presumably fixed by making $\mu_\gamma$ the exchanged particle's
mass in the case of physical mesons. In earlier calculations without
the vector and scalar exchanges the number of parameters is reduced by
assuming that the form factor parameters scale with meson mass as
$\Lambda_\gamma=\Lambda_0+\kappa\mu_\gamma$, and by adopting only two
coupling constants $g_8$ and $g_0$ for the octet and singlet
($\eta^\prime$) members of the pseudoscalar nonet. It is not clear
from Refs.~\cite{Wagenbrunn:2000sg,Glozman:2000vd} how many parameters
are used in the most recent calculations of this group.

\begin{figure}[t]
\vskip 0cm
\vbox{
\hskip 0.2cm
\epsfig{file=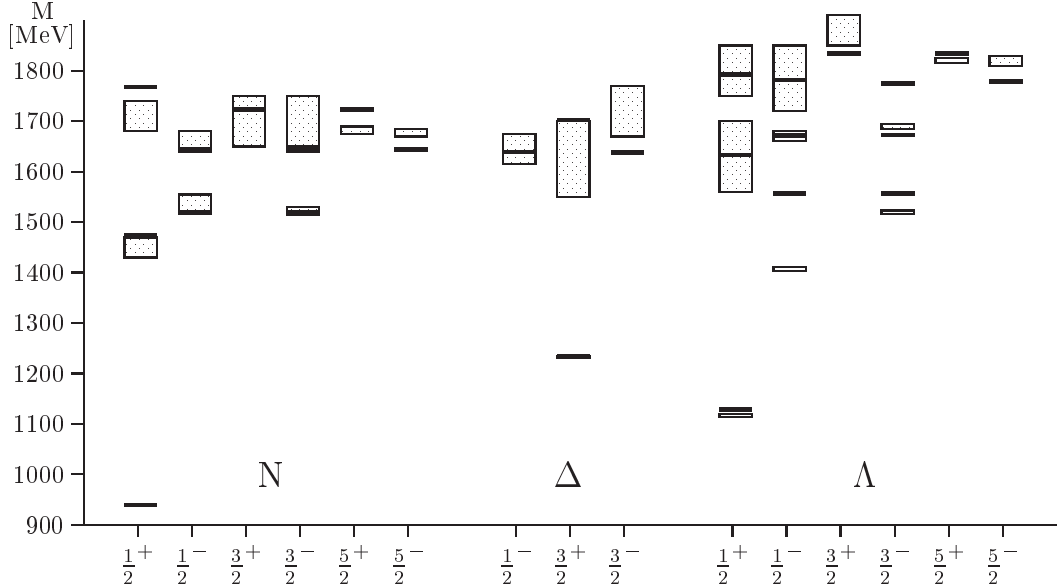,width=14cm,angle=0}}
\vskip 0.5 cm
\caption{Energy levels of low-lying $N$, $\Delta$ and $\Lambda$ baryons
from Refs.~\protect\cite{Wagenbrunn:2000sg,Glozman:2000vd}, compared to the
range of central values for resonances masses from the
PDG~\protect\cite{Caso:1998tx}.}
\label{GBEspecNDL}
\end{figure}

\begin{figure}[t]
\vskip 0cm
\vbox{
\hskip 0.2cm
\epsfig{file=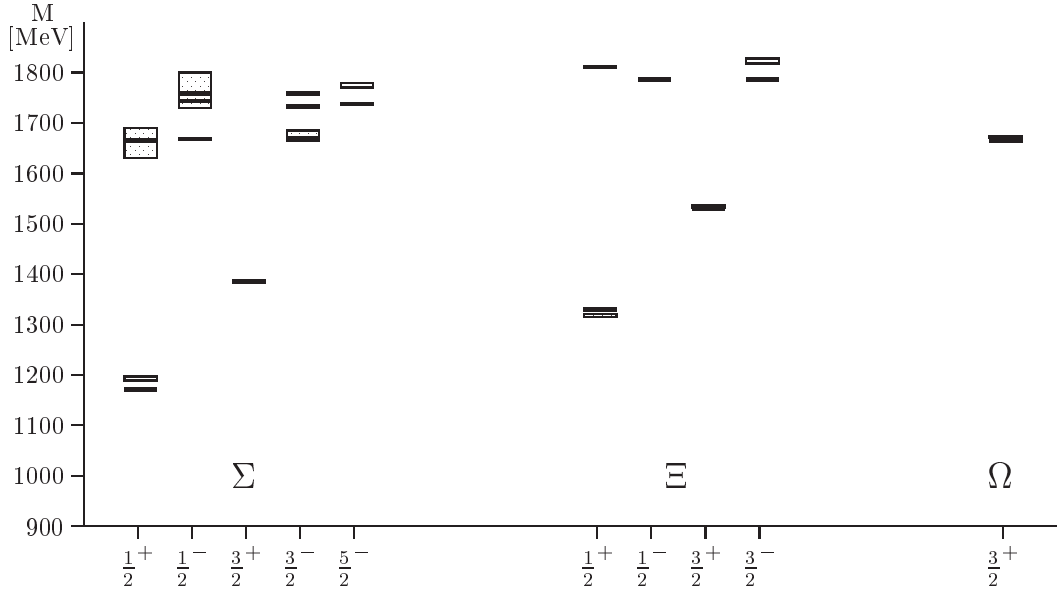,width=14cm,angle=0}}
\vskip 0.5 cm
\caption{Energy levels of low-lying $\Xi$ and $\Omega$ baryons from
Ref.~\protect\cite{Wagenbrunn:2000sg,Glozman:2000vd}. Caption as in
Fig.~\protect\ref{GBEspecNDL}.}
\label{GBEspecSXO}
\end{figure}

Potentials are now given the form expected from a nonrelativistic
reduction of the $T$-matrix element for exchange of the corresponding
meson, and matrix elements of these potentials are calculated in the
harmonic oscillator basis rather than parametrized. Note that the
momentum-dependence of the (Yukawa-like) exchange potentials is still
that given by the nonrelativistic limit, which is somewhat
inconsistent as the quarks exchanging the mesons are off-shell. The
results for the spectrum of $N$, $\Delta$ and $\Lambda$ baryons up to
spin-$\fhalf$ (predictions for $\shalf^+$ baryons are absent) are
shown in Figure~\ref{GBEspecNDL}, and those for $\Sigma$, $\Xi$ and
$\Omega$ baryons are shown in Figure~\ref{GBEspecSXO}. It is not
explained in Refs.~\cite{Wagenbrunn:2000sg,Glozman:2000vd} how model
states which are expected to appear in analyses of the data are chosen
to compare with the spectrum of such states from the PDG. Recent work
within the GBE
model~\cite{Glozman:1998ag,Glozman:1998xs,Wagenbrunn:2000sg,Glozman:2000vd}
does not include $\shalf^+$ baryon states as well as other
higher-lying missing states because the authors believe that in this
mass region both the constituent quark model and a potential model of
confinement are not adequate~\cite{Glozman:2000tk}.
An extension of the model to include strong decays to $N\pi$,
$\Delta\pi$ and $N\eta$~\cite{Glozman:1998xs} has been made using
emission of point-like (elementary) pions and etas (an elementary
emission model), and the pseudoscalar exchange potentials to describe
the masses. The resulting description of the decay widths is described
by the authors as not consistent. The authors attribute this to the
lack of meson structure, and also the lack of configuration mixing
caused by tensor interactions, both of which are shown below to be
important for a model of baryon strong decays. Recent
calculations~\cite{Wagenbrunn:Nstar2k} use the $^3P_0$ model described
below for the strong couplings and the pseudoscalar, vector and scalar
potentials to determine the masses, but have not included the tensor
interactions. The tensor force from pseudoscalar exchange and that
from pseudoscalar, vector and scalar exchanges is qualitatively
different, so exploring its consequences for strong decays in this
model is important.

Given the amount of freedom in the model to fit the spectrum, it is
perhaps not surprising that the fit is of somewhat better quality than
that of the relativized model of Ref.~\cite{Capstick:1986bm}, which
uses 13 parameters to fit the nonstrange baryon spectrum, eight of
which are the same as those used in a similar fit to meson
physics~\cite{Godfrey:1985xj}. Note that because of the special status
given to mesons, it is not clear whether a unified
picture of baryons and mesons could ever emerge in the GBE
model. Although the mass of the first recurrence of $\Delta\thalf^+$
is now above 1700 MeV (at the top end of the range quoted by the PDG)
and {\it heavier} than the negative-parity $\Delta$ states, presumably
because of the effects of the exchange of vector quantum numbers (like
that of OGE), this is still somewhat lighter than the relativized
model mass of $\simeq 1800$ MeV. The Roper resonance can be made
lighter than the $P$-wave $N$ and $\Delta$ resonances. The equivalent
$\Lambda\half^+$ state is still somewhat heavier than the
$\Lambda\half^-$--$\Lambda\thalf^-$ pair, which are predicted
degenerate at about 1550 MeV.

\vskip -12pt
\subsection{spin-orbit interactions in baryons}
\vskip -12pt
As shown above, it is inconsistent to simply ignore the spin-orbit
interactions which are associated with the one-gluon exchange
interaction postulated in several models as the source of the
hyperfine splitting in baryons. There also exist purely kinematical
spin-orbit interactions associated with Thomas precession of the quark
spins in the confining potential which must be taken into account.
Are spin-orbit interactions present in baryons, with a strength
commensurate with the vector-exchange contact interaction and the
confining interaction? Isgur and Karl~\cite{Isgur:1977ef} calculate
the size of these interactions, and show that under certain reasonable
conditions on the potentials, a cancellation of the vector and scalar
spin-orbit interactions occurs for the two-body parts of the
spin-orbit interactions. Note that this cancellation relies on the
Lorentz structure of the confining interaction being a scalar.
However, in a three-body system there are also spin-orbit
interactions involving, say, the orbital angular momentum of the 1-2
quark pair and the spin of the third quark which have no analogue in
bound states of two particles. There is no cancellation between the
three-body spin orbit interactions arising from these two
sources. Inclusion of all of these spin-orbit forces still leads to
unacceptably large spin-orbit splittings, and so Isgur and Karl leave
them out. There is also some evidence for spin-orbit splittings in the
analyses of the data, for example the splitting
$\Delta\thalf^--\Delta\half^-$ in Fig.~\ref{hyperf}. Leaving these
interactions out is unsatisfactory.

In the relativized model the contact interaction is evaluated
nonperturbatively, and the usual $\delta^3({\bf r}_{ij})$ form is
smeared out by relativistic effects and the finite size of the
constituent quarks. The perturbative evaluation of the $\delta^3({\bf
r}_{ij})$ interaction in the nonrelativistic model underestimates its
strength; in the relativized model for the same contact splitting a
value of $\alpha_s=0.6$ is required, significantly smaller than that
required in the nonrelativistic model. The result is a smaller
associated spin-orbit interaction. Some of the freedom to fit the
momentum dependence of the potentials is used to further suppress the
spin-orbit interactions relative to the contact interaction. These
effects, along with a partial cancellation of the vector and scalar
spin-orbit terms (the latter are calculated with a two-body
approximation to the string confining potential) reduce the size of the
spin-orbit interactions to acceptable levels. This calculation shows
that it is possible to construct a model based on one-gluon exchange
in which spin-orbit interactions are treated consistently and still
adequately fit the spectrum.

In these models the splitting between $\Lambda\thalf^-(1520)$ and
$\Lambda\half^-(1405)$ can arise only from a spin-orbit interaction,
{\it if\,} no mass shifts arising from decay-channel couplings [or
$qqq(\bar{q}q)$ configurations] are allowed. In the relativized model
there is very little splitting of these two states from the spin-orbit
interaction. The presence of the nearby threshold for $N\bar{K}$ decay
is expected to strongly affect the mass of
$\Lambda\half^-(1405)$. Obviously {\it any\,} model which ignores the
effects of decay-channel couplings will not be able to fully explain
this splitting. As mentioned above, the cloudy bag model calculations
of Veit, Jennings, Thomas and Barrett\cite{Veit:1984an,Veit:1985jr}
and Jennings~\cite{Jennings:1986yg} allow an unstable $\bar{K}N$ bound
state to mix with a three-quark bound state and find that the low mass
of the $\Lambda(1405)$ can be explained by a small (14\%) intensity
for the quark model bound state in the $\Lambda(1405)$. This model
predicts another $\Lambda\half^-$ state close in mass to the
$\Lambda(1520)$ in a region where one has not been seen.

In a calculation of baryon masses based on QCD sum rules,
Leinweber~\cite{Leinweber:1990hh} finds a large spin-orbit splitting
in the $\Lambda\thalf^-(1520)$--$\Lambda\half^-(1405)$ system while
maintaining a small splitting in the
$N\thalf^-(1520)$--$N\half^-(1535)$ system. A simple approximate
formula for the ratio of the masses of the $J^P=\half^-$ states is
derived that gives $\Lambda\half^-/N\half^- =
\<\bar{s}s\>/\<\bar{u}u\>$, which shows that the
$\Lambda\half^-(1405)$ becomes lighter than the $N\half^-(1535)$
because of the reduced size of the strange quark condensate in this
approach. Note that the mechanism by which the two nucleon states
remain degenerate in mass while the two $\Lambda$ states become split
is apparently complicated. Although the splittings of the above states
are described rather well, theoretical errors are substantial compared
to the spin-orbit splittings. Furthermore, because of the complexity
of this approach, only six baryon states are considered (the four
$P$-wave states above and the ground state nucleon and $\Lambda$) and
all of the nucleon states are calculated to be 50-75 MeV too light,
whereas the calculated $\Lambda$ state masses are within 25 MeV of the
physical masses. The central result is that, in this approach, the
large $\Lambda$ spin-orbit splitting is not due to coupling to the
$\bar{K}N$ scattering channel, but rather to the reduced strange quark
condensate relative to the $u$ or $d$ condensates. This is in contrast
to the cloudy-bag~\cite{Veit:1984an,Veit:1985jr,Jennings:1986yg} and
chiral potential model~\cite{Kaiser:1995eg} descriptions of the
$\Lambda(1405)$ detailed above.

The development by Glozman and Riska (see
Refs.~\cite{Glozman:1996fu,Glozman:2000vd} and references therein) of
a model for nonstrange and strangeness -1 baryon masses with hyperfine
interactions induced by Goldstone-boson-exchange (GBE) between the
quarks was partially motivated by the lack of associated spin-orbit
interactions. It is claimed that this supports the hypothesis that the
hyperfine interactions responsible for many features of the baryon
spectrum are due to GBE and not one-gluon exchange (OGE).  However, as
pointed out by Isgur~\cite{Isgur:1999jv}, it is still necessary to
confine the quarks in such a model. Glozman and Riska use harmonic
confinement, although more recent
calculations~\cite{Glozman:1998ag,Wagenbrunn:2000sg} have used linear
confinement, and as noted above such confining forces produce
spin-orbit interactions through Thomas precession. Isgur argues that,
in addition to the usual problems with the three-body spin-orbit
interactions, the cancellation of the two-body components of the
Thomas-precession spin-orbit forces which can be arranged with the OGE
hyperfine interaction will be spoiled with GBE hyperfine
interactions. This is precisely because the two-body spin-orbit
interactions which usually arise from the hyperfine interaction are
not present. Isgur points out that with OGE such a cancellation is
able to explain the small size of spin-orbit interactions in mesons.

In a recent paper on meson-like $\Lambda^*_Q$ baryons, where $Q$ is a
heavy quark, Isgur~\cite{Isgur:2000rf} shows that their
spin-independent spectra are remarkably like those of the analogous
mesons. Such states have orbital angular momentum only between the
light-quark pair and the heavy quark $Q$, so that the only spin-orbit
forces are those on the heavy quark $Q$. Spin-orbit interactions in
such states are shown to be small due to a cancellation between
one-gluon exchange and Thomas-precession spin-orbit forces which
occurs with the assumption of Lorentz scalar confinement. For such
states the three-body spin-orbit interactions from the OGE and
confining interactions conspire to produce only meson-like
quasi-two-body spin-orbit forces. Isgur also argues that the states
$^2\Sigma^*_Q$ and $^4\Lambda^*_Q$ have the same spatial wave function
as $\Lambda^*_Q= {^2\Lambda}^*_Q$, so that in these states the spin-orbit
forces on the heavy quark $Q$ (which are not the only such forces
present) also exhibit this cancellation. He concludes that a
careful reanalysis of spin-orbit splittings is required, since a
nonrelativistic solution to the size of the spin-orbit splittings in
mesons and $\Lambda^*_Q$ states is possible, although it is still
possible that relativistic effects have produced a gross enhancement
of spin-spin over spin-orbit interactions in baryons, as suggested in
Ref.~\cite{Capstick:1986bm}. 
\vskip -12pt
\subsection{diquark and collective models}
\vskip -12pt
Models of baryon structure exist which describe the nucleon (for a
review see Anselmino {\it et al.}~\cite{Anselmino:1993vg}) and its
excitations~\cite{Goldstein:1988us} in terms of a diquark and a
quark. If the diquark is tightly bound its internal excitation is
costly in energy, so that the low-lying excitations of the nucleon
will not include excitations of the diquark. Early models assumed
that, because of an attractive hyperfine interaction between a $u$ and
$d$-quark in the isospin-zero spin-zero channel, there should be a
tightly-bound isoscalar scalar diquark in the proton and other
baryons. In SU(6) language this means that, in addition to the
lightest negative-parity nonstrange excitations in the [70,1$^-$]
multiplet which have all been seen in $N\pi$ elastic scattering, the
low-lying positive-parity nonstrange excitations of the nucleon should
lie in symmetric 56-plet and mixed-symmetry 70-plet
representations. Models which treat all three quarks symmetrically
have more low-lying excitations. Light positive-parity excited states
are present in the [56$^\prime$,0$^+$], [70,0$^+$], [56,2$^+$],
[70,2$^+$], and [20,1$^+$] SU(6) multiplets, with $J^P=1/2^+$,
$3/2^+$, $5/2^+$, and $7/2^+$, for a total of twenty-one
positive-parity nonstrange excited states. Note that some
diquark models also allow for the quark and diquark to interact and
exchange a quark and so their identity in order to maintain overall
antisymmetry~\cite{Goldstein:1988us}, and may have similar numbers of
positive-parity excited states, although the [20,1$^+$] multiplet is
still excluded.

Of the twenty-one low-lying positive-parity nonstrange excited states
predicted by symmetric quark models, nine are considered well
established by the Particle Data Group (PDG)~\cite{Caso:1998tx}, there are
three tentative (one or two stars) nucleon states, and two tentative
$\Delta$ states, for a total of fourteen, of which five need
confirmation. The remaining predicted states can be defined as
missing~\cite{KI}. This definition can be expanded to include any
baryon predicted by symmetric quark models but noticeably absent in
the analyses, such as the many higher mass negative-parity states
above 1900 MeV. Although not discussed in detail here, the situation
is similar for the strange baryons $\Lambda$ and $\Sigma$, except
there are fewer excited states present in the analyses, with a few
missing low-lying negative-parity states and more missing
positive-parity states.

These states may be missing because of strong diquark clustering in
the light-quark baryons. There is some evidence from the
lattice~\cite{Leinweber:1993nr} and from an analysis of baryon strong
decays~\cite{Forsyth:1981tk} that such strong diquark clustering may
not be present. The explanation adopted here, and in the work of
others~\cite{KI}, is that such states have either small
$N\pi$ couplings in a partial wave which includes other light strongly
coupled states, or are close in mass to a more strongly coupled state,
both of which make extraction of a signal from $N\pi$ elastic
scattering difficult. This explanation can be understood using a quark
model of the spectrum and wave functions of these states and their
strong decay, which are described below. This model can then be used
to show how such states can be found.

A collective model of baryon masses, electromagnetic couplings, and
strong decays based on a spectrum-generating algebra has been
developed by Bijker, Iachello and Leviatan
(BIL)~\cite{Bijker:1994yr,Bijker:1996ii,Bijker:2000gq}. The idea is to
extend the algebraic approach, which led to the mass formulae based on
flavor-spin symmetry described above, to the description of the
spatial structure of the states. The quantum numbers of the states are
considered to be distributed spatially over a Y-shaped string-like
configuration, which is sometimes idealized as being a thin string
with a distribution of mass, charge and magnetic moments. BIL also
apply their algebraic model to a single-particle valence quark picture
for comparison purposes.

The approach to the dynamics is not the usual solution of some
Schr\"odinger-like equation, but rather bosonic quantization of the
spatial degrees of freedom, which in this case are the two relative
coordinates $\lpmb{\rho}$ and $\lpmb{\lambda}$ of
Eq.~(\ref{rholam}). This leads to six vector boson operators bilinear in
the components of these coordinates and their conjugate momenta, plus
an additional scalar boson, which generate the Lie algebra of
U(7). This means that the spectrum-generating algebra of the baryon
problem is taken to be U(7)$\otimes$SU(3)$_{\rm f}\otimes$SU(2)$_{\rm
spin}\otimes$SU(3)$_{\rm c}$.

Bases are constructed by operating with the boson operators on the
vacuum. Mass and electromagnetic coupling calculations require
construction of a complete set of basis states for representations of
U(7), and these are obtained by considering subalgebras. A basis
corresponding to two coupled three-dimensional harmonic oscillators,
corresponding to the nonrelativistic and relativized quark models, is
used for the mass calculations. A second basis corresponding to a
three-dimensional oscillator and a three-dimensional Morse oscillator
coupled together is more convenient for evaluation of the
electromagnetic coupling strengths.

All mass operators can then be expanded into elements of U(7) which
transform as irreducible representations of the rotation group SO(3)
and exchange symmetry $S_3$. Instead of expanding the Hamiltonian used
in other approaches, the mass-squared operator is expanded in terms of
operators of the algebra U(7), which is appropriate for a relativistic
system. For identical constituents, the mass-squared operator is
constrained to be symmetric under the exchange group. BIL then write
down the most general mass-squared operator which is a scalar under
rotations and the exchange group and which preserves parity, and which
is at most quadratic in the elements of U(7). For strange baryons a
more general operator which is not necessarily symmetric under the
exchange group $S_3$ is employed. The masses and wave functions
corresponding to this mass operator are then found by diagonalization
in either of the two bases described above.

By choosing coefficients of the tensor structures in the mass-squared
operator, BIL are able to make contact with harmonic oscillator quark
models, although these are usually written for the mass and not the
mass squared. They are also able to describe collective (string-like)
models with the three constituents moving in a correlated way. For the
latter model, the mass-squared operator is rewritten to describe
vibrations and rotations of the string-like configuration. The
vibrational part of the mass-squared operator has fundamental
vibrational modes which correspond to breathing and bending modes of
the strings. The rotational part does not reproduce the linear rise of
the mass squared with orbital angular momentum (linear Regge
trajectories) which is approximately reproduced by using a linear
potential and a relativistic kinetic energy operator in the usual
quark potential models. A more complicated alternate form for the
rotational part of the mass-squared operator is constructed to
reproduce this behavior, in terms of the orbital angular momentum and
its projection onto the three-fold symmetry axis of the string
configuration. This corresponds to the rotational spectrum of an oblate
top. This choice introduces vibration-rotation interactions, which are
dropped. Also the part of the rotational term in the mass-squared
operator dependent on the projection of the angular momentum onto the
symmetry axis is dropped as there is no evidence for such a term in
the spectrum. 

The resulting orbital spectrum has four parameters which are fit to
the spectrum. It describes positive-parity excitations with $L^P=0^+$
as the lightest of a group of one-phonon vibrational excitations, so
that they lie below a parity-doublet of states with $L^P=1^+,1^-$. The
spin-flavor part of the mass-squared operator is then written in terms
of six generators (Casimir operators) of the spin-flavor algebra. All
terms which are not in the diagonal part of the operator are
dropped. The operator simplifies into three terms (each multiplied by
a parameter) in the case of nonstrange baryons, which are the usual
spin-spin interaction, a flavor-dependent interaction, and `exchange'
terms which depend on the permutation symmetry of the
wave functions. Finally, there is the possibility of operators which
involve both internal and spatial degrees of freedom such as the
hyperfine tensor and spin-orbit interactions. These interactions are
not considered, which has consequences for the electromagnetic and
strong decay amplitudes calculated with the eigenstates of this
mass-squared operator.

The result of the choices outlined above, some of which have been made
with the spectrum in mind, is a mass-squared operator with seven
parameters which are fit to the nonstrange baryon spectrum. The
spin-orbit problem, described in detail above, is simply avoided by
leaving out such interactions. The `exchange' terms in the general
form of the spin-flavor dependent part of the mass operator are found
to be crucial to fitting the higher unperturbed position of the P-wave
baryons relative to the ground state and positive-parity excited
states, which was put in by hand in the Isgur-Karl model (see
Fig.~\ref{Nexpthr}) and remains a problem at the $\simeq$100 MeV level
in the relativized model. As is the case with the relativized models
of Refs.~\cite{Godfrey:1985xj,Capstick:1986bm}, there is consistency
between the slope of the Regge trajectory and of the spin-spin
interactions found here and in mesons~\cite{Iachello:1991re}.  This
model predicts more missing baryons, with different quantum
numbers and masses, than the valence quark models. The lightest of the
missing nucleon states are two states in an antisymmetric $[20,1^+]$
multiplet at 1720 MeV, with $J^P=\half^+$ and $\thalf^+$. The presence
of these states is an easily testable consequence of this model, as
their presence is predicted in addition to the model states assigned
to the resonances $N\half^+(1710)$ and $N\thalf^+(1720)$ seen in the
analyses. It is also necessary to show that these light missing states
couple relatively weakly to the $\pi N$ formation channel in a
strong-decay analysis to explain why they have not been seen. In
Ref.~\cite{Bijker:2000gq} these states are shown to have zero partial
widths into all strong decay channels considered ($N\pi$, $N\eta$,
$\Sigma K$, $\Lambda K$, and $\Delta\pi$). However, if there are
present tensor (or spin-orbit) interactions these will cause mixings
with nearby states which should allow these (presumably quite narrow)
states to be seen. It is also possible that these states decay to
channels not considered by BIL and likely to contain missing baryons,
like $N\rho$.

\begin{figure}[t]
\vskip 0cm
\vbox{
\hskip 0.2cm
\epsfig{file=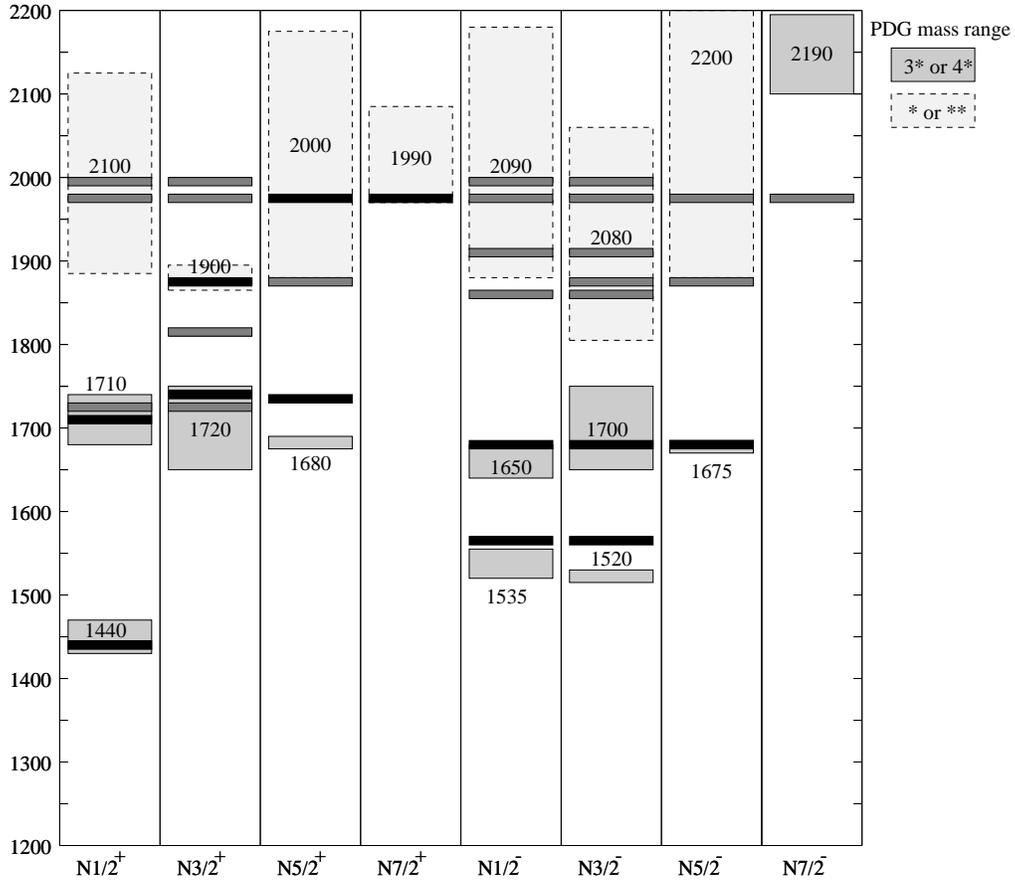,width=13.8cm,angle=0}}
\vskip 0.5 cm
\caption{Mass predictions for nucleon states from the collective
algebraic model of Bijker, Iachello, and
Leviatan~\protect\cite{Bijker:1994yr}, shown as bars, compared to the
range of central values for resonances masses from the
PDG~\protect\cite{Caso:1998tx}, which are shown as boxes. Model states which
are assigned to experimental states by virtue of their masses and
decay couplings are shown as darker-shaded bars, `missing' states as
lighter shaded bars. The ground state nucleon mass from this model is 939
MeV.}
\label{BILnucleon}
\end{figure}

\begin{figure}[t]
\vskip 0cm
\vbox{
\hskip 0.2cm
\epsfig{file=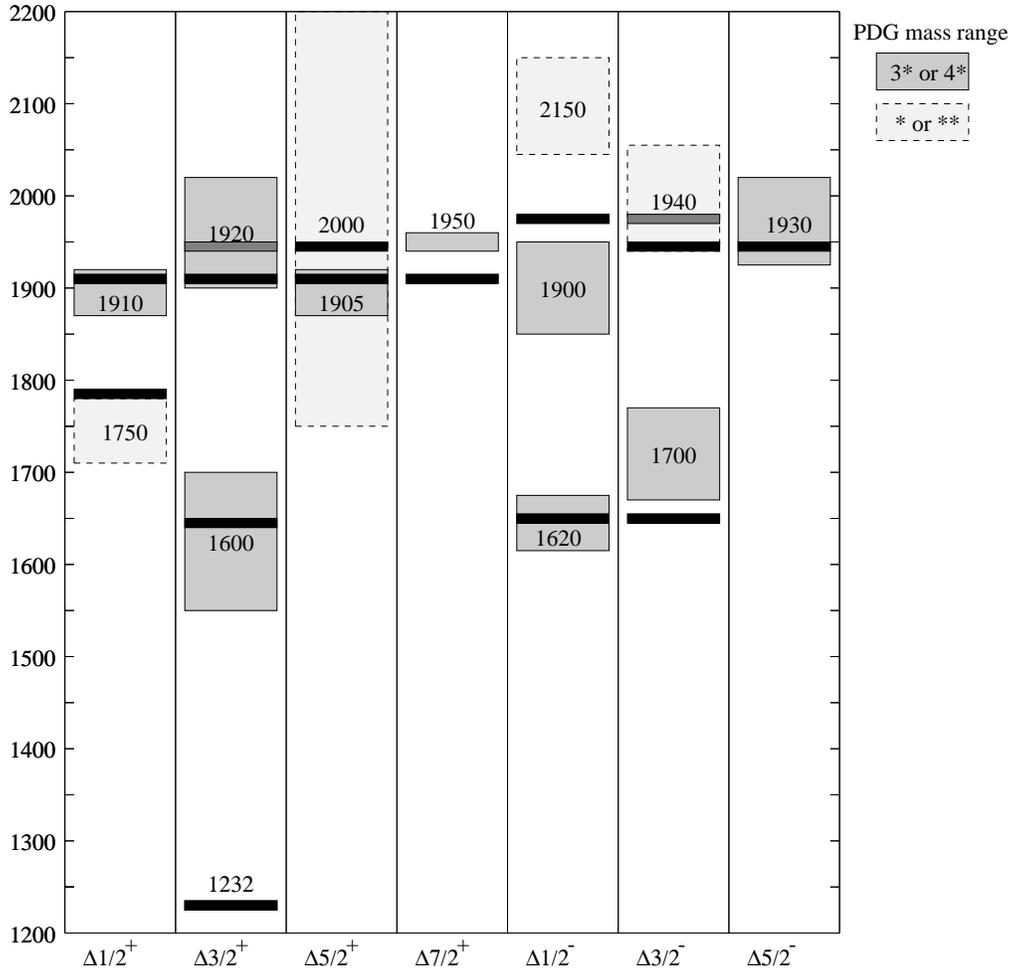,width=13.8cm,angle=0}}
\vskip 0.5 cm
\caption{Mass predictions for $\Delta$ states from the collective
algebraic model of Bijker, Iachello, and
Leviatan~\protect\cite{Bijker:1994yr}, shown as bars, compared to the
range of central values for resonances masses from the
PDG~\protect\cite{Caso:1998tx}, which are shown as boxes. Caption as
in Fig.~\protect\ref{BILnucleon}.}
\label{BILdelta}
\end{figure}

The resulting spectrum of nucleon and $\Delta$ states is shown in
Figs.~\ref{BILnucleon} and~\ref{BILdelta}. Only those model states
below 2 GeV are shown, although the plots go to higher energies to
display the errors on the states seen in the analyses. Model states
which have been assigned to resonant states from the analyses are
distinguished from `missing' states by having darker-shaded bars
representing the mass predictions. The model is able to fit the mass
of the Roper resonance, and correctly fit the centers of gravity of
the P-wave and low-lying positive-parity states, due to the additional
`exchange' terms in the mass-squared operator. The lightest
$N\shalf^-$ state is 150-200 MeV too light compared to the four star
state $N\shalf^-(2190)$, and from the less certain extractions of
masses of other multiply-excited negative-parity states, it appears
that this entire band of nucleon states is generally predicted too
light.

Although an extension of this model involving four extra parameters to
the  $\Lambda$, $\Sigma$, $\Xi$ and $\Omega$ baryons has very recently been
made by BIL~\cite{Bijker:2000gq}, space does not allow a description of the
results of this calculation here, except that the fit is of
comparable quality and that the model also predicts more missing
strange baryons than valence quark models. Strong decays in the BIL
model are described in detail below.
\vskip -12pt
\subsection{P-wave nonstrange baryons in large $N_c$ QCD}
\vskip -12pt
Recently Carlson, Carone, Goity, and
Lebed~\cite{Carlson:1998gw,Carlson:1998vx} have examined the masses of
the nonstrange P-wave baryons using a mass operator analysis in large
$N_c$ QCD. The approach is to write down all possible independent
operators which could appear in the effective mass operator, and order
those operators by their size in a $1/N_c$ expansion. The result is
that the mass operator contains, for two quark flavors, 18
spin-singlet flavor-singlet operators to order $N_c^{-2}$. As there
are seven masses and two mixing angles (those between the two
$N\half^-$ states and the two $N\thalf^-$ states), the hierarchy in
$1/N_c$ is initially used to select a set of nine operators which are
able to efficiently describe these quantities. In
Ref.~\cite{Carlson:1998gw} it is found that only a few of the
coefficients in the effective Hamiltonian turn out to be of natural
size, with the rest being small or consistent with zero. Further
operator analysis in Ref.~\cite{Carlson:1998vx} allows fits to be made
just to the masses, and the resulting fit can be used to predict the
mixing angles. An adequate fit can be made using just three operators,
although this fit does not give the best results for the mixing
angles. The set of three operators have quantum numbers consistent
with single-pion exchange between the quarks, but are not easily compatible
with other simple models such as OGE.
\vskip -12pt
\subsection{hybrid baryon masses}
\vskip -12pt
Current experiments which search for new baryons using electromagnetic
probes, such as those in Hall B at TJNAF, will also produce hybrid
baryon states. These are states which are described in quark potential
models as having explicit excitation of the gluon degrees of
freedom. Low-lying baryon states present in analyses of $\pi N$
elastic and inelastic scattering, such as the Roper resonance
$N\half^+(1440)$, have been
proposed~\cite{Li:1991sh,Li:1992yb,Kisslinger:1995yw,Kisslinger:1998xa}
as hybrid candidates. This is based on extensions of the MIT bag
model~\cite{Barnes:1983fj,Golowich:1983kx,Carlson:1983er,Duck:1983ju}
to states where a constituent gluon in the lowest energy transverse
electric mode combines with three quarks in a color octet state to
form a colorless state, and on a calculation using QCD sum
rules~\cite{Kisslinger:1995yw,Kisslinger:1998xa}. Hybrid baryons have
also been constructed recently in the large-$N_c$ limit of
QCD~\cite{Chow:1999tq}.

With the assumption that the quarks are in an S-wave spatial ground
state, and considering the mixed exchange symmetry of octet color
wave functions of the quarks, bag-model constructions show that adding
a $J^P=1^+$ gluon to three light quarks with total quark-spin 1/2
yields both $N$ ($I=\frac{1}{2}$) and $\Delta$ ($I=\frac{3}{2}$)
hybrids with $J^P=\sfrac{1}{2}^+$, $\sfrac{3}{2}^+$. Quark-spin 3/2
hybrids are $N$ states with $J^P=\sfrac{1}{2}^+$, $\sfrac{3}{2}^+$,
and $\sfrac{5}{2}^+$. Energies are estimated using the usual bag
Hamiltonian plus gluon kinetic energy, additional color-Coulomb
energy, and one-gluon exchange plus gluon-Compton O($\alpha_s$)
corrections. Mixings between $q^3$ and $q^3g$ states from gluon
radiation are evaluated. If the gluon self-energy is included, the
lightest $N$ hybrid state has $J^P=\frac{1}{2}^+$ and a mass between
that of the Roper resonance and the next observed $J^P=\sfrac{1}{2}^+$
state, $N\half^+(1710)$. A second $J^P=\frac{1}{2}^+$ $N$ hybrid and a
$J^P=\frac{3}{2}^+$ $N$ hybrid are expected to be 250 MeV heavier,
with the $\Delta$ hybrid states heavier still. A similar mass estimate
of about 1500 MeV for the lightest hybrid is attained in the QCD sum
rules calculation of~\cite{Kisslinger:1995yw,Kisslinger:1998xa}. 

These results are interesting, given the controversial nature of the
Roper resonance and its $\Delta\thalf^+(1600)$ equivalent in OGE-based
potential models, and the difficulties global models of the
electromagnetic couplings of baryons have accommodating the substantial
Roper resonance photocouplings. As the $N\half^+(1710)$ and its
photocouplings are quite well described by conventional models, if the
Roper is a hybrid then there should be another $P_{11}$ state in the
mass region from 1440-1710 MeV. Evidence for two resonances near 1440
MeV in the $P_{11}$ partial wave in $\pi N$ scattering was
cited~\cite{Arndt:1985vj}, which would indicate the presence of more
states in this energy region than required by the $q^3$ model, but
this has been interpreted as due to complications in the structure of
the $P_{11}$ partial wave in this region~\cite{Cutkosky:1990zh} (see
also Ref.~\cite{Krehl:1999km}), and not an additional physical state.

A recent calculation~\cite{Capstick:1999qq} of hybrid baryon masses in
the flux-tube model~\cite{Isgur:1983wj,Isgur:1985bm} finds that the
lightest hybrid baryons have similar good quantum numbers, but
substantially higher energies and different internal structure than
predicted using bag models. This model structure of the glue, where
the gluon degrees of freedom collectively condense into flux-tubes, is
very different from the constituent-gluon picture of the bag model and
large-$N_c$ constructions. It is based on an expansion around the
strong-coupling limit of the Hamiltonian formulation of lattice QCD,
and on the assumption that the dynamics relevant to the structure of
hybrids is that of confinement. The dynamics is treated in the
adiabatic approximation, where the quarks do not move in response to
the motion of the glue (apart from moving as a rigid body in order to
maintain the center-of-mass position). Flux lines (strings) with
energy proportional to their length play the role of the glue, which
are modeled by equal mass beads with a linear potential between
nearest neighbors~\cite{Isgur:1983wj,Isgur:1985bm}. The total mass of
all of the beads is given by the energy in the flux lines, which is
fixed by the string tension. Beads are allowed to move in a plane
perpendicular to their rest positions.

The ground state energy of this configuration of beads representing
the Y-string for definite quark positions ${\bf r}_i$ defines an
adiabatic potential $V_B({\bf r}_1,{\bf r}_2,{\bf r}_3)$ for the
quarks, which consists of the string energy $b\sum_i l_i$, where $b$
is the string tension and $l_i$ is the magnitude of the vector ${\bf
l}_i$ from the equilibrium junction position to the position of quark
$i$, plus the zero-point energy of the beads. The energy of the first
excited state defines a new adiabatic potential $V_H({\bf r}_1,{\bf
r}_2,{\bf r}_3)$. It is shown in Ref.~\cite{Capstick:1999qq} that a
reasonable approximation to the string ground state and first excited
state adiabatic surfaces may be found by allowing only the junction to
move, while the strings connecting the junction to the quarks follow
without excitation. These adiabatic surfaces are found numerically
{\it via} a variational calculation. Hybrid baryon masses are then
found by allowing the quarks to move in a confining potential given by
the linear potential $b\sum_i l_i$, plus $V_{H}-V_B$, with the rest of
the dynamics as in the relativized model calculation
of~\cite{Capstick:1986bm} except that spin-dependent terms are
neglected. When added to the spin-averaged mass of the $N$ and
$\Delta$ which is 1085 MeV, hybrids with quark orbital angular momenta
$L_q=0,1,2$ have masses 1980, 2340 and 2620 MeV
respectively. Hyperfine (contact plus tensor) interactions split the
$N$ hybrids down and the $\Delta$ hybrids up by similar amounts, so
that the $N$ hybrid mass becomes 1870 MeV. The error in this mass, due
to uncertainties in the parameters, is estimated to be less than $\pm
100$ MeV. This lightest ($L_q=0$) hybrid level is substantially higher
than the roughly 1.5 GeV estimated from bag model and QCD sum rules
calculations.

Taking into account the exchange symmetry and angular momentum of the
excited string, the calculation of Ref.~\cite{Capstick:1999qq} finds
that the lightest hybrid baryons are $N\half^+$ and $N\thalf^+$ states
with quark spin of $\half$ and masses of around 1870 MeV. The
$\Delta\half^+$, $\Delta\thalf^+$ and $\Delta\fhalf^+$ hybrids have
quark spin of $\thalf$ and masses of around 2090 MeV. The lightest
nucleon states are in the region of the `missing' $P_{11}$ and
$P_{13}$ resonances predicted by most models. Of course there will be
mixing between conventional excitations (based on the glue in its
ground state) and these hybrid states, so it is expected that the
physics of the $N\half^+$ and $N\thalf^+$ baryons in the 1700-2000 MeV
region will be complicated. Nevertheless, a careful examination of the
states in this region, with multiple formation and decay channels, may
turn up evidence for these new kinds of excitations.
\vskip -12pt
\section{Baryon electromagnetic couplings}
\begin{figure}[t]
\unitlength 0.8pt
\begin{center}
\begin{picture}(440,130)
\put(0,10){
\begin{picture}(150,100)
\put(0,30){\line(1,0){150}}
\put(10,15){$N$}
\put(128,15){$N$}
\put(0,77){
\begin{picture}(0,0)
\multiput(0,0)(10,-10){5}
{
\put(0,0){\oval(10,10)[tr]}
\put(10,0){\oval(10,10)[bl]}
}
\end{picture}
}
\put(30,65){$\gamma$}
\put(55,30){\circle*{6}}
\put(55,32){\line(1,0){40}}
\put(53,40){$N^{(*)}, \Delta^{(*)}$}
\put(95,30){\circle*{6}}
\multiput(150,85)(-15,-15){4}
{
\put(0,0){\line(-1,-1){12.5}}
}
\put(120,65){$\pi$}
\put(35,-10){(a) $s$-channel}
\end{picture}
}
\put(165,10){
\begin{picture}(140,116)
\put(0,30){\line(1,0){140}}
\put(10,15){$N$}
\put(118,15){$N$}
\put(7,115){
\begin{picture}(0,0)
\multiput(0,0)(10,-10){9}
{
\put(0,0){\oval(10,10)[tr]}
\put(10,0){\oval(10,10)[bl]}
}
\end{picture}
}
\put(35,105){$\gamma$}
\put(40,30){\circle*{6}}
\put(40,32){\line(1,0){60}}
\put(48,15){$N^{(*)}, \Delta^{(*)}$}
\put(100,30){\circle*{6}}
\multiput(126,116)(-15,-15){6}
{
\put(0,0){\line(-1,-1){12.5}}
}
\put(105,105){$\pi$}
\put(30,-10){(b) $u$-channel}
\end{picture}
}
\put(320,10){
\begin{picture}(120,130)
\put(0,30){\line(1,0){120}}
\put(10,15){$N$}
\put(108,15){$N$}
\put(-3,125){
\begin{picture}(0,0)
\multiput(0,0)(10,-10){6}
{
\put(0,0){\oval(10,10)[tr]}
\put(10,0){\oval(10,10)[bl]}
}
\end{picture}
}
\put(60,30){\circle*{6}}
\multiput(60,30)(0,10){4}
{\line(0,1){5}}
\put(64,50){$\pi, \dots$}
\put(60,70){\circle*{6}}
\multiput(120,130)(-15,-15){4}
{
\put(0,0){\line(-1,-1){12.5}}
}
\put(90,110){$\pi$}
\put(30,-10){(c) $t$-channel}
\end{picture}
}
\end{picture}
\end{center}
\caption{Born diagrams for pion photoproduction.}
\label{Born}
\end{figure}
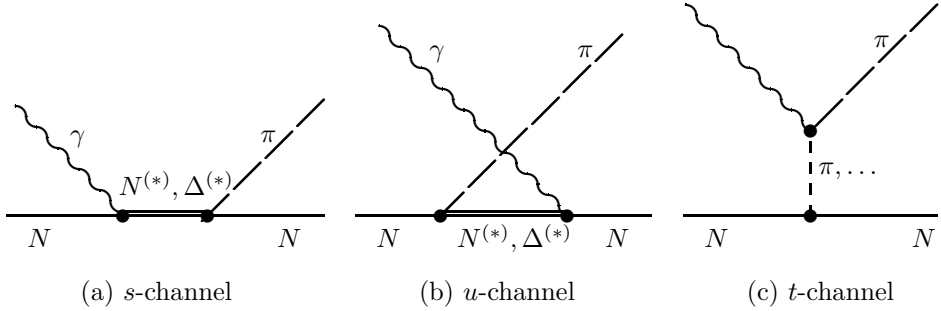

\vskip -12pt 
In Born approximation, single-pion photoproduction involves a
combination of an electromagnetic (EM) excitation process for an
excited baryon and a strong decay (in the $s$ and $u$ channels), or a
meson EM transition in the $t$ channel (see Figure~\ref{Born}). In addition, a
contact diagram is required to maintain gauge invariance in a
nonrelativistic treatment. Data for this process have been analyzed to
come up with a set of photocouplings, which give the strength of
electromagnetic transitions $\gamma N\to X$ between the nucleons and
excited baryon states. To do this, the magnitude of the strong decay
amplitudes are divided out of the product $A_{\gamma N\to X}\cdot
A_{X\to N\pi}$, to form the quantity
\beq
{A_{\gamma N\to X}\cdot A_{X\to N\pi}\over \vert A_{X\to N\pi} \vert}
\eeq
which is inclusive of the phase of the amplitude $A_{X\to N\pi}$,
which cannot be measured in $\pi N$ elastic scattering. It is these
amplitudes which are quoted as photocouplings in the
PDG~\cite{Caso:1998tx}. The calculation of these strong decay
amplitudes will be detailed below, and this can be used to find a set
of signs of these amplitudes for a given spectral model and set of
wave functions. Here the calculation of EM excitation amplitudes in
the nonrelativistic model is outlined, and a fit of such a model to
the values for these photocouplings extracted from analyses of the
data is examined.

Evaluation of the strength of electromagnetic transitions $\gamma N\to
X$ between the nucleons and excited baryon states involves finding
matrix elements of an EM transition Hamiltonian between states in the
harmonic-oscillator basis, and is similar to the evaluation of the
interquark Hamiltonian described above.

\begin{figure}[t]
\unitlength 1.0pt
\begin{center}
\begin{picture}(300,150)
\put(5,24){$N(-{\bf k})$}
\put(100,30){\oval(20,40)}
\put(92,24){$\Psi_N$}
\put(258,24){$X({\bf 0})$}
\put(200,30){\oval(20,40)}
\put(192,24){$\Psi_X$}
\put(110,40){\line(1,1){40}}
\put(110,40){\vector(1,1){22}}
\put(110,65){$p,s$}
\put(150,80){\line(1,-1){40}}
\put(150,80){\vector(1,-1){22}}
\put(172,65){$p^\prime,s^\prime$}
\put(110,30){\line(1,0){80}}
\put(110,30){\vector(1,0){42}}
\put(110,20){\line(1,0){80}}
\put(110,20){\vector(1,0){42}}
\put(50,40){\line(1,0){40}}
\put(50,40){\vector(1,0){22}}
\put(50,30){\line(1,0){40}}
\put(50,30){\vector(1,0){22}}
\put(50,20){\line(1,0){40}}
\put(50,20){\vector(1,0){22}}
\put(210,40){\line(1,0){40}}
\put(210,40){\vector(1,0){22}}
\put(210,30){\line(1,0){40}}
\put(210,30){\vector(1,0){22}}
\put(210,20){\line(1,0){40}}
\put(210,20){\vector(1,0){22}}
\put(150,80){\circle*{2}}
\put(155,100){$\gamma({\bf k})$}
\multiput(150,78)(0,10){4}
{
\put(0,5){\oval(5,5)[br]}
\put(0,5){\oval(5,5)[tr]}
\put(0,10){\oval(5,5)[bl]}
\put(0,10){\oval(5,5)[tl]}
}
\put(210,120){\vector(0,-1){40}}
\put(214,100){${\bf k}\parallel {\hat {\bf z}}$}
\end{picture}
\end{center}
\caption{The impulse approximation for the electromagnetic interaction.}
\label{impulse}
\end{figure}
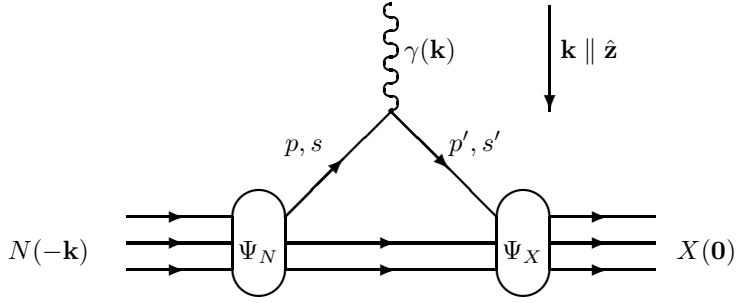

The idea is to use the impulse approximation, illustrated in
Figure~\ref{impulse}, which describes the target nucleon and the final
resonance as made up of quarks which are free while interacting with
the photon, but whose momenta, charges and spins are distributed
according to the bound-state wave functions which result from a model
of the spectrum.
\vskip -12pt
\vskip -12 pt
\subsection{nonrelativistic model}
\vskip -12pt
The EM interaction Hamiltonian can be found from a nonrelativistic
reduction of the quark current $e^{i(p^\prime-p)\cdot x}{\bar
u}(p^\prime,s^\prime)[-ie\gamma^\mu] u(p,s)$ to be
\beq
H^{\rm em}=\sum_i H^{\rm em}_i
=-\sum_i \left\{ {e_i\over 2m_i}\left[{\bf p}_i\cdot 
{\bf A}({\bf r}_i)+{\bf A}({\bf r}_i)\cdot{\bf p}_i\right]
+\lpmb{\mu}_i\cdot \lpmb{\nabla}_i \times {\bf A}({\bf r}_i)\right\},
\label{Hem}
\eeq
where $\lpmb{\mu}_i=e_i\lpmb{\sigma}_i/2m_i$ is the magnetic moment of
the $i$-th quark, $e_i$, $m_i$, $\lpmb{\sigma}_i/2$, $\pbi$ are its
charge, (constituent) mass, spin, and momentum, and ${\bf A}$ is the
photon field. The term quadraric in ${\bf A}$ is important for Compton
scattering but does not affect the tree-level quark-photon
vertex. Note that this interaction is flavor dependent, even for equal
mass quarks, through its dependence on the charge $e_i$, so amplitudes
for production from the proton and neutron contain independent
information about the structure of the initial and final baryons.

There are, in general, a pair of amplitudes $A_{1/2}$ and $A_{3/2}$
associated with photoproduction from each target, which correspond to
the two possibilities for aligning the spin of the photon and initial
baryon in the center of momentum (c.m.) frame. This is illustrated in
Figure~8 for photoproduction of a $\Delta\thalf^+$ state; note only
$A_{1/2}$ is needed to describe photoproduction of baryons $X$ with
$J_X=1/2$. These helicity amplitudes are defined in terms of helicity
states by
\beq
A^N_{\lambda}=\langle XJ_X; \rmb{0}\lambda
\vert H^{\rm em} \vert
N\half; -\rmb{k}\lambda_N\rangle,
\label{Alam1}
\eeq
where $\lambda_N=\lambda_\gamma-\lambda=1-\lambda$ if ${\bf
k}\parallel \hat{\bf z}$, and $N=p,n$.  Eq.~(\ref{Alam1}) is specialized
to the case of photoproduction of the nonstrange baryons $N$ and
$\Delta$ for which the bulk of the analyzed data exists, although the
generalization to the unequal mass case, {\it e.g.} for radiative
decays of excited hyperons~\cite{Darewych:1985dc}, is straightforward.
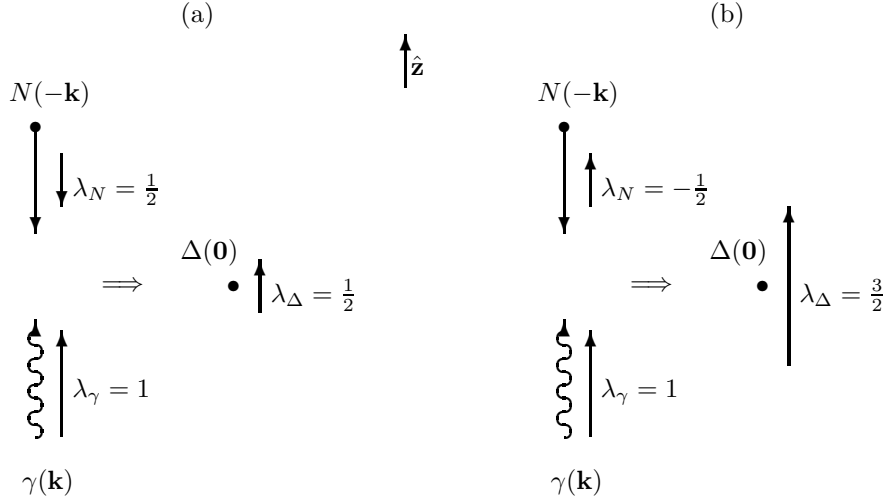
\begin{figure}[t]
\unitlength 1.0pt
\begin{center}
\begin{picture}(330,200)
\put(150,165){\vector(0,1){20}}
\put(152,170){$\hat {\bf z}$}
\put(65,190){(a)}
\put(10,150){\circle*{4}}
\put(10,150){\vector(0,-1){40}}
\put(0,160){$N(-{\bf k})$}
\put(20,140){\vector(0,-1){20}}
\put(24,124){$\lambda_N=\half$}
\multiput(10,30)(0,10){4}
{
\put(0,5){\oval(5,5)[br]}
\put(0,5){\oval(5,5)[tr]}
\put(0,10){\oval(5,5)[bl]}
\put(0,10){\oval(5,5)[tl]}
}
\put(10,72){\vector(0,1){5}}
\put(5,13){$\gamma({\bf k})$}
\put(20,33){\vector(0,1){40}}
\put(24,48){$\lambda_\gamma=1$}
\put(35,88){$\Longrightarrow$}
\put(85,90){\circle*{4}}
\put(65,100){$\Delta({\bf 0})$}
\put(95,80){\vector(0,1){20}}
\put(99,84){$\lambda_\Delta=\half$}
\put(265,190){(b)}
\put(210,150){\circle*{4}}
\put(210,150){\vector(0,-1){40}}
\put(200,160){$N(-{\bf k})$}
\put(220,120){\vector(0,1){20}}
\put(224,124){$\lambda_N=-\half$}
\multiput(210,30)(0,10){4}
{
\put(0,5){\oval(5,5)[br]}
\put(0,5){\oval(5,5)[tr]}
\put(0,10){\oval(5,5)[bl]}
\put(0,10){\oval(5,5)[tl]}
}
\put(210,72){\vector(0,1){5}}
\put(205,13){$\gamma({\bf k})$}
\put(220,33){\vector(0,1){40}}
\put(224,48){$\lambda_\gamma=1$}
\put(235,88){$\Longrightarrow$}
\put(285,90){\circle*{4}}
\put(265,100){$\Delta({\bf 0})$}
\put(295,60){\vector(0,1){60}}
\put(299,84){$\lambda_\Delta=\thalf$}
\end{picture}
\end{center}
\caption{Momenta and helicities in the center of momentum frame for
photoproduction of $\Delta\thalf^+$; (a) defines the amplitude
$A_{1/2}$, and (b) defines $A_{3/2}$.}
\label{helics}
\end{figure}
To reduce the $H^{\rm em}$ of Eq.~(\ref{Hem}) to an operator which
acts between a nucleon and a resonance wave function, first
substitute the monochromatic photon field
\beq
{\bf A}_{\bf k}({\bf r}_i)
=\sum_\lambda\sqrt{2\pi\over k_0}\left(
\lpmb{\epsilon}_{{\bf k},\lambda}e^{i{\bf k}\cdot 
{\bf r}_i}a_{{\bf k},\lambda}
+\lpmb{\epsilon}^*_{{\bf k},\lambda}e^{-i{\bf k}\cdot 
{\bf r}_i}a^{\dag}_{{\bf k},\lambda}\right),
\eeq
where $k_0=|{\bf k}|$ for real photons, and then use the first (photon
absorption) part with $\lpmb{\epsilon}_{{\bf k},+1}=-(1,i,0)/
\sqrt{2}$, and choose ${\bf k}\parallel {\hat {\bf
z}}$. It is possible, when calculating the photocouplings of the $N$
and $\Delta$ resonances, to exploit the overall symmetry of the
wave functions (without the color wave function) to make the replacement
$H^{\rm em}\to 3 H^{\rm em}_3$. This is a convenient choice, since for
the equal mass case the momentum of the third quark is just ${\bf
p}_3=\sqrt{2/3}{\bf p}_\lambda + {\bf P}/3$, where the momenta ${\bf
p}_\rho$ and ${\bf p}_\lambda$ have equivalent definitions to their
conjugate coordinates $\lpmb{\rho}$ and $\lpmb{\lambda}$, and ${\bf
P}$ is the total momentum.

The result of inserting plane waves for the center of mass motion of
the baryons and integrating over the position coordinates is
\beq
A^N_{\lambda}=3\langle X J; \lambda
\vert 
 -{e_3\over 2m_u}{1\over\sqrt{2}}\sqrt{2\pi \over k_0}
e^{-ik\sqrt{2\over 3}\lambda_z}
\left( \sqrt{2\over 3}p_{\lambda +}-k{\sigma_{3+} \over 2}\right)
\vert N\half;\,\lambda-1 \rangle.
\label{Alam3}
\eeq
Here $p_{\lambda +}=p_{\lambda x}+ip_{\lambda y}$, and $\sigma_{3+}$
raises the spin of the third quark. The expectation value in
Eq.~(\ref{Alam3}) is best carried out in momentum space, where the
recoil phase factor has the effect of shifting the ${\bf p}_\lambda$
momentum of the final-state wave function by $\sqrt{2/3}{\bf k}$.

The helicity amplitudes for photoproduction of any excited $\Delta
(I=\thalf)$ state satisfy $A^p_\lambda=A^n_\lambda$, since the
expectation values of the charge operator $e_3$ between the flavor
wave functions for the $N$ and $\Delta$ satisfy
\beqa
\langle \phi^S_{\Delta^+}\vert e_3 \vert\phi^\rho_p\rangle 
&=& \langle \phi^S_{\Delta^0}\vert e_3 \vert\phi^\rho_n\rangle =0\nonumber\\
\langle \phi^S_{\Delta^+}\vert e_3 \vert\phi^\lambda_p\rangle 
&=& \langle \phi^S_{\Delta^0}\vert e_3 \vert\phi^\lambda_n\rangle 
=-\sqrt{2}e/3,
\eeqa
where $e$ is the proton charge.

Koniuk and Isgur~\cite{KI} calculated the amplitudes of
Eq.~(\ref{Alam3}) using the Isgur-Karl
model~\cite{Isgur:1977ef,Isgur:1978xj,Isgur:1979wd} wave functions,
which include configuration mixing caused by the hyperfine
interaction. In order to compare with the helicity amplitudes quoted
in the Particle Data Group (PDG)~\cite{Caso:1998tx}, the sign of the
$N\pi$ decay amplitudes must be calculated in the model. The quality
of the $\gamma N\to N\pi$ data is poorer than the $N\pi$ elastic data,
and so a resonance $X$ must be present in $N\pi$ elastic in order to
extract its photocoupling $A^N_\lambda(X)$. As mentioned above, since
the sign of the coupling $A_{X\to N\pi}$ cannot be determined in $N\pi$
scattering, the PDG quote $A^N_\lambda(X)\cdot A_{X\to
N\pi}/|A_{X\to N\pi}|$ for the photocoupling helicity amplitude,
i.e. inclusive of the sign of $A_{X\to N\pi}$ (there are also some other
conventional signs included, see Ref.~\cite{KI}).

The results of this calculation, with signs calculated using the
elementary emission strong decay model of Ref.~\cite{KI}, are listed
for $\Delta(1232)$ and the N=1 band negative-parity nonstrange baryons
in Table~\ref{PCtab:dneg}, and for the N=2 band positive-parity
nonstrange baryons in Table~\ref{PCtab:pos}. Also shown there are the
empirical photocouplings extracted for these states from analyses of
pion photoproduction data. For most low-lying states the agreement is
quite good; for certain well determined states, such as the Roper
resonance $N\half^+(1440)$, there are sizeable discrepancies. This may
be partially due to deficiencies in the wave functions and the
nonrelativistic approximation, which is not as well justified when
applied to some photocouplings. The photon momentum is determined by
conservation of momentum in the excited baryon c.m.~frame to be
$k=(M_X^2-m_N^2)/2M_X$ at resonance, and is not small compared to the
average quark momentum when $M_X$ is substantially larger than
$M_N$. When the struck quark recoils against the other quarks in the
baryon with a relativistic momentum, the Galilean momentum shift in
Eq.~(\ref{Alam3}) cannot be accurate.
%
\begin{table}
\caption{Photoproduction amplitudes from the fits of Koniuk and Isgur
(KI)~\protect\cite{KI}, the fit $A^M_1$ of Li and Close
(LC)~\protect\cite{Li:1990qu}, the extended-string fit with $R^2=1.0$
of Bijker, Iachello and Leviatan (BIL)~\protect\cite{Bijker:1994yr},
and the relativized fit of Capstick
(SC)~\protect\cite{Capstick:1992uc} for $\Delta(1232)$ and N=1 band
baryons, compared to amplitudes extracted from analyses of the
data (PWA)~\protect\cite{Caso:1998tx}. Amplitudes are in units of $10^{-3}$
GeV$^{-\half}$; a factor of $+i$ is suppressed for all negative-parity
states. As BIL do not calculate the sign of $A_{X\to N\pi}$ the signs
of their amplitudes are to be used only to compare the relative signs
of different amplitudes for the same state.}
\label{PCtab:dneg}
\begin{tabular}{ccrrrrrr}
\multicolumn{1}{c} {State} & \multicolumn{1}{c}{$A^N_\lambda$} 
& \multicolumn{1}{c}{KI} 
& \multicolumn{1}{c}{LC}
& \multicolumn{1}{c}{BIL} 
& \multicolumn{1}{c}{SC}
& \multicolumn{1}{c}{\rm PWA}
\\
\tableline
 $\Delta\thalf^+(1232)$
  & $A^{p,n}_\half$ &    $-$103 &    $-$94 &  $-$91  & $-$108 & $-$135$\pm$  6\\
  & $A^{p,n}_\thalf$ &   $-$179 &   $-$162 & $-$157  & $-$186 & $-$255$\pm$  8\\
      $N \half^-(1535)$
  & $A^p_\half$ &  +147 &  +142 & +162/+127$^a$  &  +76 &  +90$\pm$  30\\
  & $A^n_\half$ &  $-$119 &   $-$77 & $-$112/$-$103$^a$ &  $-$63 &  $-$46$\pm$  27\\
      $N \half^-(1650)$
  & $A^p_\half$ &  +88 &   +78 &  +0/+91$^a$ &  +54 &  +53$\pm$  16\\
  & $A^n_\half$ &  $-$35 &   $-$47 & +25/$-$41$^a$ &  $-$35 &  $-$15$\pm$  21\\
 $\Delta \half^-(1620)$
  & $A^{p,n}_\half$ &  +59 &  +72 &  $-$51 &  +81 &   +27$\pm$  11\\
      $N\thalf^-(1520)$
  & $A^p_\half$ &    $-$23 &  $-$47 &   $-$43 &  $-$15 &  $-$24$\pm$  9\\
  & $A^p_\thalf$ &  +128 & +117 &  +109 &  134 & +166$\pm$  5\\
  & $A^n_\half$ &    $-$45 &  $-$75 &   $-$27 &  $-$38 &  $-$59$\pm$  9\\
  & $A^n_\thalf$ &  $-$122 & $-$127 &  $-$109 & $-$114 & $-$139$\pm$  11\\
      $N\thalf^-(1700)$
  & $A^p_\half$ &    $-$7 &  $-$16 &    0 &  $-$33 &  $-$18$\pm$  13\\
  & $A^p_\thalf$ &  +11 &  $-$42 &    0 &   $-$3 &   $-$2$\pm$  24\\
  & $A^n_\half$ &   $-$15 &  +35 &  +11 &   18 &    0$\pm$  50\\
  & $A^n_\thalf$ &  $-$76 &  +10 &  +57 &  $-$30 &   $-$3$\pm$  44\\
 $\Delta\thalf^-(1700)$
  & $A^{p,n}_\half$ &   +100 &  +81 &  $-$82 &  +82 & +104$\pm$  15\\
  & $A^{p,n}_\thalf$ &  +105 &  +58 &  $-$82 &  +68 &  +85$\pm$  22\\
      $N\fhalf^-(1675)$
  & $A^p_\half$ &    +12 &   +8 &     0  &   +2 &  +19$\pm$  8\\
  & $A^p_\thalf$ &   +16 &  +11 &     0  &   +3 &  +15$\pm$  9\\
  & $A^n_\half$ &    $-$37 &  $-$30 &   $-$33  &  $-$35 &  $-$43$\pm$  12\\
  & $A^n_\thalf$ &   $-$53 &  $-$42 &   $-$47  &  $-$51 &  $-$58$\pm$  13\\
\end{tabular}
$^a$ With the inclusion of a 38$^\circ$ mixing between the
$S_{11}$ states (not calculated in the model of BIL).
\end{table}
\begin{table}
\caption{Breit frame photoproduction amplitudes for positive parity 
(N=2 band) excited states for which there exist data. 
Caption as in Table~\protect\ref{PCtab:dneg}.}
\label{PCtab:pos}
\begin{tabular}{ccrrrrrr}
\multicolumn{1}{c} {State} & \multicolumn{1}{c}{$A^N_\lambda$} 
& \multicolumn{1}{c}{KI} 
& \multicolumn{1}{c}{LC}
& \multicolumn{1}{c}{BIL} 
& \multicolumn{1}{c}{SC}
& \multicolumn{1}{c}{\rm PWA}
\\
\tableline
 $N \half^+(1440)$
  & $A^p_\half$ &  $-$24 &   &    0 &   +4 &  $-$65$\pm$   4\\
  & $A^n_\half$ &  +16 &   &    0 &   $-$6 &  +40$\pm$  10\\
 $N \half^+(1710)$
  & $A^p_\half$ &  $-$47 &  $-$18 &  $-$22 &  +13 &   +9$\pm$  22\\
  & $A^n_\half$ &  $-$21 &  $-$22 &   +7 &  $-$11 &   $-$2$\pm$  14\\
 $\Delta \half^+(1910)$
  & $A^{p,n}_\half$ &  +59 &  $-$28 &  +17 &   $-$8 &  +3$\pm$  14\\
 $N\thalf^+(1720)$
  & $A^p_\half$ & $-$133 &  $-$68 & +118 &  $-$11 &  +18$\pm$  30\\
  & $A^p_\thalf$ & +46 &  +53 &  $-$39 &  $-$31 &  $-$19$\pm$  20\\
  & $A^n_\half$ &  +57 &   $-$4 &  $-$33 &   +4 &   +1$\pm$  15\\
  & $A^n_\thalf$ & $-$10 &  $-$33 &    0 &  +11 &  $-$29$\pm$  61\\
 $\Delta\thalf^+(1600)$
  & $A^{p,n}_\half$ &  $-$46 &  $-$38 &   0 &  +30 &  $-$23$\pm$  20\\
  & $A^{p,n}_\thalf$ & $-$16 &  $-$70 &   0 &  +51 &   $-$9$\pm$  21\\
 $\Delta\thalf^+(1920)$
  & $A^{p,n}_\half$ &     &  $-$14 &  $-$17 &  +13 &   40$\pm$  14$^a$\\
  & $A^{p,n}_\thalf$ &    &   $-$7 &  +30 &  +14 &   23$\pm$  17\\
 $N\fhalf^+(1680)$
  & $A^p_\half$ &     0 &   $-$8 &   $-$4 &  $-$38 &   $-$15$\pm$  6\\
  & $A^p_\thalf$ &  +91 & +105 &  +80 &  +56 &  +133$\pm$  12\\
  & $A^n_\half$ &   +26 &  +11 &  +40 &  +19 &   +29$\pm$  10\\
  & $A^n_\thalf$ &  $-$25 &  $-$43 &    0 &  $-$23 &   $-$33$\pm$  9\\
 $\Delta\fhalf^+(1905)$
  & $A^{p,n}_\half$ &   +8 &  +24 &  $-$11 &  +26 &  +26$\pm$  11\\
  & $A^{p,n}_\thalf$ & $-$33 &  +25 &  $-$49 &   $-$1 &  $-$45$\pm$  20\\
      $N\shalf^+(1990)$
  & $A^p_\half$ &   $-$8  &     &    0 &   $-$1 &  +15$\pm$  25$^b$\\
  & $A^p_\thalf$ &  $-$10 &     &    0 &   $-$2 &  +45$\pm$  33\\
  & $A^n_\half$ &   $-$18 &     &  +23 &  $-$15 &  $-$40$\pm$  30\\
  & $A^n_\thalf$ &  $-$23 &     &  +29 &  $-$18 & $-$147$\pm$  45\\
 $\Delta\shalf^+(1950)$
  & $A^{p,n}_\half$ &   $-$50 &   $-$28 &  +28 &  $-$33 &  $-$76$\pm$  12\\
  & $A^{p,n}_\thalf$ &  $-$69 &   $-$36 &  +36 &  $-$42 &  $-$97$\pm$  10\\
\end{tabular}
$^a$ No signs are extracted for these amplitudes.\\
$^b$ Average of two existing analyses.\\
\end{table}
\vskip -12pt
\subsection{models with relativistic corrections}
\begin{figure}[t]
\vskip 0cm
\vbox{
\hskip 0.2cm
\epsfig{file=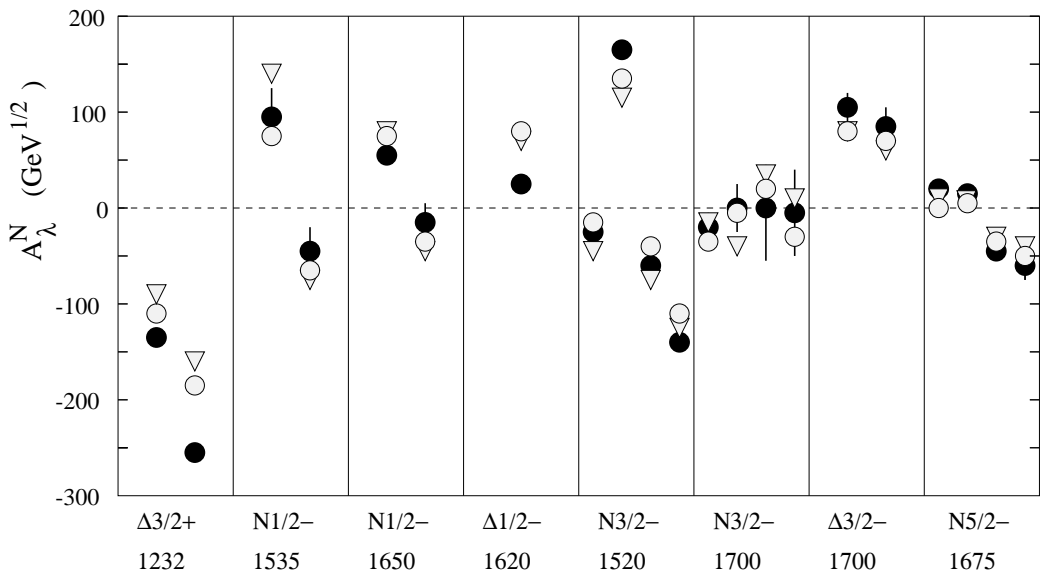,width=13.8cm,angle=0}}
\vskip 0.5 cm
\caption{Breit-frame photoproduction amplitudes for
$\Delta\thalf^+(1232)$ and the $P$-wave resonances, in units of
$10^{-3}$ GeV$^{-1/2}$. Solid circles with error bars are
photocouplings from the data analyses of
Ref.~\protect\cite{Caso:1998tx}, triangles are the fit $A_1^M$ from
Li and Close~\protect\cite{Li:1990qu}, and circles are for the relativized
calculations from Ref.~\protect\cite{Capstick:1992uc}. Amplitudes are
plotted, from left to right, in the order $A_{1/2}$, $A_{3/2}$ for
$\Delta$ states, and $A^p_{1/2}$, $A^p_{3/2}$, $A^n_{1/2}$, and
$A^n_{3/2}$ for nucleon states.}
\label{PCdneg}
\end{figure}

\begin{figure}[t]
\vskip 0cm
\vbox{
\hskip 0.2cm
\epsfig{file=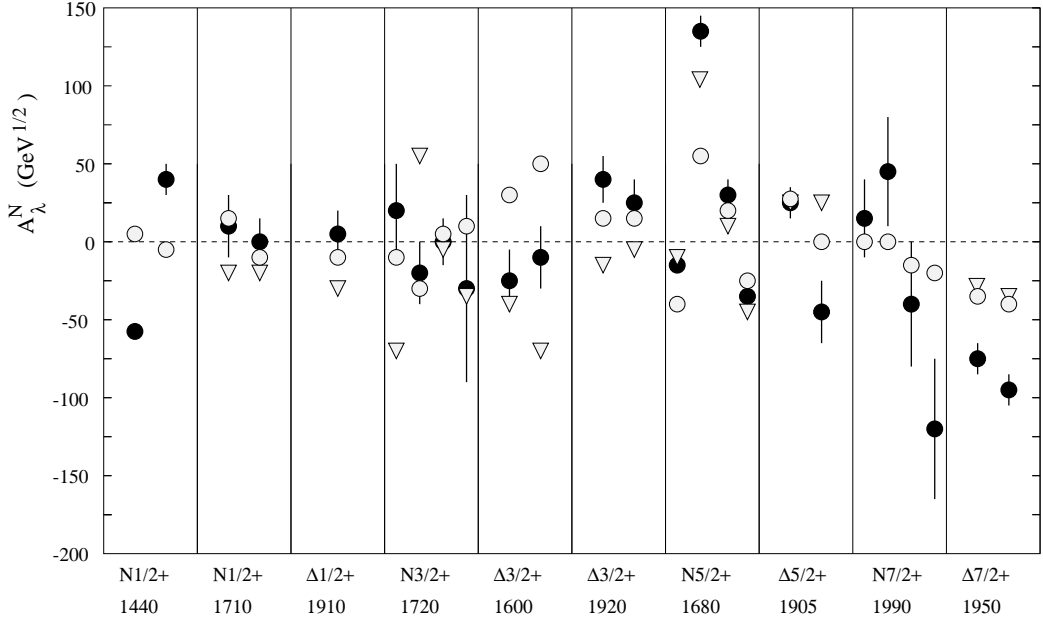,width=13.8cm,angle=0}}
\vskip 0.5 cm
\caption{Breit-frame photoproduction amplitudes for low-lying
(N=2 band) positive-parity excited states for which there exist
amplitudes extracted from the data, in units of $10^{-3}$
GeV$^{-1/2}$. Caption as in Fig.~\protect\ref{PCdneg}.}
\label{PCpos}
\end{figure}

\vskip -12pt
The nonrelativistic operator $H^{\rm em}$ adopted above is not
expanded to the same order as the interquark Hamiltonian,
O$(p^2/m^2)$. To this order spin-orbit interactions must be added, and
Brodsky and Primack~\cite{Brodsky:1968xc,Brodsky:1969ea} have shown
from the requirement of electromagnetic gauge invariance that there
are also associated two-body currents (which go beyond the simple
impulse approximation illustrated in Fig.~5). These corrections were
put together with the configuration-mixed nonrelativistic wave
functions in a calculation of the photocouplings by Close and
Li~\cite{Close:1990aj,Li:1990qu}, and using the relativized model wave
functions described above in the calculation of
Capstick~\cite{Capstick:1992xn,Capstick:1992uc}. Note that the latter
calculation does not `relativize' the O($p^2/m^2$) electromagnetic
interaction Hamiltonian by parametrizing its momentum dependence away
from the nonrelativistic limit, although the wavefunctions are
generated using a relativized inter-quark potential. These results are
compared to the empirical photocouplings extracted from pion
photoproduction analyses also shown in Figures~\ref{PCdneg}
and~\ref{PCpos}. It is clear that these additional effects improve the
comparison with the extracted photocouplings, although they clearly
are not able to account for the model's under prediction of the
amplitudes for the photoproduction of the $\Delta$ and the Roper
resonance. These difficulties may be due to the neglect of the pion
degree of freedom~\cite{Theberge:1980ye,Lee:1991pp}, as these are both
light states with strong couplings to $N\pi$. In recent
work~\cite{Capstick:1995ne,Cardarelli:1997vn} the Roper EM couplings
(along with some others) are found to be very sensitive to further
relativistic effects. Other authors~\cite{Li:1991sh,Li:1992yb} have
taken the anomalous photocouplings of the Roper resonance to be
evidence that it has a substantial amount of hybrid baryon mixed into
it.
The relativized model of Ref.~\cite{Capstick:1992xn,Capstick:1992uc}
improves the photocouplings of the other radial excitation of the
nucleon, $N\half^+(1710)$, and those of $\Delta\thalf^+(1600)$, a
radial excitation of $\Delta(1232)$, are also somewhat improved
(largely by the correct prediction of the sign of the amplitude). The
relativized model has slightly more freedom to fit the photocouplings,
as the effective quark mass $m^*$ in the electromagnetic transition
operator is allowed to vary and is fit to the photocouplings, while
the quark magnetic moments are held fixed. Close and
Li~\cite{Close:1990aj,Li:1990qu} show that $m^*$ should be thought of
as the sum of the average kinetic and scalar-binding potential energy
of the constituent quarks. Neither model is able to account for the
small photocoupling of $\Delta\half^-(1620)$ or the (relatively
precisely determined) large $A^p_{3/2}$ amplitude of $N\fhalf^+(1680)$
found in the analyses. The next generation of experiments at new
facilities such as CEBAF at the Jefferson Laboratory will examine the
pion photoproduction and electroproduction processes; the latter will
yield the $Q^2$ dependence of the Roper resonance EM couplings, for
example. It is also obvious that a model of the process $\gamma N\to
N\pi$ at 1440 MeV which includes pionic dressing of the vertex is
required to understand the Roper resonance EM couplings, since this
state couples very strongly to $N\pi$ and as a light state its decay
pion is relatively slow moving.
\vskip -12pt
\subsection{collective model}
\vskip -12pt
Bijker, Iachello and Leviatan (BIL) also calculate
photocouplings~\cite{Bijker:1994yr} in their algebraic collective
model approach. Their calculation in this model involves writing the
transition operator in terms of the constituent coordinates and
momenta, transcribing this in terms of the bosonized coordinates and
so elements of the algebra, and then algebraically evaluating their
matrix elements in the basis. As the charge and magnetic moments of
the quarks are viewed in their model as distributed along the Y-shaped
string making up the baryon, it is not simple to accomplish the first
of these steps. The approach outlined above is to look at the
nonrelativistic reduction of the coupling of point-like quarks to the
electromagnetic field. However, the resulting form is momentum
dependent and so cannot be used in this approach. A transformation is
made to coordinate-dependent terms by replacing the quark momenta
${\bf p}_i/m_i$ by $ik_0{\bf r}_i$, where $k_0$ is the photon
energy. The spin-orbit and two-body terms required for a consistent
expansion to order $({\bf p}/m)^2$ are dropped, and the two terms in
the nonrelativistic interaction Hamiltonian in Eq.~(\ref{Hem}) are then
viewed as electric and magnetic contributions. The resulting operator
can then be written in terms of the operators which make up the
spectrum-generating algebra.

The results of evaluating the photocouplings using this model are also
shown in Tables~\ref{PCtab:dneg} and~\ref{PCtab:pos}. BIL believe that
the sign of the photocouplings (which are inclusive of the sign of the
$A_{X\to N\pi}$ strong decay amplitude) cannot be determined by
calculations and so are arbitrary. The signs quoted for their
amplitudes are therefore to be used only to compare different
amplitudes for the same excited baryon. This limits the usefulness of
the comparison of their amplitudes with those extracted from the
data. Nevertheless, the fit appears degraded relative to the Li and
Close (LC) and Capstick (SC) models.  Their model also predicts a large
$A^p_\half$ amplitude for $N\thalf^+(1720)$ which is not seen in the
analyses. They are also unable to fit the substantial photocouplings
of the Roper resonance (although the mass is fit well); their results
for the photocouplings are zero. As their model does not include
configuration mixing due to hyperfine tensor interactions, which are
important in describing the photocouplings of the P-wave excited
baryons, they must introduce a mixing angle of 38$^\circ$ between the
$N\half^+(1525)$ and $N\half^+(1650)$, not calculated in their model,
in order to roughly fit the photocouplings of these states.
\vskip -12pt
\section{Strong Decay Couplings}
\vskip -12pt
\subsection{missing states and $N\pi$ couplings}
\vskip -12pt
One of the features that many of the models described in the previous sections 
have in common is that
they predict more states than have been experimentally `observed'. This would
appear to be a problem for such models, as they would clearly have failed to
describe physical reality. There are two possible interpretations to this. 

A number of authors suggest that the mismatch between the number of 
baryonic states observed,
and the number predicted in some models, is due to the fact that such models 
are using the wrong
degrees of freedom. According to these authors, the three-quark picture of a 
baryon is flawed, as a baryon in  fact consists of a diquark and a quark 
\cite{Anselmino:1993vg,Goldstein:1988us}.
The diquark is assumed to be tightly bound, so that the low-lying excitations 
of the nucleon
will not include excitations of the diquark, thus leading to fewer states in 
the excitation spectrum.
The tight binding is thought to occur in the isoscalar scalar channel, due to 
an attractive hyperfine
interaction between a $u$ and a $d$ quark. However, there is some evidence from 
the lattice~\cite{Leinweber:1993nr} and from an analysis of baryon strong
decays~\cite{Forsyth:1981tk} that such strong diquark clustering may
not be present. 

A very important consideration here is that of how the experimentally observed 
states 
are produced and observed. To date, the vast majority of known excitations of 
the nucleon have been
produced in $\pi N$ elastic scattering, with a few more being `observed' in 
$\pi
N$ inelastic processes like $\pi N\to \pi\pi N$. `Observed' is not completely
accurate, however, as all information regarding the baryon spectrum is inferred
from fits to the scattering cross sections. Thus, the most precise
interpretation of an unobserved state is that such a state is not required
to significantly improve the quality of the fit to the available scattering 
data.

Stated differently, this means that the missing states are ones that do not
provide significant contributions to the cross section of the scattering
process being examined, namely $N\pi$ elastic scattering, and they must
therefore couple weakly to this channel. Faiman and Hendry \cite{Faiman:1968js,FH2} were 
among the earliest to suggest this idea in the literature. It is therefore 
crucial that the 
observed 
pattern of $N\pi$ decay widths be reproduced by theoretical treatments, 
particularly by
those models that have a surfeit of states. Furthermore, since the missing 
states are difficult to 
detect in the $N\pi$ channel, these models should predict the channels in 
which they are most likely
to be seen. A number of models have treated these strong decays in this manner.

As with their masses, the description of the strong properties of baryons, 
namely 
the strong decay widths, coupling constants and form factors, relies
largely on phenomenological models, although some work has been done using 
effective Lagrangians. As few excited states have been
treated in the effective Lagrangian approaches, the focus of this section is 
mainly on the phenomenological approaches.

The operators responsible for strong transitions
between baryons arise from non-perturbative QCD, and are therefore essentially
unknown. The models that have been constructed assume that the mechanism of the
strong decay is either elementary meson emission, quark pair creation, string
breaking, or flux-tube breaking. These are illustrated below. The latter 
three can all be broadly called quark-pair creation models. 

The OZI-allowed~\cite{OZI1,OZI2,OZI3} strong decays of hadrons which are considered here 
have 
been examined in three classes of models described below. The `hadrodynamic' 
models, illustrated in 
Figure \ref{hadrodynamic}, in which all hadrons are 
treated as elementary point-like objects, do not lend themselves easily to 
decay calculations of the kind considered here. This is 
understandable, since each transition is described in terms of one or more 
independent 
phenomenological coupling constants, $g_{BB^\prime M}$, which means that such 
models have little predictive power. While the use of SU(2) 
or SU(3) flavor symmetry arguments give relationships among some of these 
coupling constants, the overall situation would nevertheless be largely 
unworkable. It does not appear as if there have been systematic studies 
of baryon decays using this approach.

\unitlength 1.0cm 
\vspace*{0.25in}
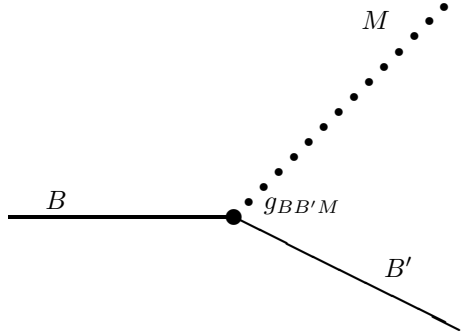
\begin{figure}
\centerline{\begin{picture}(5,5)
\put(0,2.5){\line(1,0){3}}
\multiput(3,2.5)(0.2,0.2){15}{\circle*{0.1}}
\put(3,2.5){\line(2,-1){3}}
\put(0.5,2.6){$B$}
\put(5,1.7){$B^\prime$}
\put(4.7,5){$M$}
\put(3,2.5){\circle*{0.2}}
\put(3.4,2.6){$g_{BB^\prime M}$}
\end{picture}}
\vspace*{-0.5cm}
\caption{The process $B\to B^\prime M$, as an elementary meson emission from a 
point-like baryon.}
\label{hadrodynamic}
\end{figure}

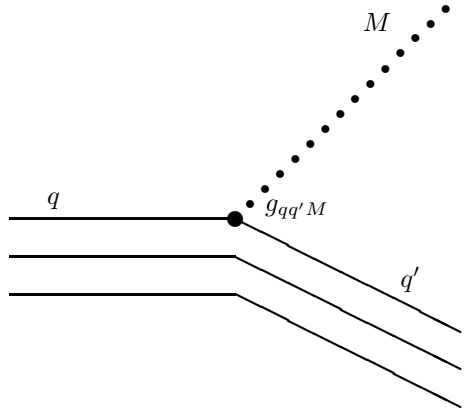
\begin{figure}
\vbox{\begin{center}
\begin{picture}(6,5)
\multiput(0,1.5)(0,0.5){3}{\line(1,0){3}}
\multiput(3,2.5)(0.2,0.2){15}{\circle*{0.1}}
\multiput(3,2.5)(0,-0.5){3}{\line(2,-1){3}}
\put(0.5,2.7){$q$}
\put(5.2,1.6){$q^\prime$}
\put(4.7,5){$M$}
\put(3,2.5){\circle*{0.2}}
\put(3.4,2.6){$g_{qq^\prime M}$}
\end{picture}
\end{center}
\caption{The process $B\to B^\prime M$, as an elementary meson emission from a 
quark.}
\label{mesonemission}
}
\end{figure}

A second class of models treats the baryons as objects with structure, but the 
decay takes place through elementary meson emission. Such an approach may be 
taken in bag models, for instance. Some potential model-based calculations, 
such
as the work of Koniuk and Isgur \cite{KI0,KI}, have used 
a similar approach. In these models, mesons are emitted from quark lines 
(Figure \ref{mesonemission}), 
and the set of $g_{BB^\prime M}$ coupling constants is replaced with a smaller 
set of 
$g_{qq^\prime M}$. In addition, SU(2) or SU(3) flavor symmetry may be used
to relate the coupling constants for mesons within a single multiplet, as well 
as those for different quarks.

\vspace*{0.25in}

Among the earliest descriptions of the strong decays of baryons is the work of 
Faiman and Hendry (FH) \cite{Faiman:1968js,FH2}. They assume that the decay occurs through 
the `de-excitation' of a single quark in the baryon. Their model is thus an 
elementary meson emission model, and their non-relativistic transition operator 
takes the form
\begin{equation}
{\cal O}=\sum_j \frac{f_q}{\mu}\left(\veca{\sigma}_j\cdot\veca{k}\right)
\left(\veca{\tau}_j\cdot 
\veca{\pi}\right) e^{i\veca{k}\cdot\veca{r}_j} \left(\frac{1}{2E_\pi}\right)^{1/2},
\end{equation}
where $\mu$ is taken to be the pion mass, $f_q$ is the quark-pion coupling 
constant, $E_\pi$ and $\veca{k}$ are the energy and momentum of the emitted 
pion, $\veca{\sigma}_j$ is the Pauli spin matrix, $\veca{\tau}_j$ is the 
isospin 
operator of the quark from which the pion is emitted, $\veca{r}_j$ is its
position, and the sum over $j$ is 
over all the quarks in the baryon. This operator was first proposed by 
Becchi and Morpurgo \cite{morpurgo1,morpurgo2,morpurgo3}.

In their work, FH find that they are able to reproduce the partial widths 
available at that time with remarkable accuracy. They propose that 
the missing resonances simply couple weakly to the $N\pi$ formation channel. 
They also suggest that the $\Delta\pi$ and other inelastic channels are 
likely to yield signals for the missing states.

Several authors use a slightly different version of the elementary
meson emission model to calculate the decay widths for processes with
a pseudoscalar meson in the final state. The elementary meson emission
models of Bijker, Iachello and Leviatan (BIL)~\cite{BIL}, Koniuk and
Isgur (KI)~\cite{KI0,KI}, and Sartor and Stancu (SAS)~\cite{sas} are
discussed here. In addition, Koniuk extends this idea to the
discussion of vector meson emission~\cite{KI2}.

The operator responsible for the emission of a pseudoscalar meson is usually 
assumed to have the form
\begin{equation}
{\cal H_P}=\sum_{i=1}^3 {\cal N} \left(g \veca{k}\cdot \veca{s}^{(i)}+
h \veca{p}_i\cdot\veca{s}^{(i)}\right)e^{-i\veca{k}\cdot\veca{r}_i} X_i^P,
 \label{emission}
\end{equation}
where $\veca{k}$ is the momentum of the emitted meson, $\veca{p}_i$ is the 
momentum of
the $i$th quark, $\veca{r}_i$ is its position, $\veca{s}$ is its spin, $g$ and 
$h$ are phenomenological constants, and the
$X_i^P$ are flavor matrices that describe the quark transitions $q_i\to
q_i^\prime+P$, where $P$ is a pseudoscalar meson. This is the simplest form 
that can be written down for this operator. The normalization constant ${\cal 
N}$ varies with the author, and has the values
\begin{eqnarray}
{\cal N} &=& \frac{1}{(2\pi)^{3/2}(2E_\pi)^{1/2}},\,\,\,\,\,\, {\rm (BIL)}
\nonumber \\
{\cal N}&=& \frac{i}{(2\pi)^{3/2}},\,\,\,\,\,\, {\rm (KI)}
\nonumber \\
{\cal N}&=& 1.\,\,\,\,\,\, {\rm (SAS)}
\nonumber
\end{eqnarray}
In the case of KI and SAS, the symmetry of the wave functions is used to 
rewrite the transition operator as
\begin{equation}
{\cal H_P}=3 {\cal N} \left(g \veca{k}\cdot \veca{s}^{(3)}+
h \veca{p}_3\cdot\veca{s}^{(3)}\right)e^{-i\veca{k}\cdot\veca{r}_3} X_3^m,
\end{equation}
where all the `3' indices refer to the third quark.

The flavor matrices $X^P$ are
\begin{eqnarray}
X^{\pi^0}&=&\lambda_3, \,\, X^{\pi^+}=-\frac{1}{\sqrt{2}}\left(\lambda_1-
i\lambda_2\right),
X^{\pi^-}=\frac{1}{\sqrt{2}}\left(\lambda_1+i\lambda_2\right),\nonumber\\
X^{K^0}&=&-\frac{1}{\sqrt{2}}\left(\lambda_6-i\lambda_7\right), \,\, 
X^{K^+}=-\frac{1}{\sqrt{2}}\left(\lambda_4-i\lambda_5\right),
X^{K^-}=\frac{1}{\sqrt{2}}\left(\lambda_4+i\lambda_5\right),\nonumber\\
X^{\eta_1}&=&\sqrt{\frac{2}{3}}{\cal I}, \,\, 
X^{\eta_8}=\lambda_8,
\end{eqnarray}
where the $\lambda_i$ are the Gell-Mann matrices, and ${\cal I}$ denotes the 
unit operator in flavor space.

The physical $\eta$ and $\eta^\prime$ mesons have the flavor compositions
\begin{equation}
\eta=\eta_8 \cos{\theta}-\eta_1\sin{\theta},\,\,\,
\eta^\prime=\eta_8 \sin{\theta}+\eta_1\cos{\theta},
 \label{eta}
\end{equation}
where
\begin{equation}
\eta_8=\frac{1}{\sqrt{6}}\left(u\bar{u}+d\bar{d}-2s\bar{s}\right),\,\,\,
\eta_1=\frac{1}{\sqrt{3}}\left(u\bar{u}+d\bar{d}+s\bar{s}\right).
\end{equation}
For the mixing angles, KI use
\begin{equation}
\eta=\frac{1}{2}\left(u\bar{u}+d\bar{d}-\sqrt{2}s\bar{s}\right),
\end{equation}
corresponding to $\theta=-9.7^\circ$, while BIL use $\theta=-23^\circ$.

\begin{figure}
\begin{center}
\begin{picture}(4,5)
\multiput(0,1.5)(0,0.5){3}{\line(1,0){3}}
\put(3.5,2.25){\line(2,-1){2.5}}
\put(3.5,2.25){\line(2,1){2.5}}
\put(3,2.5){\line(2,1){3}}
\put(3,2){\line(2,-1){3}}
\put(3,1.5){\line(2,-1){3}}
\end{picture}
\end{center}
\vskip 0.125in
\caption{The OZI allowed process $B\to B^\prime M$, in a quark pair creation 
scenario.}
\label{paircreation}
\end{figure}
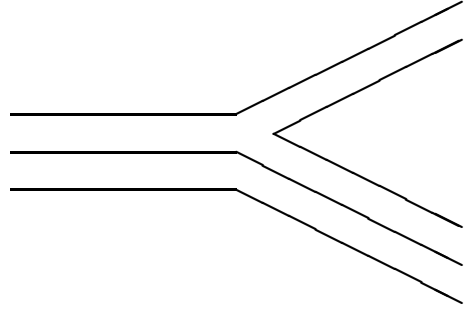

A third class of models may be referred to broadly as pair creation
models. In such models both the baryons and mesons have structure, and
the decay of the baryon, say, occurs by the creation of a
quark-antiquark pair somewhere in the hadronic medium. The created
antiquark combines with one of the quarks from the decaying baryon to
form the daughter meson, while the quark of the created pair becomes
part of the daughter baryon. This is illustrated in Figure
\ref{paircreation}.

There are several types of pair creation model. In the $\tp0$ model,
first formulated by Micu \cite{micu} and subsequently popularized by
LeYaouanc {\it et al.} \cite{ALY1,ALY2,ALY3,ALY4,ALY5,morepair}, the
quark-antiquark pair is created anywhere in space with the quantum
numbers of the QCD vacuum, namely $0^{++}$. This corresponds to the
quantum numbers $^{2S+1}L_J={\tp0}$, hence the name of the model. While
the pair, in principle, may be created very far away from the decaying
hadron, the wave function overlaps required naturally suppress such
contributions to the decay amplitude. This model has been quite
popular in descriptions of hadron decays, and has been applied to
baryon decays \cite{ALY1,ALY2,ALY3,ALY4,ALY5}, meson decays
\cite{micu,ALY1,ALY2,ALY3,ALY4,ALY5,mes}, and even the decays of
fictitious four-quark states \cite{RO1,RO2,RO3,RO4,RO5}. The
popularity of this model stems from its overall simplicity, and the
fact that, when first proposed, it could be applied to both meson and
baryon decays. In addition, it naturally provides the centrifugal
barrier to the transition amplitudes needed to describe the strong
decays.

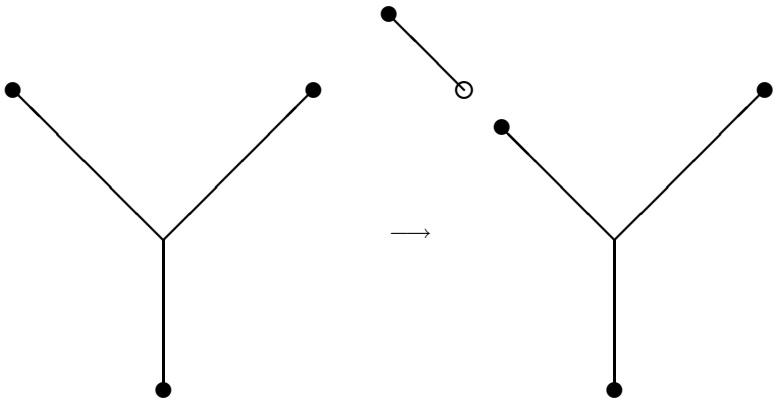
\begin{figure}
\vbox{\begin{center}
\begin{picture}(12,6)
\put(2,1){\line(0,1){2}}
\put(2,3){\line(1,1){2}}
\put(2,3){\line(-1,1){2}}
\put(2,1){\circle*{0.2}}
\put(4,5){\circle*{0.2}}
\put(0,5){\circle*{0.2}}
\put(5,3){$\longrightarrow$}
\put(8,1){\line(0,1){2}}
\put(8,3){\line(1,1){2}}
\put(8,3){\line(-1,1){1.5}}
\put(8,1){\circle*{0.2}}
\put(10,5){\circle*{0.2}}
\put(6.5,4.5){\circle*{0.2}}
\put(6.0,5.0){\circle{0.2}}
\put(6.0,5.0){\line(-1,1){1}}
\put(5.0,6.0){\circle*{0.2}}
\end{picture}
\end{center}
\vspace*{-0.5cm}
\begin{center}
\caption{The process $B\to B^\prime M$ in the string breaking picture.}
\label{stringbreaking}
\end{center}}
\end{figure}

Other pair-creation models include the string-breaking models of Dosch
and Gromes~\cite{DG}, and of Alcock, Burfitt and Cottingham
\cite{ABC}. In these models, the lines of color flux between quarks
have collapsed into a string, and the pair is created when the string
breaks. This is illustrated in Figure \ref{stringbreaking}. In the
Dosch-Gromes version of this model, the created pair have the quantum
numbers $\tp0$, while in the Alcock {\it et al.} version, the quantum
numbers of the created pair are $^3S_1$. Note that in all versions of
the $^3S_1$ model, the operator is taken to be proportional to the
scalar product of the spin of the created quark-antiquark pair, with
some vector in coordinate space. Thus, the operator is still a $0^+$
operator, which means that many aspects of the $^3S_1$ model are very
similar to those of the $\tp0$ model. Fujiwara~\cite{Fujiwara:1993yv}
also explores the strong decays of the $P$-wave baryons in a $^3S_1$
pair-creation model.

\vspace*{-0.125in}

\vspace*{-0.5in}

Recently, Ackleh, Barnes, and Swanson \cite{barnes} show that meson strong 
decays can be 
described by interactions between the quarks in the parent hadron and the 
created quark pair which are similar to the interactions used in potential 
models, namely one-gluon-exchange and confinement. They show that this leads 
to an effective decay operator which is dominantly $\tp0$, although sometimes 
because of quantum numbers the sub-dominant $^3S_1$ form plays a role. Just as 
importantly, they show that their picture of the decay process yields an 
effective pair creation strength $\gamma$ close to that used in $\tp0$ models,
when they use parameters similar to those used in potential models.

The transition operator for the $\tp0$ model is assumed to be
\beqa \label{amp1} T&=&-3\gamma\sum_{i,j} \int d \veca{p}_i\/ d
\veca{p}_j\/ \delta( \veca{p}_i\/ + \veca{p}_j\/) C_{ij}
F_{ij}\nonumber \\ &\times& \sum_m <1,m;1,-m|0,0> \chi_{ij}^m { \cal
Y\/ }_1^{-m}( \veca{p}_i\/ - \veca{p}_j\/ )b_i^{\dagger}( \veca{p}_i\/
) d_j^{\dagger}( \veca{p}_j\/ ).  \eeqa
Here, $C_{ij}$ and $F_{ij}$ are the color and flavor wave functions of the 
created pair, both assumed to be singlet, $\chi_{ij}$ is the spin triplet wave 
function of the pair, and ${\cal Y}_1(\veca{p}_i - \veca{p}_j)$ is the vector 
harmonic 
indicating that the pair is in a relative $p$-wave.

For the transition $A\to B C$, the transition amplitude $M$ is evaluated, where
\beq
M_{A\to BC}(k_0)=<B(k_0)C(-k_0)|T|A(0)>,
\eeq
and a partial decay width is calculated as
\beq
\Gamma(A\to BC)=\left|M_{A\to BC}(k_0)\right|^2\Phi(ABC),
\eeq
where $\Phi$ is the phase space for the decay. Details of the
calculation of the matrix elements of this operator, or of the
operator chosen for elementary meson emission, are omitted here for
reasons of compactness. For the phase space factor $\Phi$, a number of
options have been used. The usual prescription is
\beq
\Phi(ABC) = 2\pi 
{E_B(k_0) E_C(k_0) k_0 \over m_A},
\eeq
with $E_B(k_0)=\sqrt{ k^2_0 + m_B^2}$, $E_C(k_0)=\sqrt{ k^2_0 + m_C^2}$. This 
is a 
`semi-relativistic' prescription, since it is usually used with a matrix 
element 
calculated non-relativistically, while $E_B$ and $E_C$ have been calculated 
relativistically. A fully nonrelativistic prescription consists in using
\beq
\Phi(ABC) = 2\pi {m_B m_C k_0 \over m_A}.
\eeq
In their calculation of meson decay widths, Kokoski and Isgur
\cite{kokoski} use the prescription
\beq \label{pspace1}
\Phi(ABC) = 2\pi {\tilde m_B \tilde m_C k_0 \over \tilde m_A},
\eeq
where the $\tilde m$'s are effective hadron masses, evaluated with a
spin-independent inter-quark interaction. They argue that this is
valid in the weak-binding limit, where $\rho$ and $\pi$ are
degenerate.

In the baryon decay width calculation of Capstick and Roberts (CR)
\cite{Capstick:1993th,CR2,CR3}, there are some features that are
similar to the Kokoski-Isgur calculation of the meson decay widths:
(i) the baryon wave functions used \cite{Capstick:1986bm} are obtained
in the same spirit as the Godfrey-Isgur \cite{Godfrey:1985xj} wave
functions used in the Kokoski-Isgur calculation, and in fact, many of
the parameters of both spectroscopic calculations are chosen to be the
same or similar; (ii) the Godfrey-Isgur wave function is used for the
pion. CR argue that it makes sense, therefore, to use
Eq.~(\ref{pspace1}) in their calculation of the baryon decay
widths. For the decays of a resonance $R$ to $N \pi$, they use $\tilde
m_N=1.1$ GeV, $\tilde m_\pi$=0.72 GeV, consistent with Kokoski and
Isgur, and $\tilde m_R=m_R$.

Forsyth and Cutkosky (FC) \cite{Forsyth:1983dq} consider three versions of a pair creation 
model which, combined with their model for 
the spectrum, they fit to the Carnegie-Mellon University-Lawrence Berkeley 
Laboratory partial wave analyses 
(CMB) \cite{cmb1,cmb2}. The form that they use for the spatial part of the transition 
operator is
\begin{equation}
{\cal O}=\veca{S}\cdot\veca{V},
\end{equation}
where $\veca{S}$ is the spin of the created quark-antiquark pair, and 
$\veca{V}$ is 
some vector, chosen to be
\begin{equation}
\veca{V}=\left(G_1\veca{p}_3+G_2\veca{p}^\prime_3\right)
e^{i\veca{k}\cdot\veca{w}}.
\end{equation}
In this expression, $\veca{w}$ is the separation between the daughter baryon 
and meson, 
$\veca{p}_3$ and $\veca{p}_3^\prime$ are the momenta of the third quark in the 
parent and daughter baryon, respectively, and $\veca{k}$ is the momentum of the 
daughter meson. The form chosen for the operator means that it is a $0^+$ 
operator. This model subsumes many other pair creation models as special cases. 
As is, this is the model of Horgan \cite{horgan}, while with $G_1=-G_2$, this 
becomes the model used by Faiman and Hendry \cite{Faiman:1968js,FH2}. Setting $G_1=0$ 
results in the $\tp0$ model as formulated by the Orsay group
\cite{ALY1,ALY2,ALY3,ALY4,ALY5}.

The first of the three versions of this model examined by these authors 
utilizes the full operator, 
with both $G_1$ and $G_2$ non-zero. The second is an extension of the model 
used by KI, in which each distinct polynomial (in $k$) that arises in the 
evaluation of the spatial matrix elements is replaced by an adjustable 
parameter. In the third version, $G_1$ and $G_2$ are allowed to depend on the 
spin and orbital state of the `spectator' quarks. In this way, they 
relax the spectator conditions that are usually used in such models. They 
find that this version of the model works best. However, none of 
their results are shown here, but the interested reader is referred to the 
original work.

In selecting the results that are discussed, only those calculations
in which a number of decay channels have been treated, for a number of
baryon resonance, and for which numbers are easily available, have
been chosen. These criteria are applied both to the meson emission and
pair creation models. Among the pair-creation models, the work by Le
Yaouanc {\it et al.}  \cite{ALY1,ALY2,ALY3,ALY4,ALY5,Gavela:1978bz} of
the Orsay group (OR), CR \cite{Capstick:1993th,CR2,CR3}, SST
\cite{sst1,sst2,sst3,sst4,sst5,sst6,sst7,sst8}, and by Wagenbrunn {\it
et al.} (WPTV) \cite{Wagenbrunn:Nstar2k}, are discussed and compared
with the results obtained in the elementary meson emission models. It
must be pointed out here that although only two models for the strong
decay processes have been widely used, either one can be combined with
the wide range of models for the baryon spectrum, thus leading, in
principle, to very diverse predictions for decay widths, coupling
constants and form factors.

The results obtained for the $N\pi$ decay widths for the
models considered are shown in table \ref{nNle2t}. The first column of the 
table
identifies the state, while columns two to thirteen give the magnitudes of the 
decay amplitudes into $N\pi$ calculated in the models of OR \cite{ALY1,ALY2,ALY3,ALY4,ALY5,Gavela:1978bz}, BIL 
\cite{BIL}, SAS \cite{sas}, KI \cite{KI0,KI,KI2}, CR \cite{Capstick:1993th,CR2,CR3}, (SST)
\cite{sst1,sst2,sst3,sst4,sst5,sst6,sst7,sst8} and WPTV \cite{Wagenbrunn:Nstar2k}. 
Missing states have their 
model-predicted masses (here chosen
to be those of CR, but other models predict very similar masses) shown in 
square parentheses. The
models of BIL, SAS and KI are elementary meson emission models, while those of
OR, CR, SST and WPTV are pair creation models. The numbers obtained 
from the partial wave analyses \cite{PDG} are shown in the final column.

For all of the models chosen, the results reported are obtained by first 
fitting any unknown coupling constants ($g$ and $h$ of Eq. (\ref{emission}) 
or $\gamma$ of Eq. (\ref{amp1})) to a set of measured widths. SST explore 
three different scenarios in their article, but results from only 
two of them are presented. In one of these scenarios, they present results for 
emission of a 
point-like pion (the column labeled `Point'), while in the other, the pion has 
a finite size (the column labeled `Size'). For each scenario, they fit 
$\gamma$ to the width of the $\Delta(1232)$.

The numbers obtained by OR result from a fit in which all of the amplitudes 
that they calculate are included. These amplitudes include not just the $N\pi$
amplitudes of Table \ref{nNle2t}, but also amplitudes into the channels
$\Delta\pi$ and $N\rho$. In each channel, numbers for the decays of the
$N\half^+(1440)$ are not presented, as this resonance has been specially 
treated by these authors.

For SAS, four numbers are shown,
corresponding to four different scenarios in their model. The numbers from Set 
I result from a particular choice for two parameters that determine their 
hyperfine and tensor interactions, while Set II results from a different 
choice for these parameters. Furthermore, they examine the effects of 
configuration mixing in their wave functions, and for each of Set I and Set II, 
the column labeled U results from using unmixed wave functions, while the 
column labeled M results from using mixed wave functions. These authors 
apparently fit to the $N\pi$ widths of eighteen low-lying resonances (of which 
all but two are three- and four-star states). Neither SAS nor SST provide 
$N\pi$ amplitudes for any possible missing states.

The model of BIL involves a third parameter that arises from their description 
of the distributions (of charge, magnetic moments, etc.) within a baryon, and 
these three parameters are fit to the widths of all three- and four-star 
resonances, with the exception of the $S_{11}$ resonances. The results 
obtained in this model also depend on the quantities $\chi_1$ and $\chi_2$, 
which are related to the matrix elements for the vibrational excitations in 
this collective model. The expressions for $\chi_1$ and $\chi_2$ are
\begin{equation}
\chi_1=\frac{1-R^2}{R\sqrt{N}},\,\,\,\chi_2=\frac{\sqrt{1+R^2}}{R\sqrt{N}},
\end{equation}
where $R$ is a size parameter, and $N$ determines the size of the model space 
used in the calculation. The values of neither $N$ nor $R$ are provided in 
their articles. However, it appears that the numbers that they present are 
obtained in the limit
$N\to\infty$. In this case, the radial excitations, such as the 
$N\half^+(1440)$, do not decay to the
ground state, and the appropriate entries in the tables should be zeroes, 
instead of quantities that depend on $\chi_1$ and $\chi_2$.

KI do not explicitly calculate all of their amplitudes using Eq. 
(\ref{emission}). Instead, they identify four general amplitudes, one of which 
they term `structure independent', and three that are `structure dependent'. 
They fit these four amplitudes to the numbers obtained from the partial wave 
analyses, but it is not clear which of the experimental amplitudes are 
included in their fit.

CR fit to the $N \pi$ widths of all two-, three- and four-star resonances that 
lie in the $N\le 2$ oscillator-type bands. All of their numbers show 
`theoretical' uncertainties. These are obtained from the kinematic dependences 
of the amplitudes, which are due to the uncertainties in the 
masses of the states. For an observed state, the quoted decay amplitude is 
obtained by carrying out the calculation at the centrally reported mass of the 
state, and the uncertainty on the amplitude is obtained by evaluating the 
amplitude at the upper and lower limits given by the errors on the mass. For 
missing or unobserved states, the quoted decay amplitude is obtained using the 
theoretically predicted mass, while the errors on the amplitude are obtained 
assuming an error of 150 MeV in the mass of the state.

The wave functions used by WPTV are obtained using one version of the GBE 
model, and they use the $\tp0$ operator to
calculate the decay amplitudes. They examine a number of scenarios for the 
$N\pi$ decays. In the column labeled NR, the
Hamiltonian used is nonrelativistic, while for the column labeled SR, the 
kinetic energy operator for the
quarks is the relativistic one. In both cases, these authors fit to the 
amplitude for the decay $\Delta\to N\pi$. They
also explore the role of the pion size, and present results for two different 
pion radii. However, results for a single
pion size are presented here. It may be noted that an extended pion appears to 
be inconsistent with one of the basic
tenets of the GBE model, namely that the pion is a special, structure-less 
object that gives rise to the potential
between the quarks.

It is difficult to assess the efficacy of these models by simply examining the 
rows of numbers presented.
In the last four rows of Table \ref{nNle2t}, the $\chi^2$ values for each of 
the models is presented. This quantity is evaluated as 
\beq
\chi^2=\frac{1}{n}\sum_i^n \frac{\left(T_i-E_i\right)^2}{\sigma_i^2},
\eeq
where $T_i$ is the model prediction for the amplitude, $E_i$ is the amplitude 
extracted from the
partial wave analyses, $\sigma_i$ is the uncertainty of the extracted 
amplitude, and the sum includes
all states for which a particular model makes a prediction. A second measure of 
the efficacy of each
model, is the theoretical error $\tau$, defined by the expression
\beq
n=\sum_i^n \frac{\left(T_i-E_i\right)^2}{\sigma_i^2+\tau^2}.
\eeq
Two values each of $\chi^2$ and $\tau$ are presented. In the first set of 
numbers, all calculated
amplitudes for which there are experimental analogs are used. In the second set 
of numbers, the state $N\thalf^+(1720)$ 
is omitted, as a number
of models predict very large values for its amplitude into the $N\pi$ channel.

\squeezetable

\begin{table}[h]
\caption{ $N\pi$ decay widths for the models considered here. The first column 
identifies the state, while columns two to thirteen give the magnitude of the 
decay amplitudes into $N\pi$ calculated in the models of OR
\protect\cite{ALY1,ALY2,ALY3,ALY4,ALY5}, BIL \protect\cite{BIL}, SAS \protect\cite{sas}, KI
\protect\cite{KI,KI0}, CR \protect\cite{Capstick:1993th}, 
SST \protect\cite{sst1,sst2,sst3,sst4,sst7,sst8} and WPTV \protect\cite{Wagenbrunn:Nstar2k}, respectively. The numbers obtained from the partial wave analyses 
\protect\cite{PDG} are shown in column 14, labeled PWA. Missing states have 
their masses shown in square parentheses. The models of BIL, SAS and KI are 
elementary meson emission models, while those of OR, CR, SST and WPTV are pair 
creation models. For SAS, four numbers are shown, corresponding to four 
different scenarios in their model. Two numbers are shown for each of the SST 
and WPTV models. 
For BIL, the quantities $\chi_1$ and $\chi_2$ are defined in the text. The last four rows show the average $\chi^2$ per amplitude
(calculated using only observed states), as well as the theoretical error,
 defined in the text, for each of the models.}
\label{nNle2t}
\begin{tabular}{lrrrrrrrrrrrrr}
\multicolumn{1}{l} {state}
& \multicolumn{1}{c} {OR}
& \multicolumn{1}{c}{BIL}
& \multicolumn{4}{c} {SAS}
& \multicolumn{1}{c} {KI}
& \multicolumn{1}{c}{CR}
& \multicolumn{2}{c}{SST}
& \multicolumn{2}{c}{WPTV}
& \multicolumn{1}{c}{PWA} 
\\
\cline{4-7} \cline{10-11}\cline{12-13}\\
& & & \multicolumn{2}{c}{Set I}
&  \multicolumn{2}{c}{Set II}
& & & & & \\
\cline{4-5} \cline{6-7}\\
& & &\multicolumn{1}{c}{U} & \multicolumn{1}{c}{M} &\multicolumn{1}{c}{U} 
&\multicolumn{1}{c}{M} & & &\multicolumn{1}{c}{Point} 
& \multicolumn{1}{c}{Size} & \multicolumn{1}{c}{NR} & \multicolumn{1}{c}{SR} & \\ 
\hline
$N\half^-(1535)$ & 3.8 & 9.2 & 20.3& 14.9& 14.3& 9.4 & 5.3 & $14.4\pm 0.7$ & 18.3 & 12.9 & 11.8 & 6.3 & 8.0$\pm$2.8 \\
$N\half^-(1650)$ & 5.1 & 5.9 & 8.7& 12.7& 6.0& 8.9 & 8.7 & $10.7\pm 1.1$ & 19.6 & 2.3 & 4.0 & 7.4 & 8.7$\pm$1.9 \\
$N\thalf^-(1520)$ & 6.0 & 10.7 & 8.4& 8.5& 5.0& 5.0 & 9.2 & $10.0\pm 0.3$ & 4.5 & 8.4 & 15.1 & 6.7 & 8.3$\pm$0.9 \\
$N\thalf^-(1700)$ & 1.9 & 2.2 & 2.3& 3.4& 1.5& 2.3 & 3.6 & $6.0\pm 0.4$ & 2.3 & 4.1 & 2.0 & 1.0 & 3.2$\pm$1.3 \\
$N\fhalf^-(1675)$ & 4.1 & 5.6 & 5.0& 5.2& 3.1& 3.4 & 5.5 & $5.7\pm 0.1$ & 3.1 & 5.6 & 4.8 & 2.0 & 7.7$\pm$0.7 \\
$N\half^+(1440)$ & 8.9 & 10.4$\chi_1$ & 0.3& 0.1& 1.2& 1.0 & 6.8 & $22.2^{+0.6}_{-0.4}$ & 4.9 & 20.8 & 16.2 & 23.0 & 19.7$\pm$3.2 \\ 
$N\half^+(1710)$ & 5.1 & 13.2$\chi_2$ & 1.4& 9.1& 0.1& 6.0 & 6.7 & $3.4\pm 0.3$ & 3.5 & 1.8 & 4.0 & 7.4 & 4.7$\pm$1.2 \\
$N\half^+[1880]$ & & 4.7 & & & & & 4.4 & $3.0^{+1.1}_{-1.3}$ & & \\
$N\half^+[1975]$ & & 1.3  & & & & & 1.2 & $1.6^{+0.6}_{-0.5}$ & & \\
$N\thalf^+(1720)$ & 8.5 & 5.6 & 0.4 &10.8& 0.1& 8.6 & 6.5 & $17.1^{+0.5}_{-0.4}$ & 16.9 & 7.1 & 10.4 & 19.2 & 5.5$\pm$1.6 \\
$N\thalf^+[1870]$ & & 1.5 & & &  & & 3.2 & $5.6^{+1.9}_{-1.6}$ & & \\
$N\thalf^+[1910]$ & & 8.1 & & &  & & 1.1 & $0.2\pm 0.4$ & & \\
$N\thalf^+[1950]$ & & 8.4 & & &  & & 1.1 & $4.2^{+1.2}_{-1.1}$ & & \\
$N\thalf^+[2030]$ & & 3.9 & & &  & & 0.5 & $1.9^{+0.5}_{-0.4}$ & & \\
$N\fhalf^+(1680)$ & 6.0 & 6.4 & 7.6& 7.9& 3.9& 3.6 & 7.1 & $9.3\pm 0.2$ & 3.2 & 9.7 & 9.4 & 8.9 & 8.7$\pm$0.9 \\
$N\fhalf^+[1980]$ & & 8.2 & & &  & & 0.4 & $1.0\pm 0.1$ & & \\
$N\fhalf^+(2000)$ & & & & &  & & 1.3 & $1.3\pm 0.2$ & 0.7 & 2.0 & & & 2.0$\pm$1.2 \\
$N\shalf^+(1990)$ & & & & & & & 3.1 & $3.0\pm 0.3$ & 0.6 & 1.8 & & & $4.6\pm 1.9$ \\
$\Delta \half^-(1620)$ & 2.4 & 4.0 & 6.3& 5.8& 4.2& 3.7 & 3.3 & $4.6\pm 0.9$ & 8.3 & 0.6 & 2.5 & 2.0 & $6.5\pm 1.0$ \\
$\Delta \thalf^-(1700)$ & 3.8 & 5.2 & 4.1& 4.3& 2.7& 2.9 & 4.9 & $5.2\pm 0.5$ & 2.7 & 4.8 & 4.0 & 1.8 & $6.5\pm 2.0$ \\
$\Delta \half^+(1740)^a$ & & & & & & & 2.7 & $3.9^{+0.4}_{-0.7}$ & & & & & $4.9\pm 1.3$ \\
$\Delta \half^+(1910)$ & 5.7 & 6.5 & 0.9& 6.1& 0.4& 6.4 & 5.3 & $9.9^{+0.7}_{-0.8}$ & 7.8 & 0.7 & & & $6.6\pm 1.6$ \\
$\Delta \thalf^+(1232)$ & & 10.8 & 10.0& 10.5& 10.5& 10.9 & 11.0 & $10.2\pm 0.1$ & 10.7 & 10.7 & 11.0 & 11.0 & $10.7\pm 0.3$ \\
$\Delta \thalf^+(1600)$ & 19.0 & 10.4$\chi_1$ & 2.8& 14.0& 1.8& 10.6 & 5.4 & $6.3^{+1.9}_{-1.6}$ & 9.0 & 0.2 & 6.3 & 6.7
& $7.6\pm 2.3$ \\
$\Delta \thalf^+(1920)$ & & 4.7 & 4.0& 0.7& 3.4& 4.1 & 5.2 & $4.6\pm 0.5$ & 1.7 & 2.0 & & & $7.7\pm 2.3$ \\
$\Delta \thalf^+[1985]$ & & & & & & & 0.1 & $3.7^{+1.4}_{-1.5}$ &  &  \\
$\Delta \fhalf^+(1750)$ & & & & & & & & $2.0\pm 0.8$ \\
$\Delta \fhalf^+(1905)$ & 3.8 & 3.0 & 3.4& 2.7& 2.0& 1.1 & 4.0 & $4.3^{+0.2}_{-0.3}$ & 1.2 & 3.1 & & & $5.5\pm 2.7 $ \\
$\Delta \fhalf^+(2000)$ & & & & & & & 1.0 & $1.3^{+0.2}_{-0.3}$ & & & & & $5.3\pm 2.3$ \\
$\Delta \shalf^+(1950)$ & 9.8 & 6.7 & 7.1& 6.2& 4.2& 3.6 & 7.5 & $8.5\pm 0.1$ & 3.9 & 8.7 
& & & $9.8\pm 2.7$ \\\hline
$\chi^2_a$ & 7.2 & 3.5 (5.9) & 7.0 & 6.4 & 11.4 & 8.6 & 2.7 & 4.6 & 12.9 & 5.3 & 8.7 & 13.8 \\
$\tau_a$ & 4.1 & 2.2 (4.8) & 5.5 & 5.3 & 5.6 & 4.8 & 2.2 & 2.9 & 5.9 & 3.1 & 3.1 & 4.5 \\
$\chi^2_b$ & 7.4 & 3.8 (6.3) & 6.8 & 6.1 & 11.4 & 8.9 & 2.9 & 2.2 & 10.8 & 5.5 & 8.7 & 8.9 \\
$\tau_b$ & 4.2 & 2.3 (5.0) & 5.6 & 5.3 & 5.6 & 4.9 & 2.3 & 1.4 & 5.4 & 3.2 & 2.9 & 2.6 \\
\end{tabular}
$^a$ First $P_{31}$ state found in Ref.~\protect{\cite{MANSA}}.\\
\end{table}
\nopagebreak[4]

At first glance, all of the models seem to describe the amplitudes with similar 
kinds of success.
However, examination of the $\chi^2$ values, as well as the $\tau$ values, 
gives some indication that
a few of the models do better at describing the decays. For BIL, the amplitudes 
for the radial excitations
$N\half^+(1440)$, $N\half^+(1710)$ and $\Delta\thalf^+(1600)$ depend on two 
quantities, $\chi_1$ and $\chi_2$, which in turn
depend on $R$ and $N$. Here, $N$ is determined by the size of the model space 
that they use in obtaining the spectrum. 
In a more recent article \cite{Bijker:2000gq}, these authors treat $\chi_1$ and 
$\chi_2$ as constants with 
values of 1.0 and 0.7, respectively, values that are inconsistent with the 
large $N$ limit. 
The corresponding widths that they report are $N\pi$ widths of 108 MeV, 85 MeV 
and 108 MeV, 
respectively, for the three states in question.
When these widths are used in the calculation of $\chi^2$ and $\tau$, the 
values obtained are as shown in the 
table. However, all the rest of the BIL results appear to be obtained in the 
limit of $N\to\infty$, which means that
these three amplitudes are zero. Using these vanishing amplitudes leads to the 
values of $\chi^2$ and
$\tau$ shown in parentheses in the table.

It {\it must} be emphasized that the values of $\chi^2$ and $\tau$ are to be 
interpreted with care,
as they depend very strongly on which states are included in the calculation, 
as well as on the values
taken for the extracted amplitudes. For instance, if the $N\pi$ amplitude for 
the $N\half^+(1440)$ 
from Ref. \cite{PDG} is used, then
the quality of the models by CR, SST and WPTV would certainly appear to 
deteriorate, while all of the
elementary emission models would improve. Similarly, if the amplitudes for 
less-well-known resonances
are omitted from the calculation, some models would again improve, and others 
deteriorate. In
addition, it must be pointed out that the model of OR appears to do poorly 
when compared with many of
the others, but their $N\pi$ amplitudes were fit to the partial wave analyses 
available more than
twenty-five years ago, and many of those numbers have changed. Furthermore, 
they fit not only $N\pi$
amplitudes, but $\Delta\pi$ and $N\rho$ as well. Thus, the quality of their 
$N\pi$ predictions has
suffered.

There are significant disagreements among the model predictions for the $N\pi$ 
amplitudes of these low-lying states, and these differences 
become more apparent when other decay channels are included in the comparison. 
One of the more striking disagreements among the models is the prediction of 
SAS for the decay of the Roper resonance. Their results disagree with all the 
other models, and with the results obtained from the partial wave analyses.

The role of the kinetic energy of the quarks is seen in the comparison of the 
two 
sets of numbers obtained from the model of WPTV. In fact, changing the kinetic 
energy operator in the
Hamiltonian from the nonrelativistic to the semi-relativistic has resulted in 
differences of factors of two in some of the $N\pi$ amplitudes.

The effects of configuration mixing are clearly seen in the decay widths that 
result from the different scenarios explored by SAS. The calculated decay 
widths of the $S_{11}$ resonances, as well as those of the $P_{11}$, $P_{13}$, 
$P_{31}$ and $P_{33}$ resonances, are very different depending on whether the 
wave functions used include the effects of mixing or not. Note, too, that the 
OR and CR decay models are very similar, but the OR wave functions 
have no configuration mixing, while those of the CR model do. This difference 
accounts for much of the differences between the two sets of predictions, which
are often different by more than a factor of two. 
These differences also make it clear that predictions for the spectrum of 
states produced in a particular model are not very sensitive probes of the wave 
functions. All of the wave functions used in these calculations have been
obtained from fits to the spectrum of states that are of similar quality, yet 
the predictions for the strong 
decay widths are very different.

Another striking effect illustrated in this table is that of using a point-like 
pion in describing the decays. This is seen in comparing the two sets of 
numbers taken from the model of SST. This effect is seen most strongly in the 
numbers predicted for the  $S_{11}$ resonances, as well as those of the 
$P_{11}$, $P_{13}$, $P_{31}$ and $P_{33}$ resonances, just as with 
configuration mixing. A more striking illustration of this is the fact that 
almost all the meson emission models predict very small amplitudes for the 
$N\pi$ decay of the $P_{11}(1440)$, while the pair-creation models predict 
large amplitudes for these decays. KI predict a relatively large amplitude for 
this decay, but one must keep in mind that these authors did not calculate the 
decay amplitudes, but fit sets of amplitudes.

Perhaps the state most sensitive to the scenarios explored by the different 
authors is the Roper resonance, $N\half^+(1440)$. For point-like pions, most 
authors obtain very small $N\pi$ decay widths for this state (with the 
exception of KI, as pointed out in the previous paragraph). For pions with 
structure, the predicted partial widths are very large, very much in agreement 
with the partial wave analysis of Cutkosky and Wang \cite{Cutkosky:1990zh}. It 
must be pointed out, however, that many partial wave analyses give values of 
the order of 100 to 150 MeV for this partial width, values that would 
appear difficult to accommodate in most of the models discussed in Table 
\ref{nNle2t}.

This state has been responsible for generating many articles on its nature, 
its mass, and most recently, on the origin of the interquark potential in 
potential models. Recently, for instance, Li, Burkert and Li \cite{Li:1992yb},
 as well as Kisslinger and Li \cite{Kisslinger:1995yw},
suggest that this state might not be one of the 
usual $qqq$ states, but a hybrid, instead. SST also attempt to improve their
description of this resonance by modifying the main component of its radial 
wave
function \cite{Stancu:1990ht}. In so doing, they are able to lower its mass by about 100 MeV. Other 
authors, most recently 
Krehl {\it et al.} \cite{Krehl:1999km}, suggest that this resonance can be 
interpreted in terms of meson-baryon dynamics alone, with no need for a 
three-quark resonance. Finally, it should be noted that the recent spate of GBE models
described in previous sections
originate in part to explain the somewhat low mass of this state, as 
models relying on one-gluon exchange in the potential usually place this state 
about 100 MeV heavier. Whatever the nature of this state, it is clear that more
precise scattering data, as well as more rigorous analyses of these scattering
data, are needed to shed some light on these questions.

To a large extent, the models examined above predict relatively small
$N\pi$ widths for the missing baryons. Again, however, there are
exceptions, and these come mainly from the model by BIL. Table
\ref{nNle2t} does not show all of the missing states that they predict
\cite{Bijker:2000gq}. As pointed out in an earlier section, this model
predicts at least fourteen missing nucleon resonances below 2 GeV,
more than any other model.  In addition, six of these resonances have
appreciable couplings to the $N\pi$ formation channel, with four of
them having $N\pi$ partial widths greater than 50 MeV. In contrast
with this, four of their light missing states do not couple to {\it
any} of the channels ($N\pi$, $\Delta\pi$, $N\eta$, $\Sigma K$,
$\Lambda K$, $\Sigma^*K$) investigated by these authors. Some of the
$N\pi$ large widths for missing states are probably due to the absence
of configuration mixing in their wave functions or to the fact that
they use a point-like pion. In fact, these predictions are expected to
change if these effects are included in this
calculation. Nevertheless, the predictions from this model appear to
be difficult to reconcile with existing analyses.
\vskip -12pt
\subsection{other channels}
With the qualified success of most of these models in explaining the missing 
states (those that treat missing states), attention can now be turned to the 
question of where can these missing 
states be seen. In these models, a
number of other channels have been explored, including $N\eta$ and 
$\Delta\eta$, $\Delta\pi$,
$N\rho$ and $N\omega$, as well as some channels containing strange mesons and 
baryons, such as
$\Lambda \bar K$ and $\Sigma \bar K$. Such calculations not only test the 
models (by 
comparing their
result with whatever data are available), but also allow predictions of which 
channels are
good candidates for `resonance hunting'. In particular, isospin-selective 
channels like
$\Delta\eta$, $N\eta$ and $N\omega$, are quite interesting, as the 
partial-wave analyses
required are expected to be simpler than for some of the other channels, where 
both $N^*$ and
$\Delta^*$ states are expected to contribute.

In the pair creation models, the treatment of decays involving vector mesons 
proceeds much as treatment of decays involving pseudoscalar mesons. For 
elementary vector-meson emission, Koniuk modifies a non-relativistic form that 
has been
used for photon emission, and obtains \cite{KI2}
\begin{equation}\label{vector}
{\cal H}_V=-\frac{3i}{(2\pi)^{3/2}} \left[g \left(\veca{\varepsilon}^*\cdot 
\veca{p}_3^\prime-\veca{\varepsilon}^*\cdot \veca{k}^\prime\frac{E_3^\prime}
{E_v}
\right)+
h \veca{s}_i\cdot\veca{k}\times\veca{\varepsilon}^*\right] X_3^V.
\end{equation}
Here $\veca{\varepsilon}^*(\veca{k},\lambda)$ is the vector-meson polarization 
vector,
and $X_3^V$ are flavor matrices that describe the quark transitions $q_3\to
q_3^\prime+V$. In this sector, $\phi-\omega$ mixing can be treated in a manner 
analogous to 
$\eta-\eta^\prime$ mixing, as discussed above. With
\begin{equation}
\omega=V_8 \cos{\theta}-V_1\sin{\theta},\,\,\,
\phi=V_8 \sin{\theta}+V_1\cos{\theta},
\end{equation}
where $V_{1,8}$ denotes the singlet or octet component of the isoscalar vector 
meson, and a value of $\theta=54.7^\circ$ results if the hidden strangeness 
component of the 
$\omega$ vanishes. Note that if
\begin{equation}
g=h=\frac{e}{2m_3}\left(\lambda_3+\sqrt{\frac{1}{3}}\lambda_8\right),
\end{equation}
the vector meson emission operator above becomes the lowest order operator for 
photon emission in a non-relativistic expansion. In this expression, $\lambda_3$
and $\lambda_8$ are flavor matrices.

For the $N\rho$ and $\Delta\pi$ channels, as well as for $\Delta\eta$, one of 
the daughter hadrons produced in the decay is a broad one. BIL ignore this, 
and treat both daughter
hadrons in the narrow-width approximation. CR and SST take the width of the 
daughter hadron
into account by writing the width of the quasi-two-body decay 
$A\to (X_1X_2)_BC$ (where
$X_1$ and $X_2$ are the decay products of the hadron $B$) as
\begin{equation}
\Gamma=\int_0^{k_{\max}} dk\frac{k^2\left|M(k)\right|^2\Gamma_t(k)}
{\left(M_a-E_b(k)-E_c(k)\right)^2+\frac{\Gamma_t(k)^2}{4}},
\end{equation}
where $\Gamma_t(k)$ is the energy-dependent total width of the unstable 
daughter hadron, $B$ in this case. 

At this point it is necessary to point out some differences in the way KI 
treat the pseudoscalar emission decays, and the way Koniuk treats the vector 
emission decays. For each calculation, the coupling constants $g$ and $h$ of 
Eq. (\ref{vector}) are 
fit to available partial widths, but it is not clear which partial widths are 
used in the fit. In treating the $N\rho$ decays, Koniuk integrates over the 
line-shape of the $\rho$ meson, but the form used for the integral is not 
given. In contrast, when KI treat the $\Delta\pi$ decays, similar integrals 
over the line-shape of the $\Delta$ are not carried out. 

For the $N\eta$ channel (table \ref{neta}), results are available from
the models of BIL, KI, CR and WPTV, while only BIL and CR consider
$\Delta\eta$ (table \ref{deta}). BIL are unable to reproduce the large
couplings of the lowest lying $S_{11}$ state to the $N\eta$
channel. This may be due to the fact that their wave functions do not
include any configuration mixing from the tensor hyperfine
interaction. The fact that they use point-like mesons may also play a
role in suppressing this decay in their model. A few other significant
disagreements also appear in the $\Delta\eta$ and $\Delta\pi$
channels. In contrast, the models of CR and KI are generally in better
agreement. No partial wave analyses have yet been performed on the
$\Delta\eta$ channel, but this channel offers the possibility of
discovering some of the missing baryon resonances.  This is
particularly so for the higher lying resonances (not shown in the
table), where the available phase space is not a limitation on whether
or not the decay can proceed.

In obtaining their results for the $N\eta$ channel, WPTV refit the value of the 
$\tp0$ coupling constant in order to
match the $N\eta$ decay width of the lowest lying $S_{11}$ resonance. For both 
scenarios that they explore, this leads to
a significant change in the coupling constant. In comparison, BIL, KI and CR 
use the same value of this coupling constant
that they obtain from consideration of the $N\pi$ decays. In Table \ref{neta}, 
the results from two different scenarios
explored by WPTV are presented. In addition, the results that would have been 
obtained if the coupling constant were not
rescaled are also presented in parentheses. As is apparent from the table, 
the effect of rescaling is a very significant one.

It should be noted that except for BIL, the other models discussed present 
signs for the 
amplitudes of the decays to channels other than $N\pi$. These signs must be 
understood as being the product of the calculated sign for the $N\pi$ channel, 
with the calculated sign of the channel being discussed, $N\eta$, say, both 
calculated in the particular model. This is because the signs of the $N\pi$ 
amplitudes are inaccessible in an elastic scattering experiment, but the 
product of the signs of the $N\pi$ and $N\eta$ decay amplitudes are accessible 
in an inelastic scattering experiment. The signs of the amplitude provide a 
further means of testing the validity and success of these models. BIL state 
that all of these signs are arbitrary, and therefore do not present them.

\begin{table}[t]
\caption{$N\eta$ amplitudes for low-lying $N^*$ resonances, from the models of 
BIL \protect\cite{BIL}, CR \protect\cite{CR2}, KI \protect\cite{KI,KI0} 
and WPTV \protect\cite{Wagenbrunn:Nstar2k}. The few experimental data available are shown in the last 
column.}
\label{neta}
\begin{tabular}{lrrrrrr}
\multicolumn{1}{l} {state}
& \multicolumn{1}{c}{BIL}
& \multicolumn{1}{c} {KI}
& \multicolumn{1}{c}{CR}
& \multicolumn{2}{c}{WPTV}
& \multicolumn{1}{c}{PWA} \\
\cline{5-6}\\
&&&& \multicolumn{1}{c}{NR} & \multicolumn{1}{c}{SR} & \\
\hline
$N\half^-(1535)$ & 0.3 & +5.2 & $+14.6^{+0.7}_{-1.3}$ & (22.5) 8.0  & (12.1) 8.0  & $+8.1\pm 0.8$ \\
$N\half^-(1650)$ & 2.8 & -1.5 & $-7.8^{+0.1}_{-0.0}$ & (18.6) 6.6 & (10.1) 6.7 & $-2.4\pm 1.6$ \\
$N\thalf^-(1520)$ & 0.8 & +0.4 & $+0.4^{+2.9}_{-0.4}$ & (0.0) 0.0 & (3.9) 2.6 & \\ 
$N\thalf^-(1700)$ & 2.0 & -0.7 & $-0.2\pm 0.1$ & \\
$N\fhalf^-(1675)$ & 4.1 & -2.8 & $-2.5\pm 0.2$ & (2.8) 1.0 & (1.5) 1.0 & \\
$N\half^+(1440)$ & & & $+0.0^{+1.0}_{-0.0}$ & \\ 
$N\half^+(1710)$ & 4.1 $\chi_2$ & +2.9 & $+5.7\pm 0.3$ & (3.9) 1.4 & (13.6) 8.9 & \\
$N\half^+[1880]$ & & -0.8 & $-3.7^{+0.5}_{-0.0}$ & \\
$N\half^+[1975]$ & & & $+0.1^{+0.2}_{-0.1}$ & \\
$N\thalf^+(1720)$ & 0.5 & +1.9 & $+5.7\pm 0.3$ & (7.3) 2.6 & (9.7) 6.4 & \\
$N\thalf^+[1870]$ & & -2.9 & $-4.6\pm 0.3$ & \\
$N\thalf^+[1910]$ & & & $-0.9\pm 0.1$ & \\
$N\thalf^+[1950]$ & & & 0.0 & \\
$N\thalf^+[2030]$ & & & $+0.4\pm 0.1$ & \\
$N\fhalf^+(1680)$ & 0.7 & +0.7 & $+0.6\pm 0.1$ & (2.8) 1.0 & (1.5) 1.0 & \\
$N\fhalf^+[1980]$ & & & $-0.8\pm 0.2$ & \\
$N\fhalf^+(2000)$ & & -0.6 & $+1.9\pm 0.8$ & \\
$N\shalf^+(1990)$ & & -2.3 & $-2.2^{+0.6}_{-0.7}$ & \\
\end{tabular}
\end{table}
\begin{table}[h]
\caption{Results for the lowest-lying $\Delta$ states in
the $\Delta\eta$ channel. Results from the model calculations of CR \protect\cite{CR3} and BIL
\protect\cite{BIL}
are shown, with results from the different models presented on different rows. 
Light states with zero amplitudes are omitted from the table.}

\label{deta}
\begin{tabular}{llrrr}
\multicolumn{1}{l} {state}
& \multicolumn{1}{l}{Model}
& \multicolumn{1}{r}{$\Delta\eta$}
& \multicolumn{1}{r}{$\Delta\eta$}
& \multicolumn{1}{r}{$\sqrt{\Gamma^{\rm tot}_{\Delta\eta}}$}
\\ 
\hline
& & $s$ & $d$ & \\
\cline{3-4}\\[-5pt]

$\Delta \thalf^-(1700)$ & CR & 1.1 $^{+
 3.2}_{- 1.1}$ & 0.0 $^{+ 0.3}_{- 0.0}$ & 1.1 $^{+ 3.2}_{- 1.1}$\\[+5pt]

& & $p$ & & \\
\cline{3-3}\\[-5pt]

$\Delta \half^+(1740)$$^{\rm a}$ & CR &  3.2
    $^{+ 4.1}_{- 3.1}$ & & 3.2 $^{+ 4.1}_{- 3.1}$\\

$\Delta \half^+(1910)$ & CR &  -2.9 $\pm
 $ 0.7 & & 2.9 $\pm $ 0.7 \\
& BIL & & & 0.0\\[+5pt]

& & $p$ & $f$ & \\
\cline{3-4}\\[-5pt]

$\Delta \thalf^+(1600)$ & CR &  0.0 $^{+
0.3}_{- 0.0}$ & 0.0 $\pm $ 0.0 & 0.0 $^{+ 0.3}_{- 0.0}$\\[+5pt]

$\Delta \thalf^+(1920)$ & CR & -3.3
$\pm $ 0.9 & 0.7 $\pm $ 0.4 & 3.4 $\pm $ 0.9 \\
& BIL & & & 0.7 \\[+5pt]

$\Delta \thalf^+[1985]$ & CR &
-4.2 $^{+ 2.4}_{- 1.7}$ & - 0.7 $^{+ 0.6}_{- 1.2}$ & 4.3 $^{+ 1.9}_{-
2.5}$ \\[+5pt]

& & $p$ & $f$ & \\
\cline{3-4}\\[-5pt]

$\Delta \fhalf^+(1905)$ & CR & -0.5 $\pm
$ 0.1 & 0.6 $\pm $ 0.3 & 0.8 $\pm $ 0.3 \\
& BIL & & & 1.0 \\[+5pt]

$\Delta \fhalf^+(2000)$ & CR & -7.0 $^{+
 5.1}_{- 2.9}$ & 0.3 $^{+ 0.8}_{- 0.3}$ & 7.0 $^{+ 2.9}_{- 5.1}$ \\[+5pt]

& & $f$ & $h$ & \\ 
\cline{3-4} \\[-5pt]

$\Delta \shalf^+(1950)$ & CR & 0.9 $\pm
$ 0.1 & 0.0 $\pm $ 0.0 & 0.9 $\pm $ 0.1 \\
& BIL & & & 1.4 \\[+5pt]
\hline
\end{tabular}
$^a$ First $P_{31}$ state found in Ref.~\protect{\cite{MANSA}}.\\
\end{table}

For these channels, no authors have compared results with pointlike
and extended daughter mesons, nor with and without configuration
mixing. Given the results obtained in the $N\pi$ channel, it is to be
expected that the predictions for the quasi-two-body final states will
be significantly impacted by the inclusion or exclusion of these two
effects.  This will certainly influence which channels are good
candidates for seeking particular missing baryonic resonances.

The results obtained for the $N\rho$ and $\Delta\pi$ channels are presented in 
tables
\ref{ndpi} (for nucleons) and \ref{ddpi} (for $\Delta$s), and results there are 
from CR, BIL, KI, 
SST and OR ($\Delta\pi$), and from CR, KI, SST and OR ($N\rho$). 

\squeezetable

\begin{table}[t]
\caption{Results for the lowest-lying nucleons in the $\Delta\pi$, 
and $N\rho$ channels, for the models by CR \protect\cite{CR2}, KI \protect\cite{KI,KI0,KI2}, SST
\protect\cite{sst1,sst2,sst3,sst4,sst5,sst6,sst7,sst8}, BIL \protect\cite{BIL} and OR
\protect\cite{ALY1,ALY2,ALY3,ALY4,ALY5}. The results 
from the different models are on different rows. Also shown are the results 
from the partial wave analyses (PWA).}
\label{ndpi}
\begin{tabular}{llrrrrrrr}
\multicolumn{1}{l} {state}
& \multicolumn{1}{l}{Model}
& \multicolumn{1}{r}{$\Delta\pi$}
& \multicolumn{1}{r}{$\Delta\pi$}
& \multicolumn{1}{r}{$\sqrt{\Gamma^{\rm tot}_{\Delta\pi}}$} 
& \multicolumn{1}{r}{$N\rho$}
& \multicolumn{1}{r}{$N\rho$}
& \multicolumn{1}{r}{$N\rho$}
& \multicolumn{1}{r}{$\sqrt{\Gamma^{\rm tot}_{N\rho}}$}\\
\hline
&& $d$ & & & $s_{\textstyle{1\over2}}$ & $d_{\textstyle{3\over2}}$ & &\\
\cline{3-3}
\cline{6-7}\\[-5pt]
$N\half^-(1535)$ & CR & $+1.4\pm 0.3$ & $$ & $1.4\pm 0.3$ & $-0.7\pm 0.1$ 
& $+0.4\pm 0.1$ & $$  & $0.8^{+0.2}_{-0.1}$\\
& KI & -1.7 & & 1.7 & +6.1 & -1.6 & & 6.3 \\
& SST & -4.0 & & 4.0 & & & & 1.1 \\
& BIL & 4.8 & & 4.8 \\
& OR & +0.9 & & 0.9 \\
& PWA & $0.0$ & & $0.0$ & $-1.7\pm 0.6$ & $-1.4\pm 0.7$ & 
& $2.2\pm 0.6$\\[+5pt]

$N\half^-(1650)$ & CR & $+3.6^{+0.8}_{-0.6}$ & $$ & $3.6^{+0.8}_{-0.6}$ 
& $+0.9^{+0.3}_{-0.2}$ & $+0.4\pm 0.1$ & $$ & $1.0^{+0.3}_{-0.2}$ \\
& KI & -8.2 & & 8.2 & +9.7 & -2.7 & & 10.1 \\
& SST & -2.6 & & 2.6 & & & & 0.6\\
& BIL & 4.9 & & 4.9 \\
& OR & -6.2 & & 6.2 \\
& PWA & $+1.7\pm 0.6$ & & $1.7\pm 0.6$ & $0.0$ & $+2.2\pm 0.9$ & 
& $2.2\pm 0.9$\\[+5pt]

&& $s$ & $d$ & & $d_{\textstyle{1\over2}}$ & $s_{\textstyle{3\over2}}$ & $d_{\textstyle{3\over2}}$ &\\
\cline{3-4}
\cline{6-8}\\[-5pt]
$N\thalf^-(1520)$ & CR & $-5.7^{+3.6}_{-1.6}$ & $-1.5^{+1.3}_{-3.0}$ 
& $5.9^{+2.6}_{-3.8}$ & $-0.1^{+0.1}_{-0.3}$ & $-2.4^{+1.9}_{-6.4}$ 
& $-0.3^{+0.2}_{-1.0}$ & $2.5^{+6.5}_{-1.9}$ \\
& KI & +6.7 & +2.5 & 7.2 & -0.7 & +5.0 & +1.1 & 5.2 \\
& SST & -3.4 & +4.4 & 5.6 & +3.2 & -1.7 & -2.7 & 4.6 \\
& BIL & 1.7 & 3.0 & 3.5\\
& OR & +7.3 & +1.3 & 7.4 \\
& PWA & $-2.6\pm 0.8$ & $-4.2\pm 0.6$ & $5.0\pm 0.6$ & 
&  $-5.1\pm 0.6$ & & $5.1\pm 0.6$\\[+5pt]

$N\thalf^-(1700)$ & CR & $-27.5\pm 1.6$ & $+4.6^{+1.6}_{-1.3}$ 
& $27.9^{+1.9}_{-1.8}$ & $0.0$ & $\pm 0.1\pm 0.0$ & $-0.9^{+0.3}_{-0.6}$ 
& $0.9^{+0.6}_{-0.4}$ \\
& KI & +16.0 & -7.7 & 17.8 & +0.1 & +4.3 & +2.7 & 5.1 \\
& SST & 0.5 & 0.4 & 0.6 & & & & 3.7 \\
& BIL & 10.5 & 10.7 & 15.0 \\
& OR & +13.3 & -5.1 & 14.2 \\
& PWA & $+3.5\pm 3.6$ & $+14.1\pm 5.6$ & $14.5\pm 5.5$ & 
& $-5.6\pm 5.7$ & & $5.6\pm 5.7$\\[+5pt]

& & $d$ & $g$ & & $d_{\textstyle{1\over2}}$ & $d_{\textstyle{3\over2}}$ & $g_{\textstyle{3\over2}}$ &\\
\cline{3-4}
\cline{6-8}\\[-5pt]
$N\fhalf^-(1675)$& CR & $+5.7\pm 0.4$ & $0.0$ & $5.7\pm 0.4$ & $+0.2\pm 0.0$ 
& $-0.4\pm 0.0$ & $0.0$ & $0.5\pm 0.0$ \\
& KI & -9.3 & & 9.3 & +1.1 & +2.0 & 0.0 & 2.3 \\
& SST & -2.5 & 0.1 & 2.5 & & & & 2.0 \\
& BIL & 11.1 & 11.1 \\
& OR & +4.1 & -4.7 & 6.2 \\
& PWA & $+9.2\pm 0.3$ & $$ & $9.2\pm 0.3$ & $0.8\pm 0.4$ & $$ 
& $-0.5\pm 0.5$ & $1.0\pm 0.4$\\[+5pt]

& & $p$ & & & $p_{\textstyle{1\over2}}$ & $p_{\textstyle{3\over2}}$ & &\\
\cline{3-3}
\cline{6-7}\\[-5pt]
$N\half^+(1440)$ & CR & $+3.3^{+2.3}_{-1.8}$ & $$ & $3.3^{+2.3}_{-1.8}$ 
& $-0.3^{+0.2}_{-0.3}$ & $-0.5^{+0.3}_{-0.5}$ & $$ 
& $0.6^{+0.5}_{-0.3}$ \\
& KI & -2.4 & & 2.4 & -0.3 & -0.1 & & 0.3 \\
& SST & -10.0 & & 10.0 & & & & 1.5 \\
& BIL & 0.3$\chi_1$ & & 0.3$\chi_1$\\
& PWA & $+9.4\pm 0.8$ & $$ & $9.4\pm 0.8$ & $$ & $$ & $$ & $$\\[+5pt]

$N\half^+(1710)$ & CR & $-13.9\pm 1.5$ & $$ & $13.9\pm 1.5$ & $+0.3\pm 0.1$ 
& $-3.7^{+0.9}_{-1.2}$ & $$ & $3.7^{+1.2}_{-1.0}$ \\
& KI & +3.6 & & 3.6 & -5.5 & -2.5 & & 6.0 \\
& SST & -19.2 & & 19.2 & & & & 4.1 \\
& BIL & 8.4$\chi_2$ & & 8.4$\chi_2$ \\
& PWA & $-15.3\pm 3.6$ & $$ & $15.3\pm 3.6$ & $$ & $$ & $$ & $$\\[+5pt]

$N\half^+[1880]$ & CR & $-8.7^{+2.1}_{-0.4}$ & $$ & $8.7^{+0.4}_{-2.1}$ 
& $+2.3^{+1.7}_{-1.4}$ & $\pm 0.3^{+0.0}_{-0.1}$ & $$ & $2.3^{+1.7}_{-1.4}$ \\
& KI & 3.4 & & 3.4 & -4.6 & +1.1 & & 4.7 \\
& SST & 10.2 & & 10.2 & & & & 1.9 \\[+5pt]

$N\half^+[1975]$& CR & $-4.6^{+0.4}_{-0.2}$ & $$ & $4.6^{+0.2}_{-0.4}$ 
& $+0.7^{+0.1}_{-0.3}$ & $-2.4^{+1.0}_{-0.4}$ & $$ & $2.5^{+0.4}_{-1.1}$ \\
& KI & +1.8 & & 1.8 & -1.2 & +0.3 & & 1.3 \\
& SST & +3.2 & & 3.2 & & & & 0.3 \\[+5pt]

&& $p$ & $f$ & & $p_{\textstyle{1\over2}}$ & $p_{\textstyle{3\over2}}$ & $f_{\textstyle{3\over2}}$ &\\
\cline{3-4}
\cline{6-8}\\[-5pt]
$N\thalf^+(1720)$ & CR & $-1.7\pm 0.2$ & $-1.0^{+0.2}_{-0.3}$ & $2.0\pm 0.3$ 
& $-2.6^{+0.7}_{-0.8}$ & $+1.8^{+0.6}_{-0.5}$ & $+0.7^{+0.3}_{-0.2}$ 
& $3.3^{+1.0}_{-0.8}$ \\
& KI & +1.9 & -1.0 & 2.1 & -11.7 & +2.6 & +3.5 & 12.5 \\
& SST & +1.5 & 0.4 & 1.6 & & & & 5.2 \\
& BIL & 1.0 & 3.2 & 3.3 \\ 
& OR & +3.1 & +4.7 & 5.6 & -5.7 & -2.5 & -1.9 & 6.5 \\
& PWA & $$ & $$ & $$ & $+18.2\pm 4.6$ & $$ & $$ & $18.2\pm 4.6$\\[+5pt]

$N\thalf^+(1880)$$^{\rm a}$ & CR & $+3.8\pm 0.5$ & $-2.2^{+1.2}_{-1.5}$ 
& $4.4^{+1.3}_{-1.0}$ & $-1.4^{+0.9}_{-1.0}$ & $-1.0\pm 0.6$ 
& $+0.2^{+0.5}_{-0.2}$ & $1.8^{+1.2}_{-1.1}$ \\
& KI & -4.1 & -1.5 & 4.4 & +0.4 & +1.32 & +0.5 & 1.5 \\
& SST & -11.4 & 0.2 & 11.4 & & & & 6.1 \\
& PWA & $$ & $$ & $$ & $-14.7\pm 2.9$ & $$ & $$ 
& $14.7\pm 2.9$\\[+5pt]

$N\thalf^+[1910]$ & CR & $+16.8^{+0.2}_{-2.9}$ & $+2.9^{+1.9}_{-1.4}$ 
& $17.0^{+0.6}_{-3.0}$ & $+2.5\pm 1.3$ & $-2.2^{+1.1}_{-1.2}$ 
& $+0.5^{+0.9}_{-0.4}$ & $3.3^{+1.9}_{-1.7}$ \\
& KI & -9.4 & -0.7 & 9.4 & -3.9 & +6.3 & +3.3 & 7.1 \\
& SST & +11.5 & 0.5 & 11.5 & & & & 5.7 \\[+5pt]

$N\thalf^+[1950]$ & CR & $-5.9^{+0.7}_{-0.3}$ & $+5.1^{+3.4}_{-2.3}$ 
& $7.8^{+2.7}_{-1.9}$ & $+1.0^{+0.2}_{-0.4}$ & $+3.6^{+1.1}_{-1.6}$ 
& $+1.2^{+1.3}_{-0.8}$ & $3.9^{+1.6}_{-1.8}$ \\
& KI & -3.4 & +9.2 & 9.8 & -7.5 & +2.9 & +2.3 & 8.4 \\
& SST & +2.1 & +1.6 & 2.6 & & & & 3.3 \\[+5pt]

$N\thalf^+[2030]$ & CR & $-6.6\pm 0.4$ & $+2.5^{+1.7}_{-1.0}$ 
& $7.1^{+1.1}_{-0.7}$ & $\pm 0.1^{+0.1}_{-0.0}$ & $+2.8\pm 0.7$ 
& $+0.5^{+0.3}_{-0.4}$ & $2.9\pm 0.8$ \\
& KI & +3.4 & +4.5 & 5.6 & +0.2 & -3.4 & -1.9 & 3.9 \\
& SST & +3.5 & 0.5 & 3.5 & & & & 2.8 \\[+5pt]

&& $p$ & $f$ & & $f_{\textstyle{1\over2}}$ & $p_{\textstyle{3\over2}}$ & $f_{\textstyle{3\over2}}$ &\\
\cline{3-4}
\cline{6-8}\\[-5pt]
$N\fhalf^+(1680)$ & CR & $+1.6\pm 0.1$ & $+0.5\pm 0.1$ & $1.7\pm 0.1$ 
& $-0.2\pm 0.0$ & $-3.0^{+0.4}_{-0.5}$ & $-0.3\pm 0.1$ 
& $3.0^{+0.5}_{-0.4}$ \\
& KI & +2.0 & -0.7 & +2.1 & -1.6 & +4.0 & 1.3 & 4.5 \\
& SST & -0.7 & +0.5 & 0.9 & +3.1 & -1.3 & +2.7 & 4.3 \\
& BIL & 1.4 & 1.7 & 2.2 \\
& OR & -5.1 & -1.1 & 5.2 \\
& PWA & $-3.6\pm 0.6$ & $+1.0\pm 0.5$ & $3.7\pm 0.6$ & $$ 
& $-2.8\pm 0.7$ & $-1.7\pm 0.6$ & $3.3\pm 0.7$\\[+5pt]

$N\fhalf^+[1980]$ & CR & $+15.0^{+0.1}_{-0.3}$ & $+3.7^{+1.5}_{-1.1}$ 
& $15.5\pm 0.5$ & $-1.8^{+1.0}_{-0.8}$ & $+0.8\pm 0.1$ & $+0.8^{+0.3}_{-0.4}$ 
& $2.2^{+0.9}_{-1.0}$ \\
& KI & +4.7 & -6.5 & 8.0 & -1.6 & -8.0 & -0.7 & 8.2 \\
& SST & 0.4 & 0.4 & 0.6 & & & & 4.2 \\[+5pt]

$N\fhalf^+(2000)$ & CR & $+7.8^{+0.4}_{-0.6}$ & $-5.8^{+2.4}_{-3.9}$ 
& $9.8^{+3.0}_{-1.7}$ 
& $-0.4\pm 0.3$ & $-7.8^{+3.1}_{-0.2}$ & $-0.2\pm 0.1$ & $7.8^{+0.2}_{-3.1}$ \\
& KI & -7.0 & -4.3 & 8.2 & +1.7 & +6.6 & +4.4 & 8.1 \\
& SST & -0.4 & +0.3 & 0.5 & & & & 4.3 \\
& PWA & $+7.7\pm 5.8$ & $+1.4\pm 9.2$ & $7.9\pm 5.9$ & $$ 
& $-17.2\pm 6.2$ & $+8.5\pm 5.8$ & $19.2\pm 6.1$\\[+5pt]

&& $f$ & $h$ & & $f_{\textstyle{1\over2}}$ & $f_{\textstyle{3\over2}}$ & $h_{\textstyle{3\over2}}$ &\\
\cline{3-4}
\cline{6-8}\\[-5pt]
$N\shalf^+(1990)$ & CR & $+5.0^{+2.0}_{-1.4}$ & $0.0$ & $5.0^{+2.0}_{-1.4}$ 
& $+0.6\pm 0.3$ & $-1.0^{+0.6}_{-0.5}$ & $0.0$ & $1.2^{+0.6}_{-0.7}$ \\
& KI & -6.0 & & 6.0 & +0.8 & -4.2 & 0.0 & 4.3 \\
& SST & 0.4 & 1.0 & 1.1 & & & & 1.1 \\[+5pt]
\hline
\end{tabular}
$^a$\ \ Second $P_{13}$ found in Ref.~\cite{MANSA}.
\end{table}
%

\squeezetable
\begin{table}[t]
\caption{Results for the lowest-lying $\Delta$ states in 
the $\Delta\pi$ and $N\rho$ channels, in the models by CR \protect\cite{CR2}, KI \protect\cite{KI,KI0,KI2}, SST
\protect\cite{sst1,sst2,sst3,sst4,sst5,sst6,sst7,sst8}, BIL \protect\cite{BIL} and OR
\protect\cite{ALY1,ALY2,ALY3,ALY4,ALY5}. The results from each model are shown on different rows. 
Also shown are the results obtained from the partial wave analyses (PWA).}
\label{ddpi}
\begin{tabular}{lcrrrrrrr}
\multicolumn{1}{l} {state}
& \multicolumn{1}{r}{Model}
& \multicolumn{1}{r}{$\Delta\pi$}
& \multicolumn{1}{r}{$\Delta\pi$}
& \multicolumn{1}{r}{$\sqrt{\Gamma^{\rm tot}_{\Delta\pi}}$} 
& \multicolumn{1}{r}{$N\rho$}
& \multicolumn{1}{r}{$N\rho$}
& \multicolumn{1}{r}{$N\rho$}
& \multicolumn{1}{r}{$\sqrt{\Gamma^{\rm tot}_{N\rho}}$}\\
\hline
&& $d$ & & & $s_{\textstyle{1\over2}}$ & $d_{\textstyle{3\over2}}$ & &\\
\cline{3-3}\cline{6-7}\\[-5pt]
$\Delta \half^-(1620)$ & CR & $-4.2^{+1.3}_{-1.8}$ & $$ 
& $4.2^{+1.8}_{-1.3}$& $-3.6^{+1.3}_{-2.5}$ & $-0.3^{+0.1}_{-0.2}$ & $$ 
& $3.6^{+2.5}_{-1.3}$\\
& KI & +8.0 & & 8.0 & -7.8 & +1.7 & & 8.0 \\
& SST & -3.3 & & 3.3 & +2.5 & -3.6 & & 4.4 \\
& BIL & 9.4 & & 9.4 \\
& OR & -5.1 & & 5.1 \\
& PWA &  $-9.7\pm 1.3$ & $$ & $9.7\pm 1.3$&
$+6.2\pm 0.9$ & $-2.4\pm 0.2$ & $$ & $6.6\pm 0.8$\\[+5pt]

&& $s$ & $d$ && $d_{\textstyle{1\over2}}$ &
 $s_{\textstyle{3\over2}}$ & $d_{\textstyle{3\over2}}$ &\\
\cline{3-4}\cline{6-8}\\[-5pt]
$\Delta \thalf^-(1700)$ & CR & $+15.4^{+0.9}_{-1.0}$ 
& $+5.0^{+2.4}_{-1.8}$ & $16.2^{+1.7}_{-1.5}$& $-1.2^{+0.6}_{-1.2}$ 
& $+3.4^{+2.2}_{-1.7}$ & $+0.5^{+0.5}_{-0.2}$ & $3.6^{+2.5}_{-1.8}$\\
& KI & -10.3 & -6.3 & 12.1 & -4.2 & -16.5 & -0.9 & 17.0 \\
& SST & 0.4 & 0.2 & 0.5 & & & & 4.9 \\
& BIL & 7.4 & 9.4 & 12.0 \\
& OR & +9.8 & +3.3 & 10.3 \\
& PWA &  $+21.1\pm 4.7$ & $+5.1\pm 2.2$ 
& $21.7\pm 4.6$& $$ & $+6.8\pm 2.3$ & $$ & $6.8\pm 2.3$\\[+5pt]

&& $p$ & && $p_{\textstyle{1\over2}}$ & $p_{\textstyle{3\over2}}$ & &\\
\cline{3-3}\cline{6-7}\\[-5pt]
$\Delta \half^+(1740)$$^{\rm a}$ & CR & $+14.1^{+0.7}_{-4.5}$ & $$ 
& $14.1^{+0.7}_{-4.5}$& $-6.5^{+4.6}_{-4.1}$ & $+4.7^{+3.1}_{-3.3}$ & $$ 
& $8.0^{+5.1}_{-5.7}$\\
& KI & +7.6 & & 7.6 & -2.2 & +7.6 & & 7.9 \\
& SST & -11.7 & & 11.7 & & & & 17.1 \\
& BIL & 2.0 & & 2.0 \\ 
& PWA &  $$ & $$ & $$& $-13.8\pm 1.9$ & $$ & $$ & $13.8\pm 1.9$\\[+5pt]

$\Delta \half^+(1910)$ & CR & $-8.4^{+0.2}_{-0.1}$ & $$ 
& $8.4^{+0.1}_{-0.2}$ & $+5.6^{+0.9}_{-0.4}$ & $+2.6^{+0.4}_{-0.2}$ & $$ 
& $6.1^{+1.0}_{-0.5}$\\
& KI & -5.9 & & 5.9 & +3.7 & +4.9 & & 6.1 \\
& SST & +5.3 & & 5.3 & & & & 6.9 \\
& OR & -3.8 & & 3.8 & \\
& PWA &  $$ & $$ & $$& $+4.9\pm 1.1$ & $$ & $$ & $4.9\pm 1.1$\\[+5pt]

&& $p$ & $f$ && $p_{\textstyle{1\over2}}$ &
 $p_{\textstyle{3\over2}}$ & $f_{\textstyle{3\over2}}$ &\\
\cline{3-4}\cline{6-8}\\[-5pt]

$\Delta \thalf^+(1600)$ & CR &  $+8.4^{+3.6}_{-3.5}$ & $0.0$ 
& $8.4^{+3.6}_{-3.5}$& $+0.4^{+0.7}_{-0.3}$ & $-0.9^{+0.6}_{-1.4}$ & $0.0$ 
& $1.0^{+1.6}_{-0.6}$\\
& KI & -8.6 & -0.1 & 8.6 & +1.3 & +5.5 & +0.4 & 5.7 \\
& SST & -10.2 & -1.2 & 10.3 & & & & 2.9 \\
& BIL & 5$\chi_1$ & 0.0 & 5$\chi_1$ \\
& OR & +18.3 & +0.0& 4.1, 18.3 \\
& PWA &  $+17.0\pm 1.6$ & $$ & $17.0\pm 1.6$\\[+5pt]

$\Delta \thalf^+(1920)$ & CR &  $-8.9^{+0.3}_{-0.2}$ 
& $+4.4^{+0.8}_{-0.7}$ & $10.0\pm 0.5$& $+5.3^{+1.3}_{-0.5}$ 
& $+6.6^{+1.6}_{-0.7}$ & $-0.7^{+0.2}_{-0.4}$ & $8.5^{+2.0}_{-0.8}$\\
& KI & +3.2 & +1.4 & 3.5 & +8.1 & -6.2 & -5.5 & 11.6\\
& SST & +1.9 & -1.3 & 2.3 & & & & 5.2 \\
& BIL & 3.9 & 3.7 & 5.4 \\
& PWA & $-11.2\pm 1.7$ & $$ & $11.2\pm 1.7$\\[+5pt]

$\Delta \thalf^+[1985]$ & CR & $-9.2^{+0.8}_{-0.6}$ & 
$-3.2^{+1.4}_{-2.2}$ & $9.7^{+1.4}_{-1.2}$& $-6.3^{+2.6}_{-0.5}$ 
& $+3.2^{+0.5}_{-1.4}$ & $-2.2^{+1.5}_{-1.6}$ & $7.4^{+1.2}_{-3.2}$\\
& KI & +0.5 & -7.7 & 7.7 & +5.5 & -1.8 & +1.3 & 5.9 \\
& SST & & & & & & & 9.1 \\[+5pt]

 && $p$ & $f$ && $f_{\textstyle{1\over2}}$ 
 & $p_{\textstyle{3\over2}}$ & $f_{\textstyle{3\over2}}$ &\\
\cline{3-4}\cline{6-8}\\[-5pt]
$\Delta \fhalf^+(1750)$$^{\rm b}$ \\
& PWA &  $+8.4\pm 3.6$ & $+11.0\pm 2.9$ 
& $13.9\pm 3.2$& $$ & $-7.4\pm 1.9$ & $$ & $7.4\pm 1.9$\\[+5pt]

$\Delta \fhalf^+(1905)$ & CR & $-1.5\pm 0.0$ & $+4.7\pm 0.6$ 
& $4.9^{+0.6}_{-0.5}$ & $-0.7\pm 0.2$ & $+6.3^{+0.8}_{-0.4}$ 
& $-0.7^{+0.1}_{-0.2}$ & $6.4^{+0.8}_{-0.4}$\\
& KI & -3.2 & -5.5 & 6.4 & +0.1 & +2.1 & +6.4 & 6.7 \\
& SST & +1.1 & +2.0 & 2.3 & & & & 5.1 \\
& BIL & 4.2 & 5.2 & 6.7 \\
& OR & -7.6 & +3.8 & 8.5 & +0.3 & -6.6 & +1.3 & 6.7 \\
& PWA &  $-2.0\pm 2.5$ & $+1.4\pm 1.4$ 
& $2.4\pm 2.2$& $$ & $+16.8\pm 1.3$ & $$ & $16.8\pm 1.3$\\[+5pt]

$\Delta \fhalf^+(2000)$ & CR & $-14.0^{+1.6}_{-0.1}$ 
& $+1.5^{+1.5}_{-0.8}$ & $14.1^{+0.4}_{-1.6}$& $+2.6^{+2.8}_{-2.1}$ 
& $+3.1\pm 1.2$ & $-3.1^{+2.4}_{-3.2}$ & $5.1^{+4.2}_{-3.0}$\\
& KI & +6.2 & -1.4 & 6.4 & -7.2 & -17.8 & -4.6 & 19.7 \\
& SST & +3.8 & +4.3 & 5.7 & & & & 8.9 \\
& PWA &  $$ & $$ & $$\\[+5pt]

& & $f$ & $h$ && $f_{\textstyle{1\over2}}$ &
 $f_{\textstyle{3\over2}}$ & $h_{\textstyle{3\over2}}$ &\\
\cline{3-4}\cline{6-8}\\[-5pt]
$\Delta \shalf^+(1950)$ & CR & $+4.8\pm 0.2$ & $0.0$ 
& $4.8\pm 0.2$& $+1.3\pm 0.1$ & $-2.3\pm 0.2$ & $0.0$ & $2.6\pm 0.2$\\
& KI & -5.5 & 0.0 & 5.5 & +4.7 & +8.2& 0.0 & 9.4 \\
& SST & +2.4 & +0.9 & 2.6 & & & & 4.5 \\
& BIL & 6.0 & & 6.0 \\ 
& OR & +5.7 & & 5.7 & +1.1 & +1.9 & 0.0 & 2.2 \\
& PWA & $+7.4\pm 0.7$ & $$ & $7.4\pm 0.7$& $$ & 
$+11.4\pm 0.5$ & $$ & $11.4\pm 0.5$\\[+5pt]
\hline
\end{tabular}
$^a$ First $P_{31}$ state found in Ref.~\cite{MANSA}.\\
$^b$ Ref.~\cite{MANSA} finds two $F_{35}$ states; this one and 
$\Delta(1905)F_{35}$.
\end{table}
%
As with the $N\pi$, $N\eta$ and $\Delta\eta$ channels, there are some
differences among the predictions for the $N\rho$ and $\Delta\pi$
channels obtained from the different models. As is the case with the
channels discussed earlier, these differences are probably mainly due
to the treatment of the final state mesons in the decays, as well as
due to the absence or presence of configuration mixing in the wave
functions of the parent and daughter baryons.

A few comments on some of the so-called missing states are relevant here. If 
the hypothesis
that the missing states are unobserved because they couple weakly to the $N\pi$ 
formation
channel is correct, then it may be possible to see some of these missing states 
in other
channels, provided that the formation channel is something other than $N\pi$. 
The results
shown in the last few tables illustrate that this may indeed be the case for 
some of the
predicted states. One example is the $N\half^+[1880]$ state. This state, 
according to the
models of CR and SST, should couple quite strongly to the $\Delta\pi$ channel. 
The states 
$N\thalf^+[1910]$ and $\Delta \thalf^+[1985]$, according to the models of CR, 
KI and SST, 
should couple strongly to either or both the $N\rho$ and $\Delta\pi$ channels. 
These states
should be observable in a process like $\gamma N\to N\pi\pi$, provided that the 
corresponding
photocouplings of the states are large. The $N\eta$, $\Delta\eta$ and 
$N\omega$ channels, the
latter of which has not been discussed in this article, also appear to be good 
candidates for
seeking missing resonances. Note that the weakly coupling state
$\Delta\fhalf^+(1750)$ reported by Manley and Saleski \cite{MANSA} is not 
easily accommodated in any of the models discussed here.

All of the results reviewed here have been for the lowest lying states in 
the baryon 
spectrum, states with masses less than 2 GeV. Most of these models predict many
states above 2 GeV, few of which have been observed.
However, note that not many of the models discuss the strong decays of 
these higher lying
states. Nevertheless, the status of models and their 
predictions can be illustrated by limiting the discussion to 
states below 2 GeV. Furthermore, all of
these models become less trustworthy as the masses of the states increase.

One channel that is of some interest, but which has not been treated in this 
article, is the 
$N(\pi\pi)_S$ channel, in which the two pions are in an isoscalar state.
Stassart \cite{Stassart:1992mu}
and OR treat this by the approximation of two-body decays to a
fictitious $\sigma$ meson. OR find
that, for all of the states they investigate, the couplings to this channel
are very small. Stassart, on the other hand, finds that two states, namely
$N\half^-(1650)$ and $N\thalf^-(1700)$, have sizeable couplings to this channel
($\ge$ 45 MeV).

None of the decays discussed so far involve the strange quark. A study
of the strange decays of non-strange baryons illustrates the
importance of SU(3) breaking effects. In the work of CR, the
probability for creating an $s\bar s$ pair is the same as for a light
pair, so that SU(3) breaking effects arise from the wave functions,
and from phase space. This is similar to the work of KI, but since the
latter do not calculate amplitudes, but fit them globally, there are
no SU(3) breaking effects in their amplitudes. Thus, for those
authors, the only SU(3) breaking effects arise from phase space
considerations. The results from the models of KI and CR, for the
$\Lambda K$ and $\Sigma K$ channels are shown in Table~\ref{lambdak}.

Not surprisingly, most of the amplitudes are predicted to be quite
small, due largely to the high threshold for these channels. Somewhat
surprisingly, the model of CR shows that, even for heavier states (not
shown), the vast majority of amplitudes for all of the final states
considered are quite small. This is to be contrasted with the
analogous decays to non-strange channels, where many of the amplitudes
become quite large. This difference can only be attributed to the
different mass of the strange quark.

\squeezetable

\begin{table}[t]
\caption{Results for the lowest-lying nucleon and $\Delta$ states in the 
$\Lambda K$ 
and $\Sigma K$ channels, for the models by CR \protect\cite{CR3} and KI \protect\cite{KI,KI0}. The 
results from the different models are on different rows. Also shown are the results 
from the partial wave analyses (PWA).}
\label{lambdak}
\begin{tabular}{llrr}
\multicolumn{1}{l} {state}
& \multicolumn{1}{l}{Model}
& \multicolumn{1}{r}{$\Lambda K$}
& \multicolumn{1}{r}{$\Sigma K$}\\
\hline
$N\half^-(1650)$ & CR & $-5.2 ^{+ 1.4}_{- 0.5}$ \\
& KI & -3.0 & $\simeq -2.0$ \\
& PWA & -3.3 $\pm $ 1.0 & $\simeq$ 2.7 $\pm $ 1.8 \\[+5pt]

$N\thalf^-(1520)$ & CR & 0.0 $^{+ 0.0}_{- 0.9}$  & \\
& PWA & 0.0 $\pm $ 0.0 \\[+5pt]

$N\thalf^-(1700)$ & CR & $-0.4 \pm $ 0.2 & 0.0 $^{+
  0.3}_{- 0.0}$  \\
& KI & -0.2 & -small\\
& PWA & $- 0.4 \pm $ 0.3 & $<$ 0.5 \\[+5pt]

$N\fhalf^-(1675)$& CR & 0.0 $\pm $ 0.0  \\
& KI & 0.1 & -small\\
& PWA & 0.4 $\pm $ 0.3 & $<$ 0.1 \\[+5pt]

$N\half^+(1710)$ & CR & $-2.8 \pm $ 0.6 & $1.1 ^{+0.9}_{-1.1}$  \\
& KI & -2.1 & 0.8 \\
& PWA & $+4.7 \pm $ 3.7 & $\simeq - 1.1 \pm $ 1.4 \\[+5pt]

$N\half^+[1880]$ & CR & $-0.1 \pm $ 0.1 & -3.7$^{+ 2.4}_{- 1.2}$  \\
& KI & -1.4 & -1.7 \\[+5pt]

$N\half^+[1975]$& CR & $-1.1 ^{+ 0.3}_{- 0.2}$ & -0.6
  $\pm $ 0.1  \\[+5pt]

$N\thalf^+(1720)$ & CR & $-4.3 ^{+ 0.8}_{- 0.7}$ & $0.3 \pm $
  0.3  \\
& KI & -1.7 & small \\
& PWA & $-3.2 \pm $ 1.8 & $\simeq$ 2.2 $\pm $ 1.1 \\[+5pt]

$N\thalf^+(1880)$ & CR & $-0.9 ^{+ 0.4}_{- 0.1}$ & -7.0$^{+4.9}_{-2.5}$ \\
& KI & & -3.3 \\[+5pt]

$N\thalf^+[1910]$ & CR & 0.0 $\pm $ 0.0 & -2.5 $^{+
  1.3}_{- 0.8}$  \\[+5pt]

$N\thalf^+[1950]$ & CR & $-1.9 ^{+ 0.5}_{- 0.2}$ & -1.4
  $^{+ 0.6}_{- 0.3}$  \\[+5pt]

$N\thalf^+[2030]$ & CR & $-0.9 \pm $ 0.2 & 0.0 $\pm $
  0.0  \\[+5pt]

$N\fhalf^+(1680)$ & CR & $-0.1 \pm $ 0.0 \\
& KI & -0.1 & -small \\
& PWA & $\simeq$ 0.1 $\pm $ 0.1 \\[+5pt]

$N\fhalf^+[1980]$ & CR & 0.0 $\pm $ 0.0 & -0.4 $\pm $
  0.3 \\
& KI & -3.2 \\[+5pt]

$N\fhalf^+(2000)$ & CR & $-0.5 \pm $ 0.3 & $0.6 ^{+
  0.6}_{- 0.4}$  \\
& KI & 0.9 & -0.7 \\
& PWA & & $\simeq$ 2.5 $\pm $ 2.2 \\[+5pt]

$N\shalf^+(1990)$ & CR & 0.0 $\pm $ 0.0  & -1.1 $^{+
  0.5}_{- 0.7}$  \\
& PWA & $\simeq$ 1.5 $\pm $ 2.4  & $\simeq$ 2.9 $\pm $ 2.2 \\[+5pt]

$\Delta \half^-(1620)$ & CR & & 0.1 $^{+
  0.6}_{- 0.1}$  \\
& KI & & -small \\[+5pt]

$\Delta \thalf^-(1700)$ & CR &  &$\simeq$ 0.2 $\pm $
   0.1  \\
& KI & & -small \\[+5pt]

$\Delta \half^+(1740)$ & CR &  & -2.9 $^{+ 2.9}_{- 1.4}$ \\
& KI & & -1.3 \\[+5pt]

$\Delta \half^+(1910)$ & CR & & -6.9 $^{+0.7}_{- 0.6}$  \\
& KI & & -3.4 \\
& PWA & & $<$ 1.0 \\[+5pt]

$\Delta \thalf^+(1600)$ & CR &  & 0.0 $^{+
  0.0}_{- 1.1}$ \\
& KI & & -1.9 \\
& PWA &  & $\simeq$ 1.1 $\pm $ 0.9 \\[+5pt]

$\Delta \thalf^+(1920)$ & CR & & -3.3 $\pm
  $ 0.3  \\
& KI & & -3.2 \\
& PWA &  & $\simeq$-2.2 $\pm $ 1.2\\[+5pt]

$\Delta \thalf^+[1985]$ & CR & & -3.2 $^{+ 0.9}_{- 0.3}$ \\[+5pt]

$\Delta \fhalf^+(1905)$ & CR & & -0.4 $\pm $
  0.1 \\
& KI & & $\pm 2\pm 1$ \\
& PWA &  & $\simeq - 0.9 \pm $
   0.3  \\[+5pt]

$\Delta \fhalf^+(2000)$ & CR & & -0.2 $^{+
  0.2}_{- 0.3}$  \\[+5pt]

$\Delta \shalf^+(1950)$ & CR &  & -1.2 $\pm
  $ 0.1 \\
& KI & & -1.9 \\
& PWA & & $\simeq$-1.5 $\pm $
   0.4 \\[+5pt]
\hline
\end{tabular}
\end{table}
%
There has been at least one attempt to modify the form of the $\tp0$
operator used in the description of the strong decays of baryons. CR
\cite{Capstick:1993th,CR2,CR3} replace Eq. (\ref{amp1}) with
\beqa \label{amp2}
T&=&-3\gamma\sum_{i,j} \int d \pbi\/ d \pbj\/ \delta( \pbi\/ +
\pbj\/) C_{ij} F_{ij}\nonumber e^{-\lambda^2 \left(\pbi-\pbj\right)^2/2}\\
&\times& \sum_m <1,m;1,-m|0,0> \chi_{ij}^m { \cal Y\/ }_1^{-m}( \pbi\/
- \pbj\/ )b_i^{\dagger}( \pbi\/ ) d_j^{\dagger}( \pbj\/ ),
\eeqa
but find that the best fit to the partial wave analyses is obtained
with $\lambda=0$.  Silvestre-Brac and Gignoux \cite{gignoux}, in
exploring the effects of hadronic loops on the masses of the $P$-wave
baryons, also use such a form, and find that the modification is
indispensable, with $\lambda\ne 0$.

Cano {\it et al.} \cite{canoetal} examine the effects of including higher order terms in
$p_i/E_i$ and $p_i/m_i$ expansions of both the elementary meson emission and 
$\tp0$ models. In their treatment, the meson emission operator is not the one
discussed in Eq. (\ref{emission}), but arises from the two-component reduction 
of the pseudovector quark current $\bar{q}\gamma_\mu\gamma_5q$. The 
two-component
operator is then treated in both $p_i/m_i$ and $p_i/E_i$ expansions. For the 
$\tp0$
operator, they use their results from the elementary meson emission amplitude 
to deduce the replacement
\begin{equation}
\begin{array}{rcl}
 &   {\cal Y}_{1}^{m} \left[ -\frac{4}{3} \veca{k} - \sqrt{\frac{2}{3}} \left(
\veca{p}_{\xi_{2}} + (\veca{p}_{\xi_{2}} + \sqrt{\frac{2}{3}} \veca{k})
\right) \right]  & \\ 
\rightarrow  &  \sqrt{\frac{m_{\pi}}{\omega_{\pi}}} 
{\cal Y}_{1}^{m} \left[ 
- \veca{k} \left( 1 + \frac{\omega_{\pi}}{6 E'}\right) 
 + \frac{\omega_{\pi}}{2}
\left( - \sqrt{\frac{2}{3}} \right) 
\left( \frac{\veca{p}_{\xi_{2}}}{E_i} +  
\frac{\veca{p}_{\xi_{2}} + \sqrt{\frac{2}{3}} \veca{k}}{E^\prime_i} \right) 
\right],
\end{array} 
\end{equation}
where $E_i$ and $E_i^\prime$ are the quark energies, and $\omega_\pi$ is the 
energy of the emitted pion.

The results that they obtain in a number of scenarios are shown in
Table~\ref{tablecano}. In this table, the first column identifies the
particular decay. Note that for the $\Delta\pi$ channel, the width of
the $\Delta$ is ignored. Columns two to five result from considering
only two-body forces in the Hamiltonian leading to the baryon
spectrum, while columns six to nine show the effects of including
three-body forces. Column ten contains the partial widths listed in
\cite{PDG}. Columns two and six result from the elementary meson
emission model, at lowest order in the $p_i/m_i$ expansion. These
numbers are therefore closest in spirit to the model results of BIL,
KI and SAS. Columns three and seven also arise from the elementary
meson emission model, but with all orders in the $p_i/E_i$ expansion
included. Columns four and eight show the results obtained using the
$\tp0$ model ({\it cf.} OR, SST and CR), while columns five and nine
arise from the modified $\tp0$ model.

The numbers shown in this table illustrate three effects very clearly.
One of these is the role of the pion size, as hinted at in the
comparison of the models of SAS, BIL and KI with those of SST and
CR. Here, the effect is clear, as the wave functions are the same for
the different calculations. The second effect is that of terms in the
potential, specifically the three-body force in this case.  The effect
of the relativistic modification mentioned above is also clear. One
aspect that Cano {\it et al.} do not explore is the effect of these
modifications on the ratios of partial widths, such as $S/D$ or $P/F$
ratios, in the $\Delta\pi$ channel. Such ratios are expected to be
quite sensitive to the details in the potential, and in the decay
operator~\cite{barnes}.

\begin{table}
\caption{$N\pi$ and $\Delta\pi$ decay widths from the model by Cano
{\it et al.} \protect\cite{canoetal}. The first column identifies the particular
decay. Columns two to five result from considering only two-body forces in the
Hamiltonian leading to the baryon spectrum, while columns six to nine show the
effects of including three-body forces. Column ten contains the partial widths
listed in \protect\cite{PDG}. Columns two and six result from the elementary 
meson emission model, at lowest order in the $P/m$ expansion. Columns
three and seven also arise from the elementary meson emission model, but with 
all
orders in the $P/E$ expansion included. Columns four and eight show the results
obtained using the $\tp0$ model, while columns five
and nine arise from the modified $\tp0$ model.}
\begin{center}
\tabcolsep 3pt
\begin{tabular}{lrrrrrrrrr}
\multicolumn{1}{l} {Decay}
& \multicolumn{4}{c}{Without three-body forces}
& \multicolumn{4}{c}{With three-body forces}
& \multicolumn{1}{r}{PWA}\\
\cline{2-5}\cline{6-9}\\[-5pt]
& $(p/m)$ & \multicolumn{1}{c}{\scriptsize All} & $^{3}P_{0}$ & R$^{3}P_{0}$
& $(p/m)$ & \multicolumn{1}{c}{\scriptsize All} & $^{3}P_{0}$ & R$^{3}P_{0}$ & \\
\hline
\rule{0pt}{4.5ex}
$\Delta(1232) \rightarrow N \pi$ & 
 79.6 & 75.8 & 167 & 83.9 & 72.1 & 75.7 & 210 & 106 & 115--125 \\ [2.ex]
\rule{0pt}{2.5ex}
$N(1440) \rightarrow N \pi$ & 
 3.4 & 13.1 & 452 & 73.5 & 0.2 & 15.3 & 1076 & 236 & 210--245 \\ [2.ex]
\rule{0pt}{2.5ex}
$N(1440) \rightarrow \Delta \pi$ & 
 7.1 & 9.3 & 66.5 & 20.7 & 17.6 & 30.9 & 228 & 106  & 70--105 \\ [2.ex]
\rule{0pt}{2.5ex}
$\Delta(1600) \rightarrow N \pi$ & 
20.1 & 8.0 & 19.8 & 0.3 & 94.1 & 50.4 & 0.5 & 8.8 & 35--88 \\ [2.ex]
\rule{0pt}{2.5ex}
$\Delta(1600) \rightarrow \Delta \pi$ & 
2.9 & 9.3 & 255 & 41.9 & 0.1 & 5.9 & 498 & 93.5  & 140--245 \\ [2.ex]
\rule{0pt}{2.5ex}
$N(1520) \rightarrow N \pi$ & 
61.8 & 57.7 & 268 & 92.0 & 22.3 & 31.5 & 319 & 100 & 60--72  \\ [2.ex]
\rule{0pt}{2.5ex}
$N(1520) \rightarrow \Delta \pi$ & 
78.0 & 45.1 & 532 & 17.1 & 56.1 & 44.8 & 999 & 34.4 & 18--30 \\ [2.ex]
\rule{0pt}{2.5ex}
$N(1535) \rightarrow N \pi$ & 
240 & 101 & 429 & 0.2 & 149 & 82.8 & 464 & 0.5 & 53--83 \\ [2.ex]
\rule{0pt}{2.5ex}
$N(1535) \rightarrow \Delta \pi$ & 
9.7 & 10.3 & 28.1 & 15.7 & 8.3 & 12.6 & 74.0 & 37.7 & $< $1.5 \\ [2.ex]
\rule{0pt}{2.5ex}
$N(1650) \rightarrow N \pi$ & 
47.9 & 20.9 & 49.1 & 1.1 & 24.9 & 15.8 & 44.2 & 0.8 & 90--120 \\ [2.ex]
\rule{0pt}{2.5ex}
$N(1650) \rightarrow \Delta \pi$ & 
12.4 & 12.8 & 49.3 & 20.9 & 10.0 & 13.6 & 109 & 43.1 & 4--11 \\ [2.ex]
\rule{0pt}{2.5ex}
$N(1700) \rightarrow N \pi$ & 
4.1 & 3.2 & 11.5 & 3.2 & 1.4 & 1.5 & 11.2 & 2.8 & 5--15 \\ [2.ex]
\rule{0pt}{2.5ex}
$N(1700) \rightarrow \Delta \pi$ & 
383 & 222 & 1643 & 114 & 220 & 185 & 2417 & 217 & 81--393 \\ [2.ex]
\hline
\end{tabular}
\vskip 2cm
\end{center}
\label{tablecano}
\end{table}

\nopagebreak
\vskip -12pt
\section{Effects of strong couplings on the spectrum}
\vskip -12pt
In addition to their decay widths, the strong couplings of the 
baryons are expected to have an impact on the baryonic mass spectrum. These
couplings can provide
`self energy' contributions to the masses of the baryons, as illustrated in 
figure \ref{self-energy}. Furthermore, contributions to the
mixing between baryons with the same quantum numbers can arise from these
couplings, as illustrated in figure \ref{mixing}.

\vspace*{-1.0in}

\unitlength 0.04cm

\begin{figure}
\begin{center}
\begin{picture}(200,150)
\Line(0,50)(200,50)1
\DashCArc(100,50)(40,0,180)41
\put(10,30){$B$}
\put(90,30){$B^\prime$}
\put(150,30){$B$}
\end{picture}
\end{center}
\vspace*{0.5in}
\caption{The contribution to the self energy of a baryon from a meson
loop. The dashed arc represents a meson.\label{self-energy}}
\end{figure}
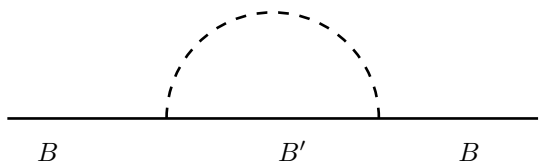

\vspace*{-1.0in}

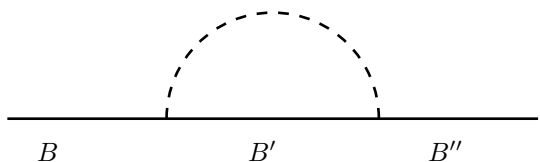
\begin{figure}
\begin{center}
\begin{picture}(200,150)
\Line(0,50)(200,50)1
\DashCArc(100,50)(40,0,180)41
\put(10,30){$B$}
\put(80,30){$B^\prime$}
\put(140,30){$B^{\prime\prime}$}
\end{picture}
\end{center}
\vspace*{-0.5in}
\caption{The contribution to the mixings of baryons with the same
quantum numbers from a meson loop. The dashed arc represents a meson.\label{mixing}}
\end{figure}

Most of the work that discusses the effects of the decay channels, real or
virtual, on the spectrum of baryons do so for a few baryon states, usually the
ground state baryons, and perhaps a few of the lower-lying baryonic excitations.
None of the authors cited in the sub-sections on strong decays examine
this problem. A number of other authors variously examine the
contributions of the loop effects to the masses, widths, mixing angles and
electromagnetic couplings of some of these states.

One of the earliest attempts to incorporate the effects of the strong couplings
of baryons into their mass spectrum is made by Faiman \cite{faiman}. He uses
arguments based
on those originated by Katz and Lipkin \cite{lipkinkatz} for $\omega-\phi$ 
mixing, to examine the
mixings and mass shifts generated by the couplings of baryons to some of their
real or virtual decay channels. In their model, these mixings are generated in
the absence of any other intra-multiplet mixing and splitting scenarios. States
with the same isospin, strangeness, total angular momentum and parity start off
being degenerate. 

Katz and Lipkin argue that the approximate decoupling of the
$\phi$ from the $\rho\pi$ channel is a natural consequence of $\rho\pi$ being
the dominant intermediate state under the hypothesis that two-meson (PP or PV,
where P represents a pseudoscalar, and V a vector) exchange is the mechanism
responsible for mixing. In a similar vein, Faiman argues that the two hadron
channels such as $N\pi$ are the dominant intermediate states under the
hypothesis that baryon-meson exchange is the dominant mechanism responsible for
mixing among baryons. He notes that among the known $S_{11}$ resonances, the
state at 1535 MeV does not couple strongly to $N\pi$, while the state at 1700 
MeV
does, and among the $D_{13}$ resonances, the state at 1520 MeV couples strongly
to $N\pi$, while the state at 1700 MeV does not. He generalizes this argument 
to
examine the splittings among a number of pairs and triplets of states, both
strange and non-strange.
spin-independent, but momentum dependent
contributions such as the Darwin-term and the orbit-orbit interaction
seems to be necessary
One interesting aspect of the work by Faiman is that he uses his
mixing scenario to predict the existence of states that should exist, but which
should decouple from formation channels. Apart from the $S_{11}$ and $D_{13}$
resonances mentioned above, he notes that there should be a $D_{03}$ state at
1750 MeV that decouples from the $N\bar K$ formation channel. This prediction 
is in
striking agreement with subsequent quark model predictions that place a
state with the appropriate quantum numbers at 1770 MeV. He also notes that 
among
the $S_{01}$ states, the $\Lambda$(1405) is either {\it not} a normal
three-quark state, or if it is, there {\it should} be another $S_{01}$
intermediate in mass between the $\Lambda(1405)$ and the $\Lambda(1670)$. 

Nogami and Ohtsuka \cite{nogamiohtsuka} incorporate pion effects into their 
quark model by examining
the effects of pion loops. They include these loops by using elementary pion
emission, and look at the effects on the masses of a few states, as well as
the magnetic moments of the ground state baryons. In a separate article, in the 
framework of a
similar model, Horacsek, Iwamura and Nogami \cite{horacsek} also examine 
baryon self energies.

A number of other authors, including Brack and Bhaduri \cite{brackbhaduri}, 
and Guiasu and Koniuk \cite{koniukguiasu},
examine the effects of pion loops on quantities such as magnetic moments
and self energies. Both of these models treat the pion as an elementary
particle. Other authors who consider the effects of meson loops include
Zenczykowski \cite{zenczy1} and collaborators \cite{zenczy2,zenczy3}, Silvestre-Brac 
and Gignoux \cite{gignoux}, Blask, Huber and Metsch \cite{blask1}, and Fujiwara \cite{fujiwara3}.

In many of these articles, the crucial point is that the propagator for a
baryon is written as 
\begin{equation} \label{looppropagator}
G(E)=\frac{1}{E-M_R-\Sigma_R(E)},
\end{equation}
where $M_R$ is the `bare' mass of the resonance, and $\Sigma_R(E)$ is the
self-energy contribution from the hadron loop. This usually takes the form
\begin{equation}
\Sigma_R(E)=\sum_B\int d^3p\frac{V^\dag_{\pi RN}(p)V_{\pi RN}(p)}
{E-E_R(p)-E_\pi(p)},
\end{equation}
where $E_\pi(p)=\sqrt{p^2+m_\pi^2}$, $E_R(p)=\sqrt{p^2+M_R^2}$, and 
$V_{\pi RN}$
is the energy-dependent strong interaction vertex function for pion emission
from the resonance $R$, resulting in the ground-state N. The form written above
arises when the loop corrections are treated in first-order time-ordered
perturbation theory.

One of the more comprehensive studies of these meson loop effects is carried
out by Zenczykowski \cite{zenczy1}. His work is essentially the baryonic 
extension of the
unitarised quark model (UQM) \cite{uqm1,uqm2,uqm3,uqm4} that was developed to treat mesons.
 In the
UQM, the observed splittings and mixings among the mesons are thought to be
dominantly due to the influence of the decay channels. In this framework, the
mass shifts are written in terms of a dispersion integral
\begin{equation}\label{dispersion}
m_A^2-\left(m_A^0\right)^2=\sum_iw_i^A\int_{S_{\rm
thr}^{i}}\frac{\rho(s,m_B,m_M)}{m_A^2-s}ds,
\end{equation}
where $m_A^0$ is the bare baryon mass, $m_A$ its physical mass, and
$\rho(s,m_B,m_M)$ is the spectral function, assumed to have the form
\begin{equation}
\rho\left(s, m_B,m_M\right)=\rho\left(\sqrt{s}-(m_B+m_M)\right).
\end{equation}
Here, $m_B$ and $m_M$ are the masses of the baryon and meson, respectively, that
comprise the channel $i$, and the sum over $i$ is, in principle, the sum over
all possible baryon-meson intermediate channels, open and closed. The weights 
$w_i^A$ are numerical constants that arise from the symmetries of the
particular set of decay channels being considered.

The further assumptions of this model are
\begin{enumerate}
\item SU(3) breaking in the bare hadron spectrum occurs through quark mass
terms, mesons are mixed ideally, and one-gluon exchange is ignored;

\item Flavor-spin(-space) symmetry relations exist for the decay vertex 
functions;

\item Phenomenological expressions for the spectral functions which are chosen
to describe the decay widths correctly;

\item unitarity and analyticity.

\end{enumerate}

In a simplified scenario explored by Zenczykowski, a linearized version of the
unitarised model yields a number of equations that relate the mass
splittings among the baryons to those among the mesons. Furthermore, the
Gell-Mann-Okubo mass relation for baryons is also recovered. In obtaining
these results, he uses all possible two-body intermediate states that can be 
composed from the ground state octet and decuplet of baryons, taken with the 
ground state vector and pseudoscalar nonets of mesons.

In a `more realistic' model, Zenczykowski uses the amplitudes generated in the
$\tp0$ model, or in the meson-emission model of Koniuk and Isgur \cite{KI}, as 
the
spectral functions in Eq. (\ref{dispersion}). In this scenario, he finds that
the contribution to the $\Delta N$ splitting, for instance, that arises from
meson loop effects, is quite large, accounting for 60 - 90 \% of the measured
splitting. He finds similar results in other channels, and concludes that, in
the spectroscopic model, the effects of the one-gluon exchange part of the
potential should be much weaker, with $\alpha_s$ as small as 0.2 or 0.3. He 
goes
on to examine mixing effects in the $P$-wave baryons, and there he finds that
the loop effects are again quite large, with typical mass shifts being of the
order of 700 MeV.

Silvestre-Brac and Gignoux \cite{gignoux} investigate loop effects in the
$P$-wave baryons by considering the equations
\begin{eqnarray} \label{gignoux}
E_i&=&E_i^0+\Sigma_i(E_i), \nonumber\\
\Sigma_i(E_i)&=&\lim_{\epsilon\to 0} \sum_c\int_0^\infty \frac{k^2dk
\left|V_{ic}(k)\right|^2}{E-E_c(k)+i\epsilon},
\end{eqnarray}
where $V_{ic}$ is the interaction that couples the baryon $i$ to the decay
channel $c$. They use a modified version of the
$\tp0$ model for the $V_{ic}$. These authors also find that the mass shifts
associated with these unitary or loop effects are quite large, again of the
order of several hundreds of MeV.

At this point it is appropriate to comment on the recent work of Geiger and
Isgur on $\rho-\omega$ mixing, and on the more general question of loop effects
in hadron spectra \cite{geigerisgur1,geigerisgur2,geigerisgur3,geigerisgur4,geigerisgur5}. These authors find that when a few 
channels
are included in their calculation, the contributions to the $\rho-\omega$ mass
difference are typically large, of the order of a few hundreds of MeV. When the
`infinite' towers of possible intermediate states are included, the
contributions from different sectors essentially cancel, leaving only a small
net $\rho-\omega$ mass difference. In the closure approximation, all of the
contributions cancel exactly. Furthermore, they show that this kind of 
cancellation also occurs with mesons with other quantum numbers, with the 
exception of scalar mesons.

In a similar vein, Pichowsky and collaborators \cite{Pichowsky:1999mu} show 
that, 
in a covariant model based on the
Schwinger-Dyson equation of QCD which assumes exact SU(3)-flavor
symmetry, the contributions to the self energies of the $\rho$ and
$\omega$ due to several pseudoscalar-pseudoscalar and
pseudoscalar-vector meson loops are at most $10\%$ of the bare
mass. The result for the mass shift of the 
$\rho$ meson due to the two-pion loop agrees with a previous
calculation of this quantity using an effective chiral Lagrangian
approach~\cite{Leinweber:1994yw}. They find that such contributions decrease
rapidly as the mass of the intermediate mesons increases beyond $m_{\rho}/2$. 
They compare the mass shifts due to several two-meson intermediate states 
with those from Geiger and Isgur's more extensive nonrelativistic
study of the $\rho$-$\omega$ mass splitting, and find these to be
smaller, especially for intermediate states involving higher-mass
mesons. A net mass splitting of $m_{\omega} - m_{\rho}\approx 25$~MeV is
found from the $\pi\pi$, $K\bar{K}$, $\omega\pi$, $\rho\pi$,
$\omega\eta$, $\rho\eta$ and $K^* K$ channels. These results suggest that a 
more complete calculation should exhibit rapid convergence as the number of
two-meson intermediate states is increased by including states with
higher masses. This implies that inclusion of two-meson loops
into the vector-meson self-energy yields a small correction to
the predominant valence quark-antiquark structure of the vector meson.

Isgur \cite{isguradiabatic} uses the results of his work with Geiger to suggest 
that the net effect of
considering these meson loops is to renormalize the string tension between the
quarks, provided that a full complement of states is included in the loops.
Thus, if the `physical' string tension is used from the
outset, there is, in essence, no need to include {\it any} meson loop 
corrections, as
these have already been taken into account.
Perhaps a similar mechanism is realized in the baryon sector. What the study of
Geiger and Isgur does not address are the possible effects on the widths of the
states, as well as the mixing that might arise from these effects. Their 
results
for the masses lead one to speculate that similar cancellations might occur 
when
the effects on the mixings and widths are considered.

If the results of Zenczykowski, or of any similar calculation, are
taken at face value, then there is an outstanding puzzle regarding the
source of hyperfine and spin-orbit interactions within baryons, as
well as that of the mixing between baryons. The possible sources that
can give rise to these splittings and mixings in the literature are
one-gluon exchange, instanton induced interactions, meson exchange,
and meson loops. This is a question that can only be answered by
confronting the models with more precise experimental data.
\vskip -12pt
\section{Summary}
\vskip -12pt
It is not clear what conclusion to draw from arguments surrounding
the nature of the hyperfine interaction in baryons. The OGE-based
model has significantly less freedom to fit the spectrum. The fit
appears reasonable, but splittings between states with wave functions
predominantly in a given harmonic-oscillator band are predicted better
than the corresponding band centers of mass (with about a 50 MeV
error). The GBE-based model in either form considered here has
significantly more freedom to fit the spectrum and is able to place
certain radial excitations, notably the states corresponding to the
Roper resonance $N\half^+(1440)$ and $\Lambda\half^+(1600)$, below
most of the $P$-wave baryons of the same flavor. While it is perhaps
natural to extend the GBE model to include effective vector and
scalar-meson exchanges because of multiple-pion exchange, the special
nature of the pion and to a lesser degree the other pseudoscalar
mesons due to chiral symmetry does not extend to vector and scalar
mesons. It is entirely possible to generate effects similar to those
due to the exchange of gluons by the exchange of a nonet of vector
mesons. The GBE model does not (as shown below) solve the puzzle of
the small size of the spin-orbit interactions in baryons. Also, it
appears that a description of strong decays in this model requires the
use of a model with a final-state pion with structure, which is
somewhat inconsistent given the basis of the model.

In practical terms, neither of these models of the baryon spectrum is
relativistic and neither is QCD, and although some aspects of both
pictures are doubtless reflected in the masses and properties of
excited baryons, one could argue that OGE is the simplest and most
economical model which describes the spectrum. It is also the case
that the OGE-based model has been applied with success to the
description of a wider range of strong couplings than the GBE
model. Although the exchange current effects on magnetic moments are
considered in the GBE model, baryon resonance photocouplings are not,
and the successful description of these was and continues to be an
advantage of the OGE-based models. The apparently clearest evidence
for GBE is the mass of the Roper resonance $N\half^+(1440)$. It is
likely that couplings to decay channels (the equivalent of
$qqq(\bar{q}q)$ Fock-space components in the wave functions) shift
many baryon masses, as one could naively expect shifts of the order of
the width of the state involved. An example is the
$\Lambda\half^-(1405)$, which is predicted degenerate with the model
state assigned to $\Lambda\thalf^-(1520)$ using both the OGE and GBE
models. The Roper resonance is special~\cite{Krehl:1999km} because it
has a very strong coupling to the $\pi N$ channel (with a width of
150-550 MeV and a 60-70\% $\pi N$ branch), and the extraction of its
mass is complicated by the onset of the $\pi\pi N$ channel. Until the
effects of such decay-channel couplings are comprehensively calculated
(existing calculations are referred to below), a 100 MeV error in the
mass of the Roper resonance, or any baryon, should not carry that much
significance.

Operator analysis of the P-wave nonstrange baryons masses in large
$N_c$ QCD leads to the suggestion that the underlying dynamics in this
part of the spectrum may be related to pion exchange and not gluon
exchange between the quarks, although the minimal set of operators
leading to this conclusion do not give the best fit to the mixing
angles in the $N\half^-$ and $N\thalf^-$ sectors. This result is
interesting; however, it may not be possible to draw firm conclusions
about interquark dynamics until this calculation can be extended to
the entire spectrum. The mass splittings in this restricted sector are
relatively small, so that reasonable uncertainties in the masses
of the states which are fit could mean substantial uncertainties in
the identity of the operators responsible for the
splittings. Nevertheless, this is an interesting new development which
deserves further attention.

Recent flux-tube model calculations of hybrid baryon masses find the
lightest hybrid baryon states to be $N\half^+$ and $N\thalf^+$ states,
with masses in the 1800-1950 region, which coincides with the mass
range of the missing $P_{11}$ and $P_{13}$ resonances predicted by
most models. Calculations of these masses in the bag model agree on
the flavor, total spin and parity of the lightest states, although the
predicted masses are significantly lighter at about 1500 MeV. This is
due to the different description of the gluonic degrees of freedom in
these approaches. Mixing between conventional excitations and hybrid
states can be expected to complicate the physics of the lighter
$N\half^+$ and $N\thalf^+$ baryons. Evidence for these new kinds of
excitations could take the form of a surfeit of states found in
multichannel analyses of the scattering data in the mass range of
1700-2000 MeV in the $P_{11}$ and $P_{13}$ partial waves, or the
identification of one or more states with anomalous strong and
electromagnetic decay signatures. As outlined above, the Roper
resonance appears to have anomalous photocouplings, but this may be
due to pionic dressing of the transition vertex, given the large
$N\pi$ decay width of this state.

The collective-model calculation of Bijker, Iachello and Leviatan
(BIL) demonstrates that it is possible to make a reasonable fit to the
spectrum using a spectrum-generating algebra which includes the
spatial degrees of freedom. It also predicts the existence of many
missing baryons, including a pair of light positive-parity nucleon
states at 1720 MeV. Some of the missing nonstrange states have
substantial couplings to the $\pi N$ formation channel (see
Table~XVIII in Ref.~\cite{Bijker:2000gq}) and so perhaps should have
already been seen. The relation between this model and the properties
of low-energy QCD is less clear than it is for other models. Given the
essential character of the model, which is to write down the most
general mass-squared operator consistent with the symmetries of the
system, it is necessary to make several choices in order to have a
model which can be compared with experiment. Although these choices
are loosely based on the assumption of a string picture of the spatial
structure of the baryon, it is not clear, for example, why the
spectrum of an oblate top is related to that of baryons. It is natural
that these choices are made with the spectrum in mind. The spin-orbit
problem is avoided by simply leaving out such interactions, and the
mixings caused by tensor interactions which are required to fit
certain photocoupling and strong decay amplitudes are not
included. The calculation of photocouplings in this model does not
appear to improve on that of the OGE-based models.

It is obvious that the spin-orbit puzzle in baryons is still a matter
of active research, and it seems equally obvious that a solution will
ultimately require the description of baryons in a relativistic
framework. For example, if the Lorentz structure of the confining
interaction between quarks is not scalar~\cite{Parramore:1995fu}, then
the cancellation described above will not hold. Recent progress in
describing baryons in a model based on the Schwinger-Dyson
Bethe-Salpeter approach~\cite{Oettel:1999gc} and in unquenched lattice
gauge calculations~\cite{Aoki:2000yr,Allton:1999gi} show promise, but
until these calculations can be extended to the $P$-wave baryons, the
nature of spin-orbit interactions is likely to remain a puzzle. Data
pertaining to the structure of $P$-wave baryons, especially the
electromagnetic structure of $\Lambda(1405)$ and $\Lambda(1520)$,
would be very useful.

There are many aspects of baryon spectroscopy that have not been
addressed in this article, for lack of space. Strong coupling
constants, weak decay form factors, strong form factors, are three key
topics that have not been discussed. Furthermore, even within the
topics that have been discussed, the work of many authors has been
omitted from the discussion because of lack of space.

One very striking effect has emerged in the discussion presented
above. All of the models described are capable of reproducing the mass
spectrum of the baryons reasonably well, often with emphasis on
different aspects. However, in many cases, the predictions for the
strong and electromagnetic decay amplitudes are quite diverse, and
these differences can be traced back to a number of sources. The
kinetic energy operator in the Hamiltonian, the absence or presence of
a tensor interaction, as well as of three-body forces, and the size of
the pion, have all significantly modified the predictions for the
decay amplitudes. It is clear that any model that purports to describe
hadron spectroscopy must reproduce not only the mass spectrum of the
states, but also other quantities such as the decay
amplitudes. Detailed consequences of the wave functions are much more
sensitive to the nature of the Hamiltonian than the spectrum.

It must be emphasized that none of the models described do what can be
termed an excellent job of describing what is known about baryon strong
decays. The main features seem to be well described, but many of the
details are simply incorrect, and this is illustrated by the fact
that, in the $N\pi$ channel alone, the {\it best} value of the average
$\chi^2$ per decay amplitude is larger than two.  Much work still
needs to be done to understand the baryon spectrum, and it is clear
that calculations of the spectrum and of the strong and
electromagnetic decays must be done simultaneously. In particular, a
systematic study of the effects of strong decay channels on the masses
and widths of all baryon states must be carried out before many important
issues in baryon spectroscopy can be resolved.
\vskip -12pt
\section{Acknowledgements}
\vskip -12pt 
The authors are grateful to D.~Leinweber for his help with
understanding recent lattice QCD and QCD sum rules calculations, and
to L.~Ya~Glozman, W.~Plessas and R.F.~Wagenbrunn for supplying
information relating to the GBE models. WR acknowledges the support of the
National Science Foundation.
This work was supported by the U.S.~Department of Energy under
Contract DE-FG02-86ER40273 (SC), and by the U.S.~Department of Energy
under Contract No.\ DE-AC05-84ER40150 and Contract No.\
DE-FG02-97ER41028 (WR).
\vskip -12pt
\section{Appendix}
\vskip -12pt
\subsection{N=2 band symmetrized states}
\vskip -12pt
Here the construction of symmetrized basis wave functions to represent
the low-lying positive-parity excited states of the non-strange
baryons is detailed~\cite{Isgur:1979wd}. First it is necessary to
construct symmetrized spatial basis wave functions at the $N=2$
level. For example, there are three $L^P=0^+$ states, two made from
$l_\rho=0\otimes l_\lambda=0$ with one radial node in either of the
two oscillators,
\beqa
\psi^{S^\prime}_{00}&=&\sqrt{2\over 3}
{\alpha^5\over \pi^{3\over 2}} {1\over \sqrt{2}}
\left(\rho^2-{3\over 2\alpha^2}+\lambda^2-{3\over 2\alpha^2}\right)
e^{-{\alpha^2}(\rho^2+\lambda^2)/2} \nonumber\\
\psi^{M^\lambda}_{00}&=&\sqrt{2\over 3}
{\alpha^5\over \pi^{3\over 2}} {1\over \sqrt{2}}
\left(\rho^2-\lambda^2\right)
e^{-{\alpha^2}(\rho^2+\lambda^2)/2},
\eeqa
[where $\thalf-(\alpha \rho)^2$ is the Laguerre polynomial
$L_n^{l+{1\over 2}}=L_0^{1\over 2}$] and one from $l_\rho=1\otimes
l_\lambda=1$ with no radial nodes,
\beq
\psi^{M^\rho}_{00}=-{2\over \sqrt{3}}
{\alpha^5\over \pi^{3\over 2}}\lpmb{\rho}\cdot\lpmb{\lambda}
e^{-{\alpha^2}(\rho^2+\lambda^2)/2}.
\eeq
The radial excitation $\psi^{S^\prime}$ is $S$ under $S_3$ because it
has the structure
$\lpmb{\rho}\cdot\lpmb{\rho}+\lpmb{\lambda}\cdot\lpmb{\lambda}$.
It is straightforward to see that the $\psi^{M^\lambda}_{00}$ and 
$\psi^{M^\rho}_{00}$ wave functions also have the advertised $S_3$
symmetry.

It is also possible to construct an $L^P=1^+$ state from $l_\rho=1\otimes
l_\lambda=1$,
\beq
\psi^A_{11}=-
{\alpha^5\over \pi^{3\over 2}}\left(\rho_+\lambda_0-\rho_0\lambda_+\right)
e^{-{\alpha^2}(\rho^2+\lambda^2)/2},
\eeq
where only the top state with $|L,M\rangle=|1,1\rangle$ is written
down. There are also the $L^P=2^+$ states constructed from
$l_\rho=0\otimes l_\lambda=2$, $l_\rho=2\otimes l_\lambda=0$, and
$l_\rho=1\otimes l_\lambda=1$,
\beqa
\psi^S_{22}&=&{1\over \sqrt{2}}
{\alpha^5\over \pi^{3\over 2}}\left(\rho_+^2+\lambda_+^2\right)
e^{-{\alpha^2}(\rho^2+\lambda^2)/2} \nonumber\\
\psi^{M^\lambda}_{22}&=&{1\over \sqrt{2}}
{\alpha^5\over \pi^{3\over 2}}\left(\rho_+^2-\lambda_+^2\right)
e^{-{\alpha^2}(\rho^2+\lambda^2)/2} \nonumber\\
\psi^{M^\rho}_{22}&=&{\alpha^5\over \pi^{3\over 2}}
\rho_+\lambda_+ 
e^{-{\alpha^2}(\rho^2+\lambda^2)/2},
\eeqa
where once again only the top states are shown.

It is then straightforward but rather tedious to find the fully
symmetrized wave functions (which therefore represent nonstrange
states) which occur at the $N=2$ level in the oscillator. Note that,
from Eq.~(\ref{NDphi}), there are only symmetric and mixed-symmetry
flavor wave functions in the nonstrange case. These are
\beqa
[56^\prime,0^+]:&&\vert N^2S_{S^\prime}\half^+\rangle=C_A\psi^{S^\prime}
{1\over \sqrt{2}}(\phi^\rho_N\chi^\rho_\half+
\phi^\lambda_N\chi^\lambda_\half)\nonumber\\ 
&&\vert \Delta^4S_{S^\prime}\thalf^+\rangle=
C_A\phi^S_\Delta\psi^{S^\prime}\chi^S_\thalf\nonumber\\[0pt] 
%
[70,0^+]:&&\vert N^4S_M\thalf^+\rangle=C_A\chi^S_\thalf
{1\over \sqrt{2}}(\phi^\rho_N\psi^\rho_{00}+
\phi^\lambda_N\psi^\lambda_{00})\nonumber\\
&&\vert \Delta^2S_M\half^+\rangle=
C_A\phi^S_\Delta(\psi^\rho_{00}\chi^\rho_\half+
\psi^\lambda_{00}\chi^\lambda_\half)\nonumber\\
&&\vert N^2S_M \half^+\rangle=C_A {1\over 2}
\left\{\phi^\rho_N[\psi^\rho_{00} \chi^\lambda_\half 
+ \psi^\lambda_{00} \chi^\rho_\half] 
+\phi^\lambda_N[\psi^\rho_{00} \chi^\rho_\half 
- \psi^\lambda_{00} \chi^\lambda_\half]\right\}\nonumber\\[0pt]
[56,2^+]:&&\vert \Delta^4D_S(\half^+,\thalf^+,\fhalf^+,\shalf^+)\rangle=
C_A\phi^S_\Delta\psi^S_{2M}\chi^S_\thalf,\nonumber\\
&&\vert N^2D_S(\thalf^+,\fhalf^+)\rangle=C_A\psi^S_{2M}
{1\over \sqrt{2}}(\phi^\rho_N\chi^\rho_\half+
\phi^\lambda_N\chi^\lambda_\half)\nonumber\\[0pt]
[70,2^+]:&&\vert N^4D_M(\half^+,\thalf^+,\fhalf^+,\shalf^+)\rangle=
C_A\chi^S_\thalf{1\over \sqrt{2}}(\phi^\rho_N\psi^\rho_{2M}+
\phi^\lambda_N\psi^\lambda_{2M})\nonumber\\
&&\vert \Delta^2D_M(\thalf^+,\fhalf^+)\rangle=
C_A\phi^S_\Delta(\psi^\rho_{2M}\chi^\rho_\half+
\psi^\lambda_{2M}\chi^\lambda_\half)\nonumber\\
&&\vert N^2D_M(\thalf^+,\fhalf^+)\rangle=C_A {1\over 2}
\left\{\phi^\rho_N[\psi^\rho_{2M} \chi^\lambda_\half + 
\psi^\lambda_{2M} \chi^\rho_\half] 
+\phi^\lambda_N[\psi^\rho_{2M} \chi^\rho_\half 
- \psi^\lambda_{2M} \chi^\lambda_\half]\right\}\nonumber\\[0pt]
[20,1^+]:&&\vert N^2P_A(\half^+,\thalf^+)\rangle=C_A\psi^A_{1M}
{1\over \sqrt{2}}(\phi^\rho_N\chi^\lambda_\half-
\phi^\lambda_N\chi^\rho_\half),
\eeqa
where the SU(6) supermultiplet classification of these states is
listed for reference. The notation $(\half^+,\thalf^+,...)$ lists all
of the possible $J^P$ values from the ${\bf L}+{\bf S}$ coupling. 

In the case of states with unequal mass quarks the procedure for
constructing the basis proceeds in analogy to that outlined above,
except now the simpler requirement that the states are antisymmetric
under exchange of only the two equal-mass quarks is imposed.


\begin{references}
\bibitem{PLambda} If we impose SU(6), then we need a symmetrized
wavefunction for this $\Lambda$ state; since the flavor wavefunction
may be totally antisymmetric if it contains three distinguishable
symbols $u$, $d$, and $s$, this is 
\hfill \break 
$\vert
\Lambda^2P_M(\half^-,\thalf^-)\rangle=C_A\phi^A_\Lambda
(\psi^\rho_{1M}\chi^\lambda_\half-\psi^\lambda_{1M}\chi^\rho_\half)$.
\end{references}
\end{document}